\documentclass[fleqn,usenatbib]{mnras}
\usepackage{hyperref}
\usepackage[T1]{fontenc}
\usepackage{ae,aecompl}
\usepackage{graphicx}	% Including figure files
\usepackage{amsmath}	% Advanced maths commands
\usepackage{amssymb}	% Extra maths symbols
\usepackage{pdflscape}
\usepackage{xcolor}
\usepackage{mathrsfs}
\usepackage{soul}
\usepackage{longtable}

\newcommand{\pk}{\textit{Planck}}

\newcommand{\polang}{\psi}

\newcommand{\irht}{I_{\rm RHT}}
\newcommand{\angrht}{\theta}
\newcommand{\sigrht}{\sigma_{\theta}}
\newcommand{\angcat}{\theta_{\rm cat}}

\newcommand{\devrht}{\delta \bar{\theta}}
\newcommand{\devup}{\Delta \bar{\theta}_{\rm up}}
\newcommand{\devdown}{\Delta \bar{\theta}_{\rm down}}
\newcommand{\herschel}{\textit{Herschel}}
\newcommand{\angb}{\phi}
\newcommand{\sigangb}{\sigma_{\phi}}

\newcommand{\qbkg}{Q_{{\rm bkg}, i}}
\newcommand{\ubkg}{U_{{\rm bkg}, i}}
\newcommand{\qfil}{Q_{{\rm fil}, i}}
\newcommand{\ufil}{U_{{\rm fil}, i}}
\newcommand{\ndet}{137}
\newcommand{\nuni}{90}

\newcommand{\nhighstd}{26\%}
\newcommand{\deltanh}{\Delta N_{\rm H}}
\newcommand{\nhbkglim}{1.2 \times 10^{21}}
\newcommand{\nhbkghigh}{N_{{\rm H,bkg}}^{\rm high}}
\newcommand{\nhbkglow}{N_{{\rm H,bkg}}^{\rm low}}
\newcommand{\dnhlow}{\Delta N_{\rm H}^{\rm low}}
\newcommand{\dnhhigh}{\Delta N_{\rm H}^{\rm high}}
\newcommand{\dnhlim}{4 \times 10^{20}}
\newcommand{\nboot}{5,\!000}
\newcommand{\voldenslim}{1200 \, \rm{cm^{-3}}}
\newcommand{\voldenslimtwo}{2900 \, \rm{cm^{-3}}}
\newcommand{\filament}{filament}%$-$clump}

\title[Statistical analysis of the interplay between interstellar magnetic
fields and filaments]{Statistical analysis of the interplay between
interstellar magnetic fields and filaments hosting Planck Galactic Cold Clumps}
\author[D. Alina et al.]{
D. Alina$^{1,2}$\thanks{E-mail: dana.alina@nu.edu.kz}, 
I. Ristorcelli$^{2}$, 
L. Montier$^{2}$, 
E. Abdikamalov$^{1}$, 
M. Juvela$^{3}$, \newauthor
K. Ferri\`ere$^{2}$, 
J.-Ph. Bernard$^{2}$,   
and E. R. Micelotta$^{3}$
\\
$^{1}$ Department  of  Physics,  School  of  Science  and  Technology, Nazarbayev University, Astana 010000, Kazakhstan\\
$^{2}$ IRAP, Universit\'{e} de Toulouse, CNRS, UPS, CNES, Toulouse, France \\
$^{3}$ Department of Physics, PO Box 64, University of Helsinki, 00014, Helsinki, Finland
}

\date{Accepted 2019 February 15. Received 2019 March 14; in original form 2017 December 26}

\pubyear{2019}
\begin{document}
%\thepage
%\label{firstpage}
\maketitle

\begin{abstract}
We present a statistical study of the relative orientation in the plane of the sky between interstellar magnetic fields and filaments hosting cold clumps. 
For the first time, we consider both the density of the environment and the density contrast between the filaments and their environment. 
Moreover, we geometrically distinguish between the clumps and the remaining portions of the filaments.
We infer the magnetic field orientations in the filaments and in their environment from the Stokes parameters\footnote{Based on observations obtained with Planck (\url{http://www.esa.int/Planck}),an ESA science mission with instruments and contributions directly fundedby ESA Member States, NASA, and Canada.}, assuming optically thin conditions.
Thus, we analyze the relative orientations between filaments, embedded clumps, and internal and background magnetic fields, depending on the filament environment and evolutionary stages.
We recover the previously observed trend for filaments in low column density environments to be aligned parallel to the background magnetic field;
however, we find that this trend is significant only for low contrast filaments,  whereas high contrast filaments tend to be randomly 
orientated with respect to the background magnetic field.
Filaments in high column density environments do not globally
show any preferential orientation, although low contrast filaments alone tend to have perpendicular relative orientation with respect
to the background magnetic field.
For a subsample of nearby filaments, for which volume densities can be derived, we find a clear transition in the relative orientation
with increasing density, at $n_{\rm H} \sim 10^{3}~{\rm cm}^{-3}$, 
changing from mostly parallel to mostly perpendicular in the off-clump portions of filaments 
and from even to bimodal in the clumps.
Our results confirm a strong interplay between interstellar magnetic fields 
and filaments during their formation and evolution.
\end{abstract}

%%%%%%%%%%%%%%%%%%%%%%%%%%%%%%%%%%%%%%%%%%%%%%%%%%

\begin{keywords}
ISM: dust, magnetic fields -- polarization -- methods: statistical
\end{keywords}

%%%%%%%%%%%%%%%%%%%%%%%%%%%%%%%%%%%%%%%%%%%%%%%%%%

%%%%%%%%%%%%%%%%% BODY OF PAPER %%%%%%%%%%%%%%%%%%

\section{Introduction}

First observed in extinction by interstellar dust particles \citep{schneider1979} and then in dust and CO emission \citep{abergel1994,falgarone2001}, filamentary structures in molecular clouds recently became central to many studies. 
There is growing evidence that filaments play a fundamental role in the onset of star formation for low and high-mass stars. 
The unprecedented angular resolution and sensitivity of the Herschel \citep{pilbratt2010} maps of dust emission in the far-infrared (FIR) wavelength range made it possible to discover a network of filamentary structures ubiquitous in a wide range of environments of the ISM \citep{molinari2010,motte2010,menshchikov2010, miville-deschenes2010,juvela2012}. 
Moreover, it was shown that prestellar cores and protostars form primarily in the densest and gravitationally-bound filaments \citep{andre2010, polychroni2013, konyves2015}. 
Investigating the origin and evolution of filaments is then crucial to better understand the early stages of star formation. 

MHD simulations show that a hierarchy of sheets and filaments can form as a result
of shock-compression and shear flows in supersonic turbulence, at the same time as the collapse or fragmentation of gravitationally unstable structures \citep{andre2014,li2014}. 
Magnetic fields are believed to play a key role in the formation of structures, but their interplay with turbulence and self-gravity still needs to be better understood at different spatial scales and evolutionary stages. 

Observations of starlight polarization revealed a connection between the magnetic
field geometry and the dense filamentary structures, finding predominantly
perpendicular relative orientations between the mean field and the long axes of the filaments \citep{goodman1990, pereyra2004, franco2010, chapman2011, alves2008}.
Studies of the magnetic field orientations derived from observations of 
starlight polarization are however limited to small regions and depend on the background star distribution. 
Alternatively, measurements of dust polarized thermal emission can also be used to
trace the magnetic field geometry and allow us to probe denser media. 
This method was shown to be the most efficient way to reveal magnetic field structures in molecular clouds \citep{crutcher2012}.
Such observations toward small bright regions of molecular clouds are performed from ground-based telescopes and their high angular resolution is well suited to studies of prestellar cores \citep{dotson2000, matthews2009, ward-thompson2000, tang2009, tassis2009,pattle2017,liu2018}.
Massive dense cores with distances of a few kiloparsecs have been resolved using interferometric observations such as SMA, revealing the structure of the magnetic field at 0.1-0.01 pc scales \citep{zhang2014, koch2014, ching2017}. 
The major axes of dense cores are found to be either parallel or perpendicular to
the mean magnetic field at parsec scales, although at much smaller scale, typical of accretion disks, the relative orientation appears more random. 
These studies support the hypothesis that magnetic fields play an important role during the collapse and fragmentation of filaments and clumps, and the formation of dense cores. 
At smaller scales, kinematics arising from gravitational collapse would dominate over magnetic fields.

At the scales of filamentary structures, parallel relative orientation was 
observed between the mean magnetic field and either low-density sub-filaments \citep{sugitani2011} or the striation features revealed with Herschel in Taurus \citep{palmeirim2013}, Musca \citep{cox2016}, and in the high-Galactic latitude cloud L1642 \citep{malinen2016}. 
A bimodal distribution of relative orientations was both observed in filamentary molecular clouds of the Gould Belt \citep{li2013} and predicted from MHD simulations \citep{nakamura2008, soler2013, li2016}.

The {\pk}\footnote{Planck (\url{http://www.esa.int/Planck}) is an ESA science mission with instruments and contributions directly funded by ESA Member States, NASA, and Canada.}
satellite survey has recently provided the first all-sky map of dust polarized emission \citep{planck2014-XIX, planck2018-XII}. 
With unprecedented sensitivity measurements at 353 GHz and a resolution of
$5\arcmin$, these data are uniquely suited to probe the dust polarized emission over a large range of column densities, at intermediate scales between molecular clouds and cold cores. 
Studying the relative orientation between the magnetic field and the column density structures at $15\arcmin$ resolution and over most of the sky, \cite{planck2014-XXXII} found that most of the elongated structures in the diffuse ISM have preferential parallel relative orientation with the magnetic field, while perpendicular structures appear in molecular clouds. 
In a detailed study of ten nearby molecular clouds of the Gould Belt,
\cite{planck2014-XXXV} showed that, as gas column density increases, the relative
orientation changes from mostly parallel or having no preference to mostly perpendicular at high column densities.     
A similar trend was found in the Vela C molecular cloud observed with BLAST-POL at higher angular resolution ($3\arcmin$), with a sharp transition in $N_H$ characterizing the change of relative orientation \citep{soler2013}. 
Simulations show that such a change could be related to the degree of magnetization of the cloud, particularly when magnetic energy is in equipartition with turbulent energy  \citep{hennebelle2013, soler2017, chen2016}. 
In a detailed study of three nearby filaments, \cite{planck2014-XXXIII} showed
that the mean magnetic fields inside the filaments are
coherent along their lengths and do not have the same orientations as in the background. 
Moreover, the relative orientation between magnetic fields inside the filaments and in the 
background was shown to be different in the three studied cases. 
All these studies suggest that the magnetic field must play an important role in the assembly of interstellar matter and the formation of dense structures. 
Further statistical observations are required to characterize the range of scales
and densities for which the magnetic field has a significant impact.
To better understand the early stages of star formation, it is particularly important to study whether and how the properties of the ambient magnetic field are related to the filament evolution and prestellar core formation.

The Planck catalogue of Galactic Cold Clumps is built from the {\pk} full all-sky survey \citep{planck2016-XXVIII}. 
The catalogue contains more than 13,000 cold sources distributed across the whole sky, from the Galactic plane to high latitudes, and are mainly situated in molecular clouds. 
The sources span a broad range in temperature, mean density, mass, and size. 
Because of the limited angular resolution, the most compact and nearby sources have a linear diameter of $\sim 0.1$ pc, though at large distances, many sources have an intrinsic size of tens of parsec. 
As revealed by higher angular resolution observations performed with
Herschel follow-up on a representative subsample of 350 Planck Galactic Cold
Clumps (PGCC) targets, these clumps are characterized by the presence of substructures.
The clumps often contain colder and denser individual starless or prestellar cores and young protostellar objects still embedded in their colder surrounding cloud \citep{juvela2010,juvela2011,Juvela2012a,montillaud2015,planck2011-7.7a}. 
As shown with a statistical analysis based on molecular dense tracers, most of the {\pk} clumps correspond to an early stage of star formation \citep{wu2012, liu2014, tatematsu2017,liu2017,liu2016}. 
Furthermore, it was found that a substantial fraction of them are embedded in filamentary structures \citep{juvela2012}. 

In the present work, our goal is to investigate the possible interplay between magnetic fields and filaments hosting PGCCs. 
Having selected a sample of these filaments surrounded by an ordered magnetic
field, we perform a statistical analysis of the relative orientations in the plane of the sky between the filaments, their embedded clumps, and the magnetic fields inside and
around the filaments.

This paper is organized as follows. 
In Section~\ref{sec:data}, we present the {\pk} data, the PGCC catalogue and a description of the methods used for the filament detection and the separation of the polarization Stokes parameters ($I$, $Q$, $U$) between those intrinsic to the filaments and those of their surroundings. 
In Section~\ref{sec:fil_selection}, we present the properties of the filaments selected for this statistical study, and the subsamples built according either to the column densities 
of the filament environment or to the density contrast of the filaments. 
In Section~\ref{sec:results} we describe the results obtained on the relative orientation between the
filamentary structures and the magnetic fields in the background and inside the filaments. 
We discuss the implications of these results in Section~\ref{sec:discussion}.
Finally, Section~\ref{sec:conclusion} presents our conclusions and perspectives.

%%%%% ========== %%%%%
\section{Data and methodology}
%%%%% ========== %%%%%
\label{sec:data}
%--------------------------------------------------------------------------%
\subsection{{\pk} data used and PGCC catalogue}
%--------------------------------------------------------------------------%
In our analysis, we use the {\pk} $353$ GHz full mission data, which is a part of the Planck Product 2015 release.
In order to improve the signal-to-noise ratio (SNR), the data is smoothed from the
nominal resolution of the maps ($5'$) to a resolution of $7'$. 
The smoothing is performed using a convolution with a Gaussian kernel of $5'$. 

The {\pk} all-sky Galactic Cold Clump catalogue \citep{planck2016-XXVIII} contains 13,188 sources detected using a dedicated method  \citep{montier2010,planck2011-7.7b}.
The method consists of a color-based detection of the excess emission in the cold residual maps after subtraction of the warm component traced by IRAS 100\,$\mu$m observations.
Clumps were fitted using elliptical Gaussian functions, and the catalogue contains
the resulting semi-major and semi-minor axes and the position angles of the ellipses.
With the exception of some sources in the Large Magellanic Cloud and in the Small Magellanic Cloud, the catalogue excludes extragalactic sources \citep{planck2016-XXVIII} and contains nearby cores, clumps, and distant molecular clouds.
This makes the catalogue particularly useful in studies related to the early stages of star formation.
The {\pk} data are in HEALPix format, with a resolution parameter 
$N_{\rm side} =  2048$.
We extract small $2\degr$ square maps centered on the positions of the clumps from the catalogue.
Such maps are called minimaps hereafter. 
The pixel size is 1'. The Shannon-Nyquist criterium for sampling is thus satisfied. The slight oversampling does not affect our further analysis as it is based on histograms, and the uncertainties are estimated for independent pixels.

We select cold clumps from the PGCC catalogue using the following criteria:
\begin{itemize}
\item Clumps are within the Galactic latitude range $2{\degr} \le |b| \le 60{\degr}$ 
to avoid confusion along the line of sight (LOS) in the Galactic plane.
\item Clumps are detected with accurate flux density estimates in the {\pk} 545 and 857 GHz channels as well as in the IRIS 3 THz band.
This corresponds to the PGCC catalogue flag {\tt FLUX\_QUALITY=1}.
We also include fainter sources: clumps for which there is no detection in the IRIS 3 THz band (flag {\tt FLUX\_QUALITY=2}).
\item Clumps are detected with a SNR $\ge 8$ in the three highest {\pk} frequency bands (857, 545 and 353 GHz).
\item Clumps have an average SNR on the polarized intensity ($P = \sqrt{Q^2+U^2}$) SNR($P$) $\ge 2$.
We use SNR($P$) as it reflects the joint SNR of the Stokes parameters $Q$ and $U$ used in the estimation of the polarization angle (see Equation~\ref{eq:psi}).
SNR($P$) is estimated as $P/\sigma_P$, where $\sigma_P = \sqrt{\sigma_Q \, \sigma_U}$.
Such an estimate of the uncertainty in the polarized intensity 
was discussed by \cite{Montier1}.
SNR($P$) is calculated in a disk of 10$\arcmin$ radius around the position of the
clumps and in an annulus with inner and outer radii of $30\arcmin$ and $50\arcmin$. 
We require $\left\langle P/\sigma_P\right\rangle \ge 2$ in both regions.
\end{itemize}
This yields a selection of 3,743 clumps.

%--------------------------------------------------------------------------%
\subsection{Filament detection method}
\label{sec:method}
\begin{figure}
\begin{center}
\includegraphics[width= 4.5 cm]{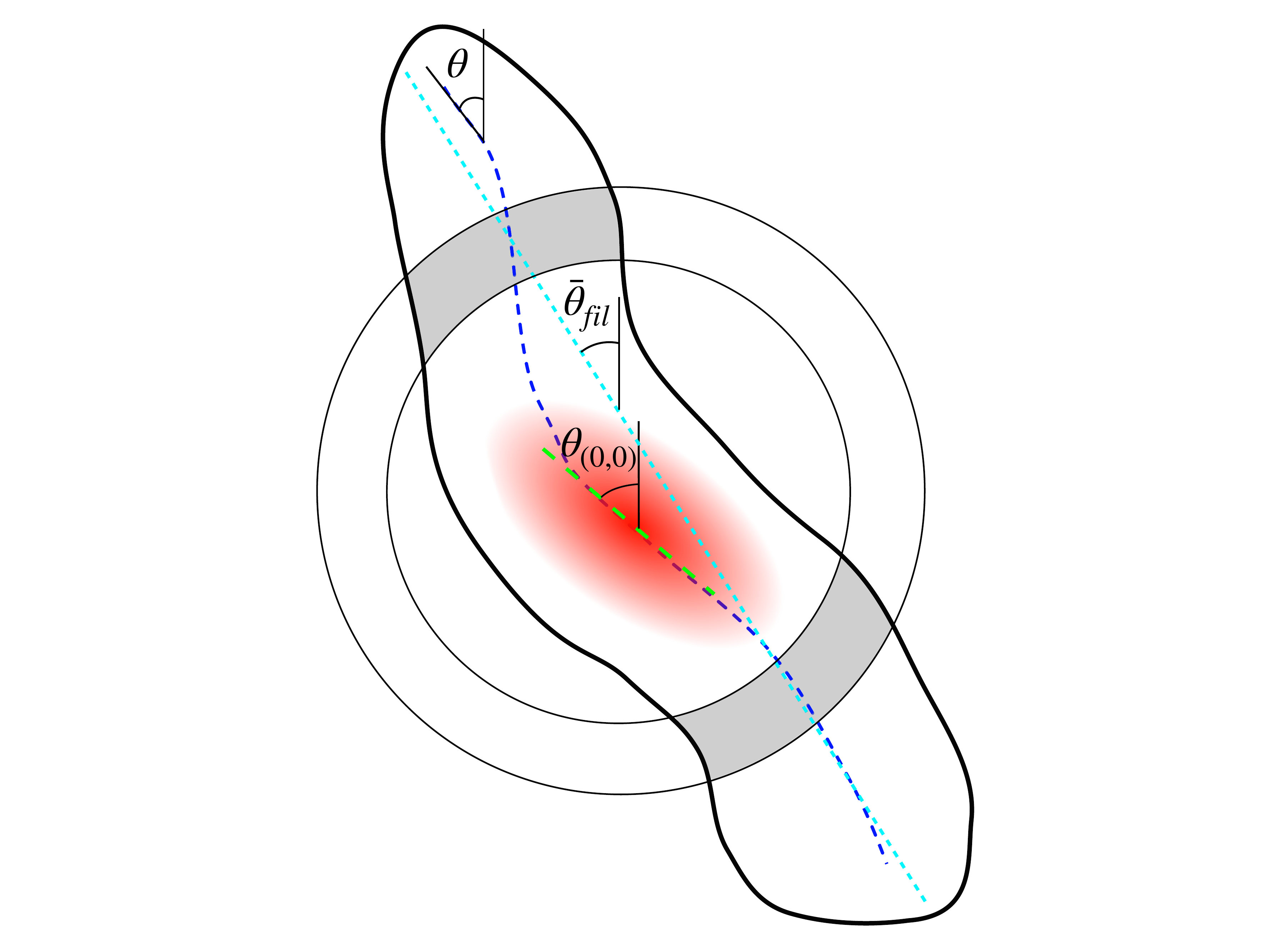}
\caption{
Schematics showing a filament with an embedded clump.
The red area shows the clump ellipse. 
The green and light blue dashed lines show 
the SupRHT orientation at the center of the clump (given by $\angrht_{0,0}$)
and the average orientation of the filament 
(given by $\bar{\angrht}_{\rm fil}$), respectively. 
The angles $\devup$ and $\devdown$, defined in the main text,
are computed over the gray areas.}
\label{fig:scheme1}
\end{center}
\end{figure}

%--------------------------------------------------------------------------%
Several methods exist to detect filamentary structures \citep{soler2013,sousbie2011,planck2013-p06b}. 
Among those, we opted for the SupRHT method, which is an improved version of the Rolling Hough Transform proposed by \cite{clark2014}.
It is a machine vision method that allows one to determine whether each pixel in an image belongs to a curve of a given shape.
First, a map is smoothed using a two-dimensional "top-hat" kernel. 
Second, a bitmask is built by subtracting the smoothed map from the original one and by thresholding the result at 0.
Then, an input kernel of a given shape is applied.
For each orientation of the input kernel, the method counts the number of pixels falling inside the kernel. 
The orientation angle $\angrht$ in each pixel is defined as the average
value of the SupRHT histogram after thresholding above a given value, set to $0.7$ in our study.
$\angrht$ is counted positively in the direction from North to East and ranges between $-90\degr$ and $90\degr$ (cf. Figure~\ref{fig:scheme1}).
All the angles used in our analysis follow the same convention.
The intensity $\irht$ is defined as the sum of the histogram over all orientations.
In the case where no kernel-like structure is detected at a given position, $\irht = 0$.
The main advantage of this method is that it is independent of the intensity of the structure with respect to the background intensity, as a top-hat filtering is performed before calculations.
Another advantage is that it provides an estimate of the uncertainty in the angle ($\sigrht$).
A disadvantage of the method is that it detects only structures with a shape similar to the input kernel.

We use the 353 GHz {\pk} minimaps, which we convolve with a bar kernel of length $k_{\rm s} = 31'$ and width $k_{\rm w} = 6'$ as an input to the method.
The selection of the kernel's dimension is described in Appendix \ref{sec:app_kernel}.   

We run the intensity minimaps centered on the pre-selected PGCCs through the SupRHT procedure.
The raw SupRHT maps contain several overlapping filamentary structures. 
We isolate only the filamentary structure associated with the PGCC in the center of each minimap.
Hereafter, we refer to such filamentary structures as filaments.
We build a mask to isolate the clump ellipses defined by the semi-major and
semi-minor axes provided in the PGCC catalogue and the position angles
$\angrht_{0,0}$ in the center of the clumps.
We also build a mask to separate the area covered by clumps (which we refer to as the clump area) 
from the area covered by the rest of the filament (which we refer to as 
the {\filament} area).
The number of pixels in a filament ranges from $\sim180$ to $\sim1200$ with an average of $\sim500$.
The number of pixels in a clump ranges from $\sim50$ to $\sim500$ with an average of $\sim 200$, as shown in the histograms in Figure \ref{fig:pix_hist}.

For the clump in the center of each minimap, we fit the histograms of the orientation angle, $\angrht$, inside the clump ellipse 
with a Gaussian function. 
We normalize the histograms to one and call
them distribution functions (DFs).
We derive the position of the peak, $\bar{\theta}_{\rm clump}$, 
and the standard deviation, $\sigma'_{\theta, \, {\rm clump}}$, of the DFs.
The reason why we use a Gaussian fit is to obtain a good approximation
of the average orientation angle over most pixels in the clump,
%%of the average orientation angle in the clump,
that is not affected by the possible presence of pixels with a separate distribution.
We define $\devrht$ as the difference between $\bar{\theta}_{\rm clump}$ and the SupRHT
orientation angle at the center of the clump, 
$\angrht_{0,0}$.
The parameters $\sigma_{\theta, \, {\rm clump}}$ and $\devrht$ allow us to ensure
that the angle $\angrht$ does not change much over the area of the clump ellipse,
so that the pixels belong to the same filament.

We also calculate the average angles, 
$\bar{\theta}_{\rm up}$ and $\bar{\theta}_{\rm down}$, 
in the two regions of the filament contained between two circles centered on 
the clump center and with radii equal to 1.5 and 2 times the clump's semi-major axis 
(see gray areas in Figure \ref{fig:scheme1}).
The absolute differences between these angles and $\angrht_{0,0}$
are called $\devup$ and $\devdown$.
These parameters allow us to control the apparent length of the filaments and to put constraints on their curvature.

We apply the following criteria to the above parameters for the detection of linear filaments in our study:
\begin{itemize}
\item $\irht \neq 0$ at the center of each minimap;
\item $\sigrht < 1{\degr}$ at the center 
of each minimap;
\item $\devrht < 5{\degr} $;
\item $\sigma'_{\theta,clump} < 5{\degr} $;
\item $\devdown < 15{\degr} $ and $\devup < 15{\degr} $;
\item if a filament contains several PGCCs and thus is detected several times, 
we retain only one minimap, 
but we consider the contributions of all the PGCCs in the filament.
\end{itemize}
This yields a pre-selection of $\ndet$ fields.

%--------------------------------------------------------------------------%
\subsection{Coherence of the background polarization angle}
\label{sec:uniformity}
%--------------------------------------------------------------------------%
The polarization signal towards a filament comes from the filament itself and both the background and foreground interstellar medium.
For convenience, we call the background and foreground contributions as the "background".
We separate the background contribution from the observed signal using the following simple approach.
We assume that the emission observed in the immediate surrounding of the filament is the same as the background emission at the location of the filament.
Thus, the total signal in a pixel $i$ towards the filament is
\begin{equation}
X_i = X_{\rm fil,i}+X_{\rm bkg,i} \, ,
\label{eq:bkg_contribution}
\end{equation}
where $X_i = \{I_i, \, Q_i, \, U_i\}$ and the "fil" and "bkg" subscripts stand for "filament" and "background" respectively.
Such an approach is only valid when the background is the same on both sides of a filament and for optically thin media which is the case at the {\pk} 353 GHz channel.

\begin{figure*}
\center
\begin{tabular}{cc}
\includegraphics[width=4.7cm]{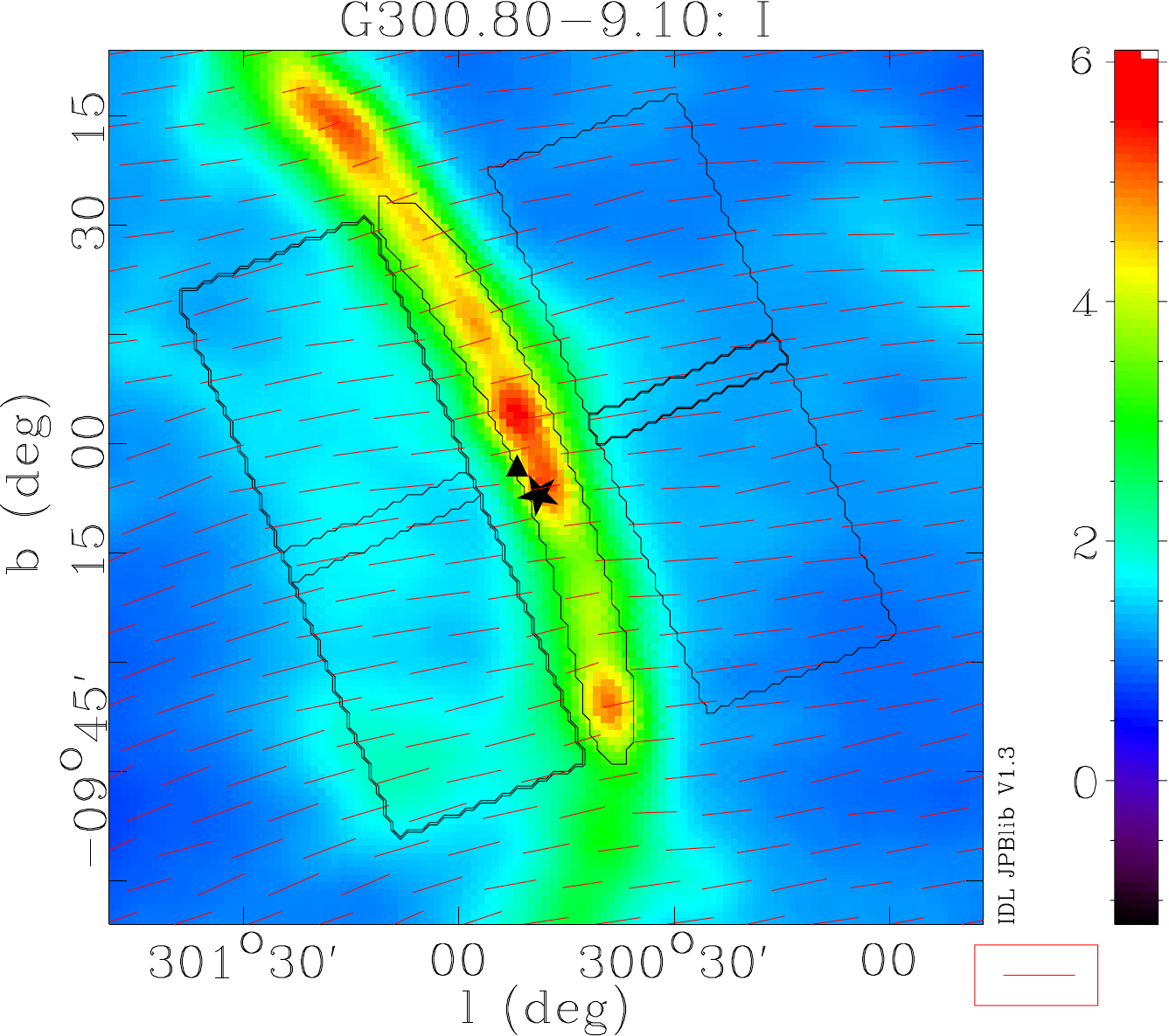}&
\includegraphics[width=6cm]{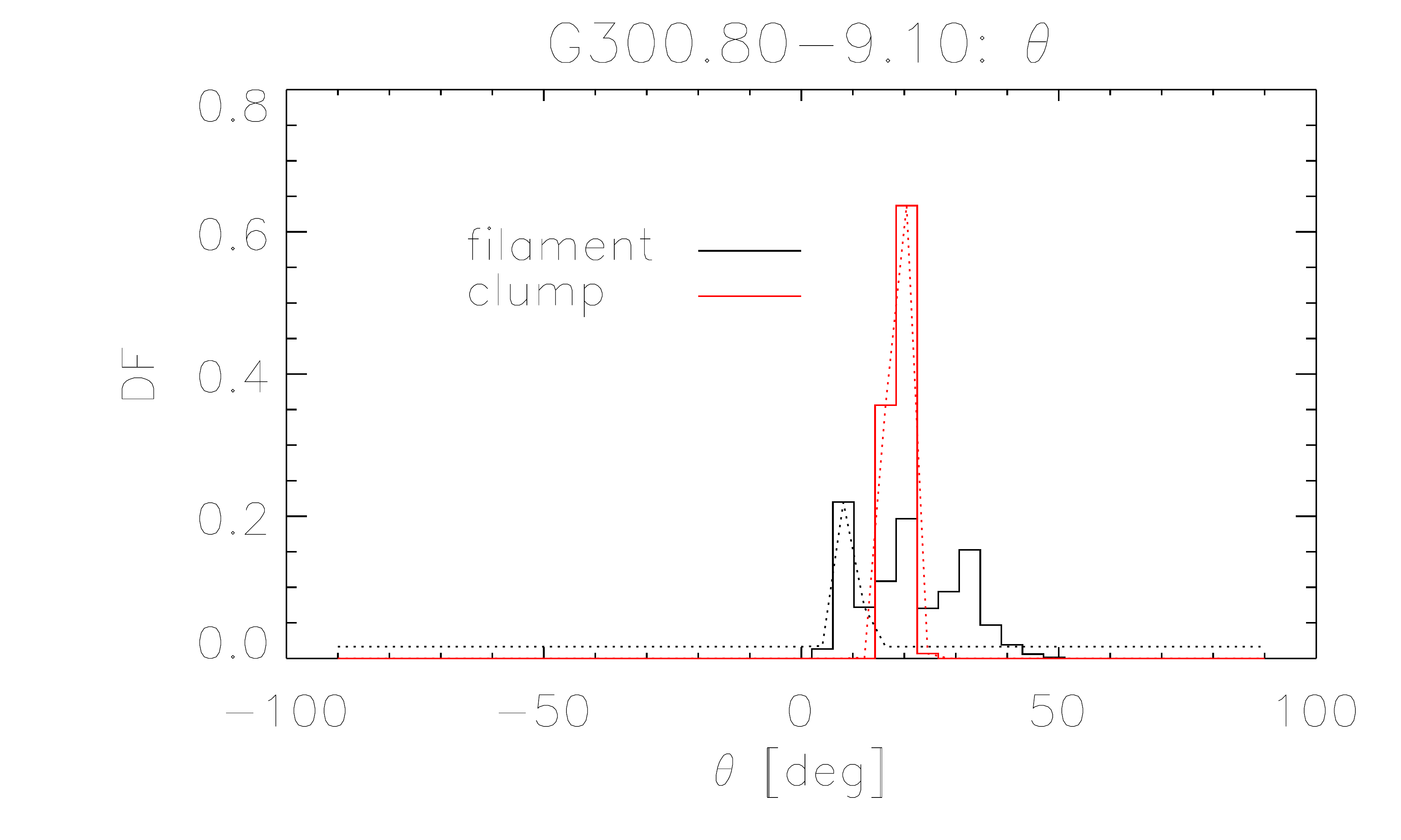} \\ 

\includegraphics[width=4.7cm]{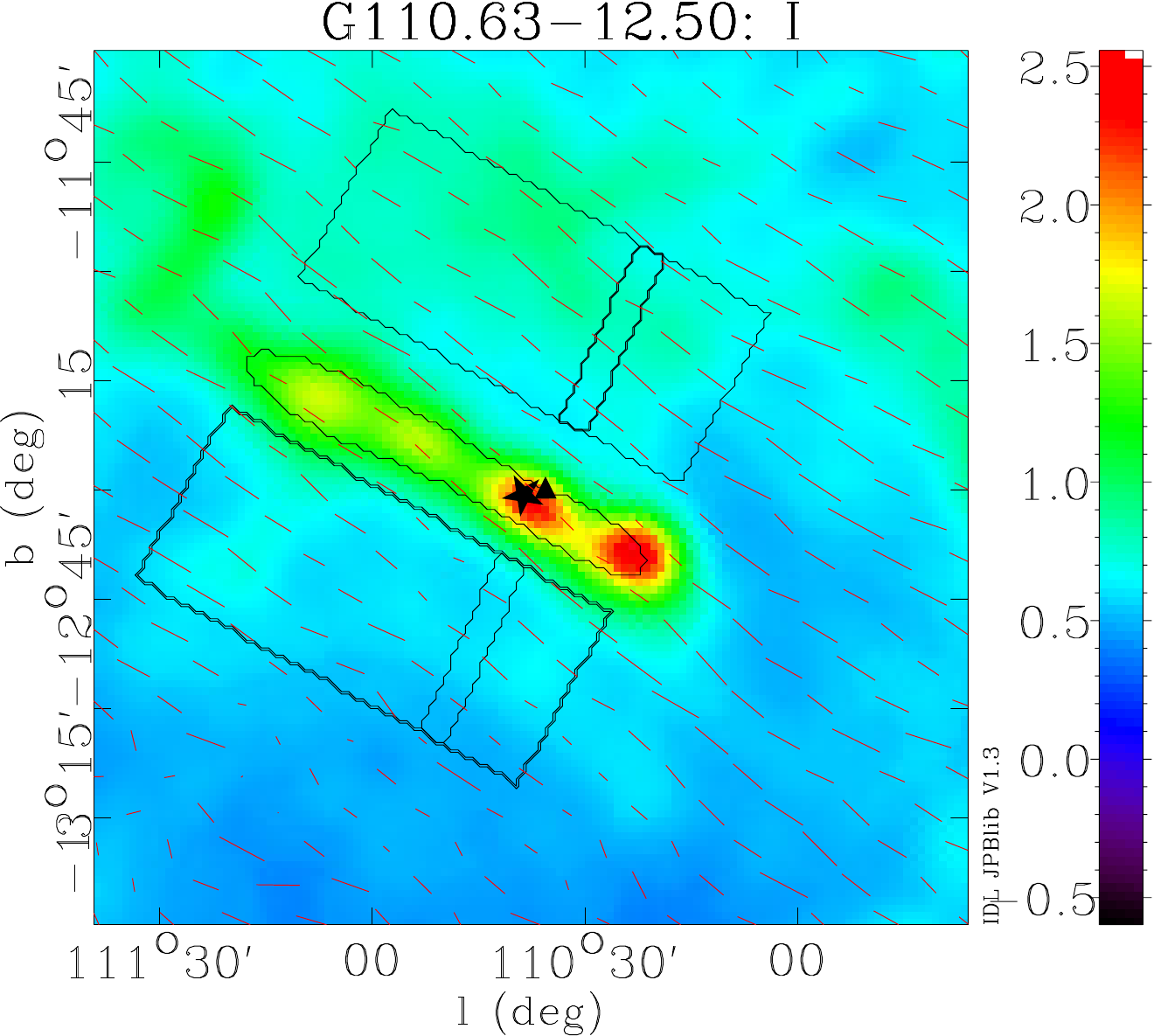} &
\includegraphics[width=6cm]{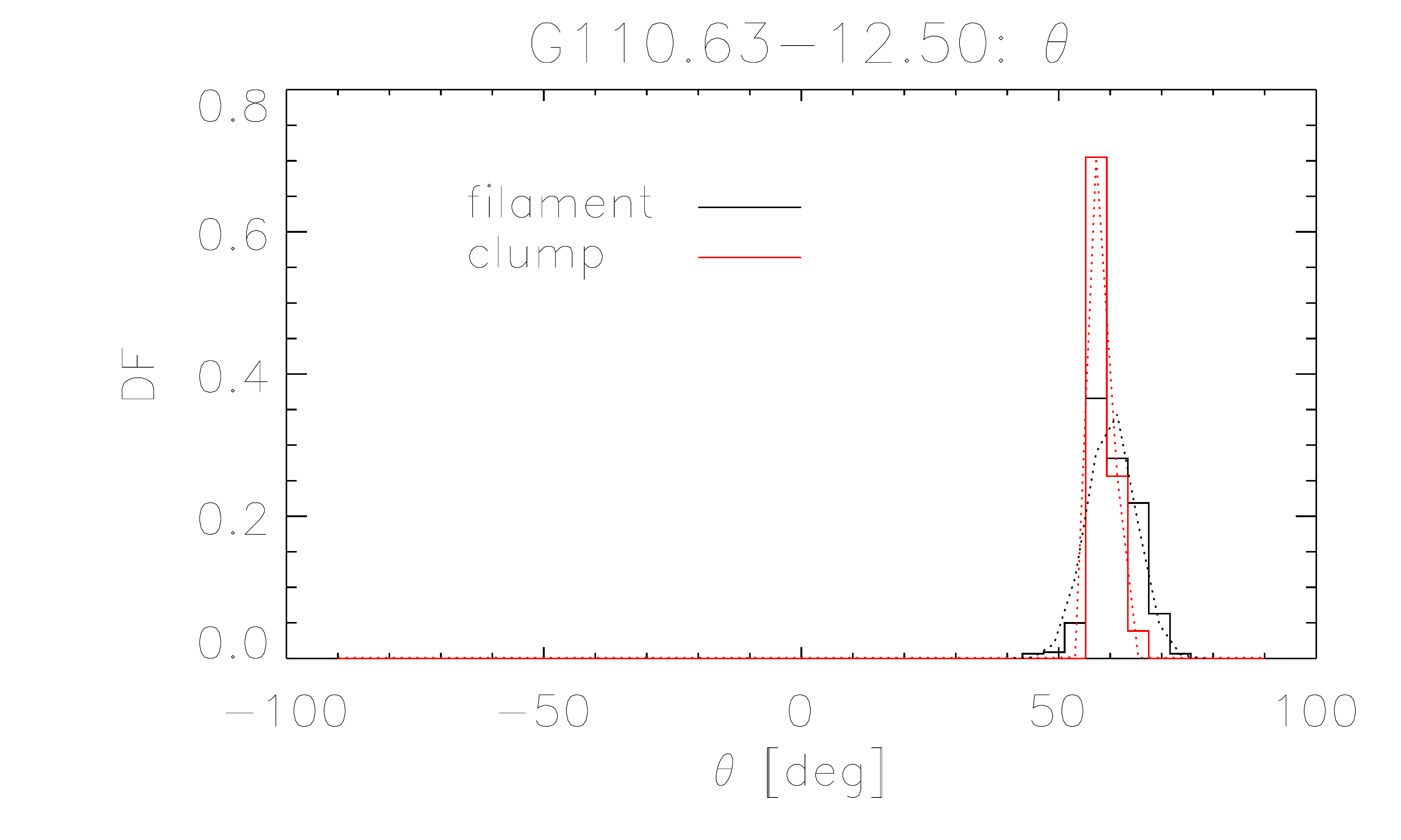} \\ 

\includegraphics[width=4.7cm]{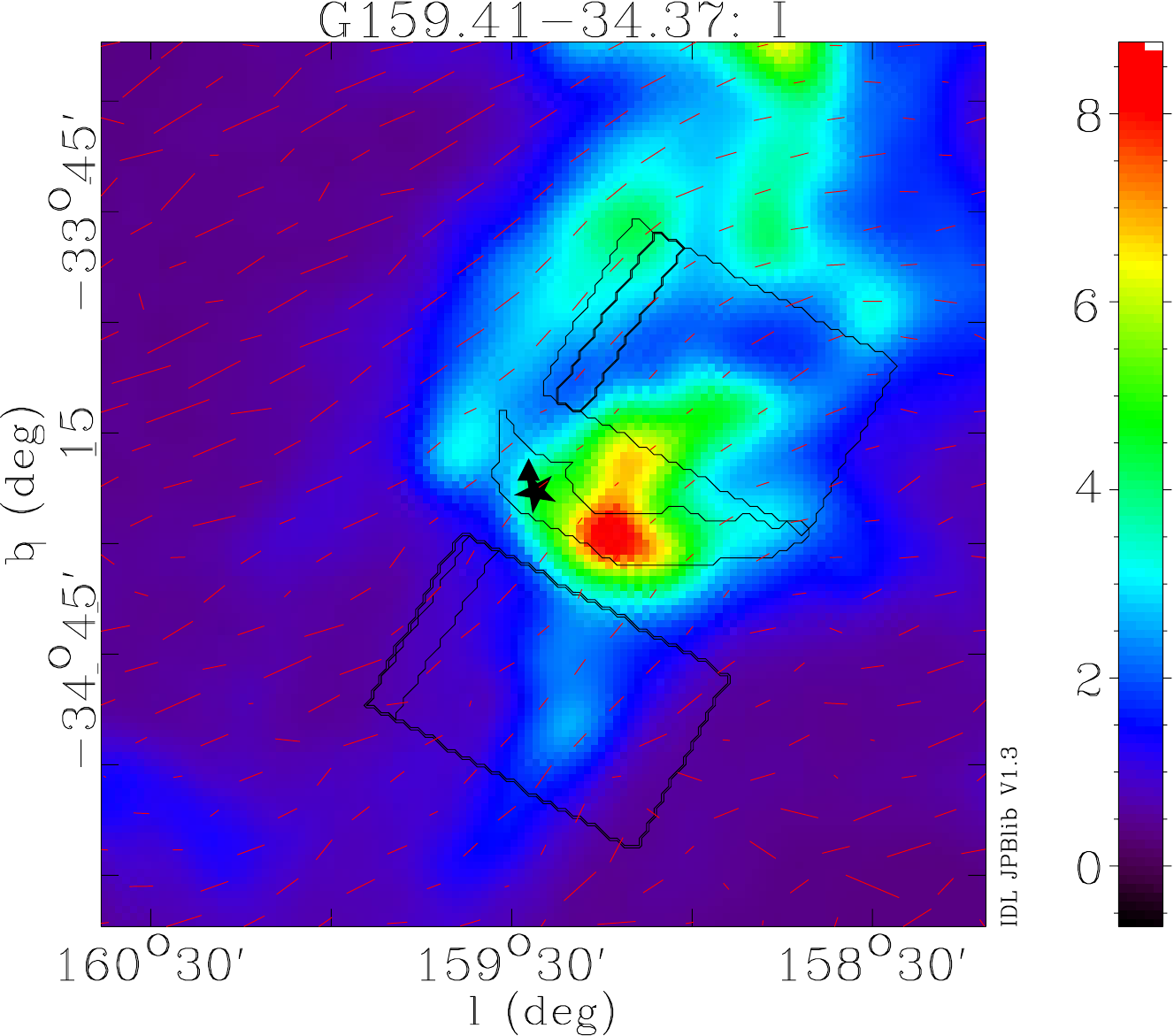} &
\includegraphics[width=6cm]{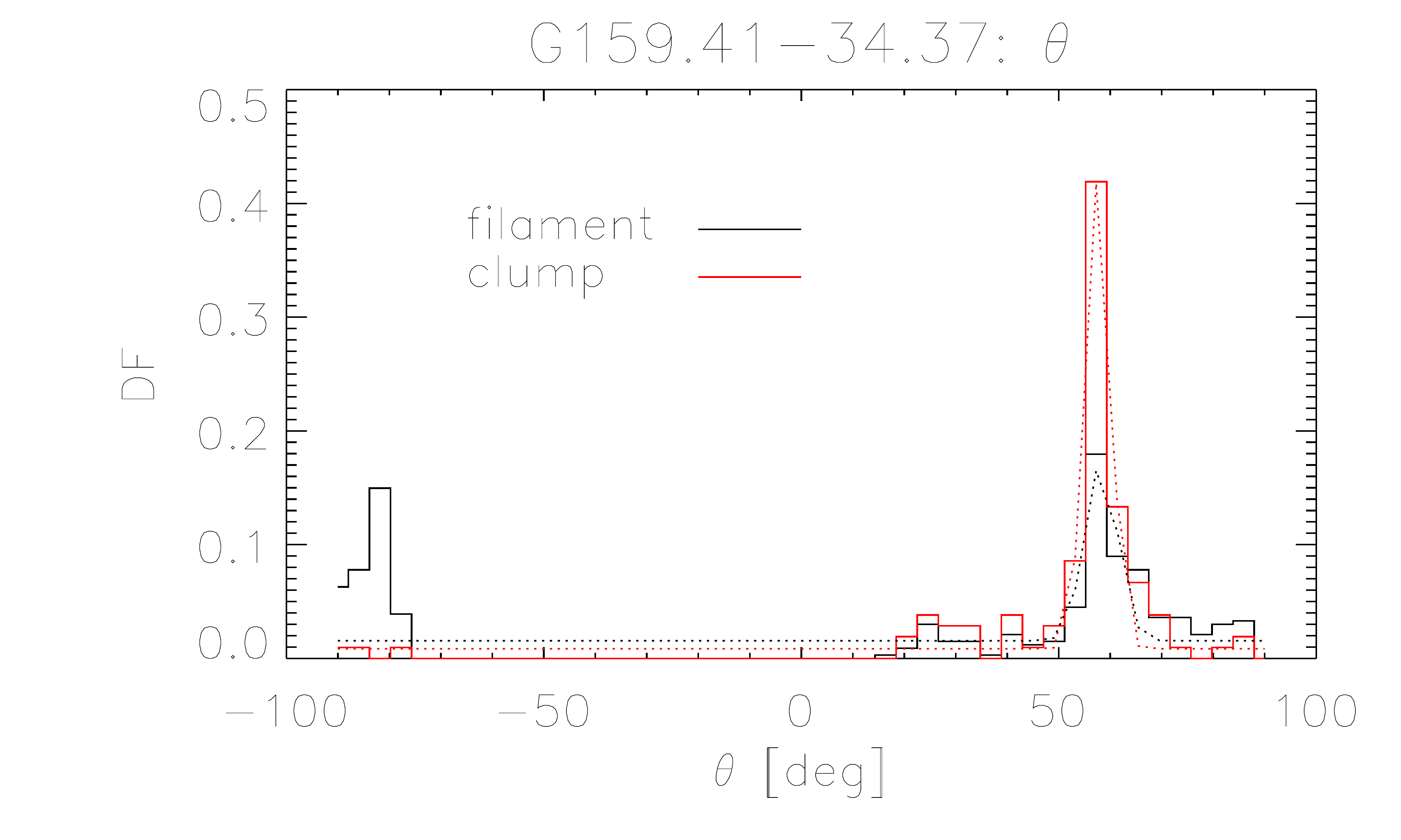} \\ 

\includegraphics[width=4.7cm]{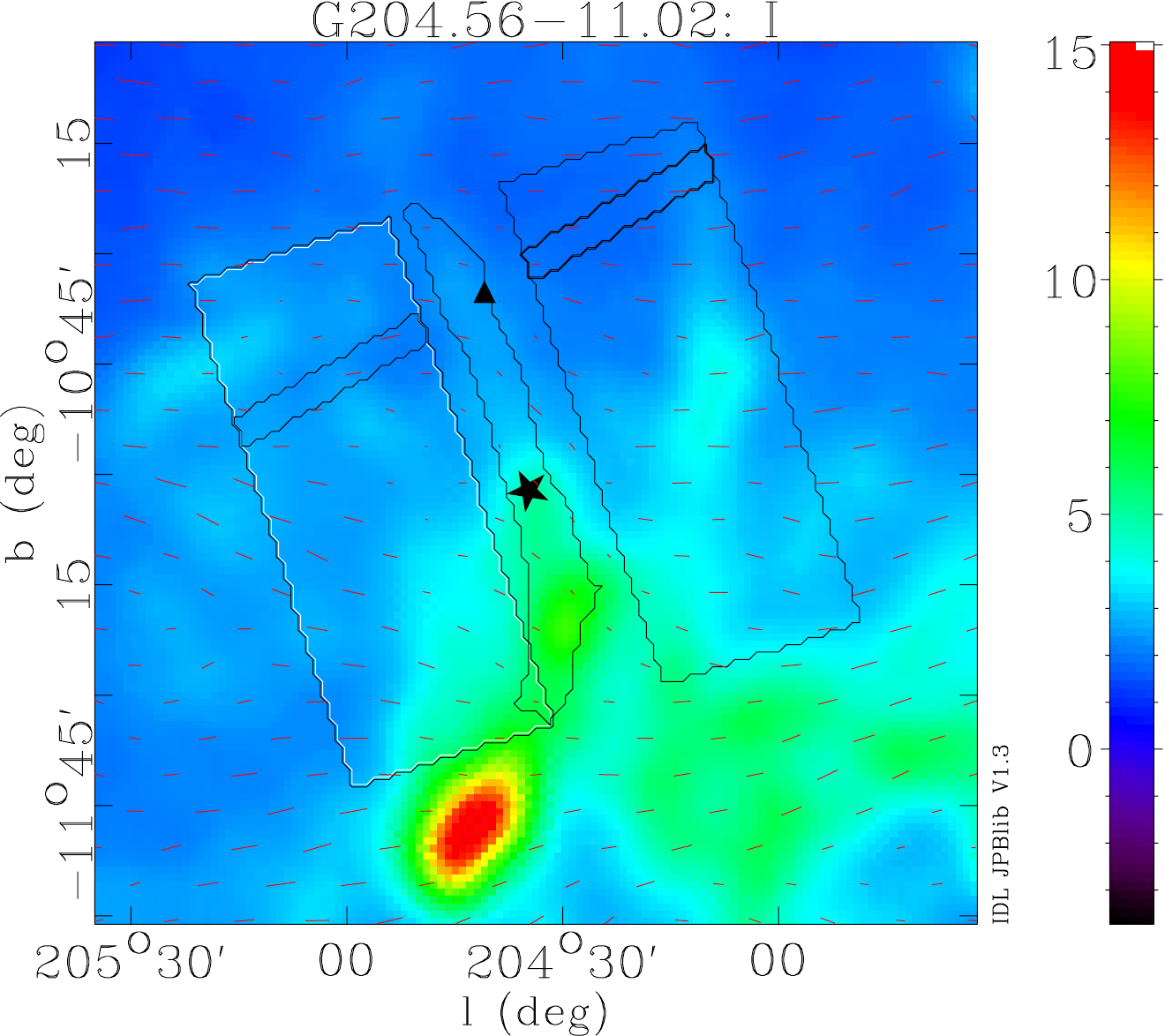} &
\includegraphics[width=6cm]{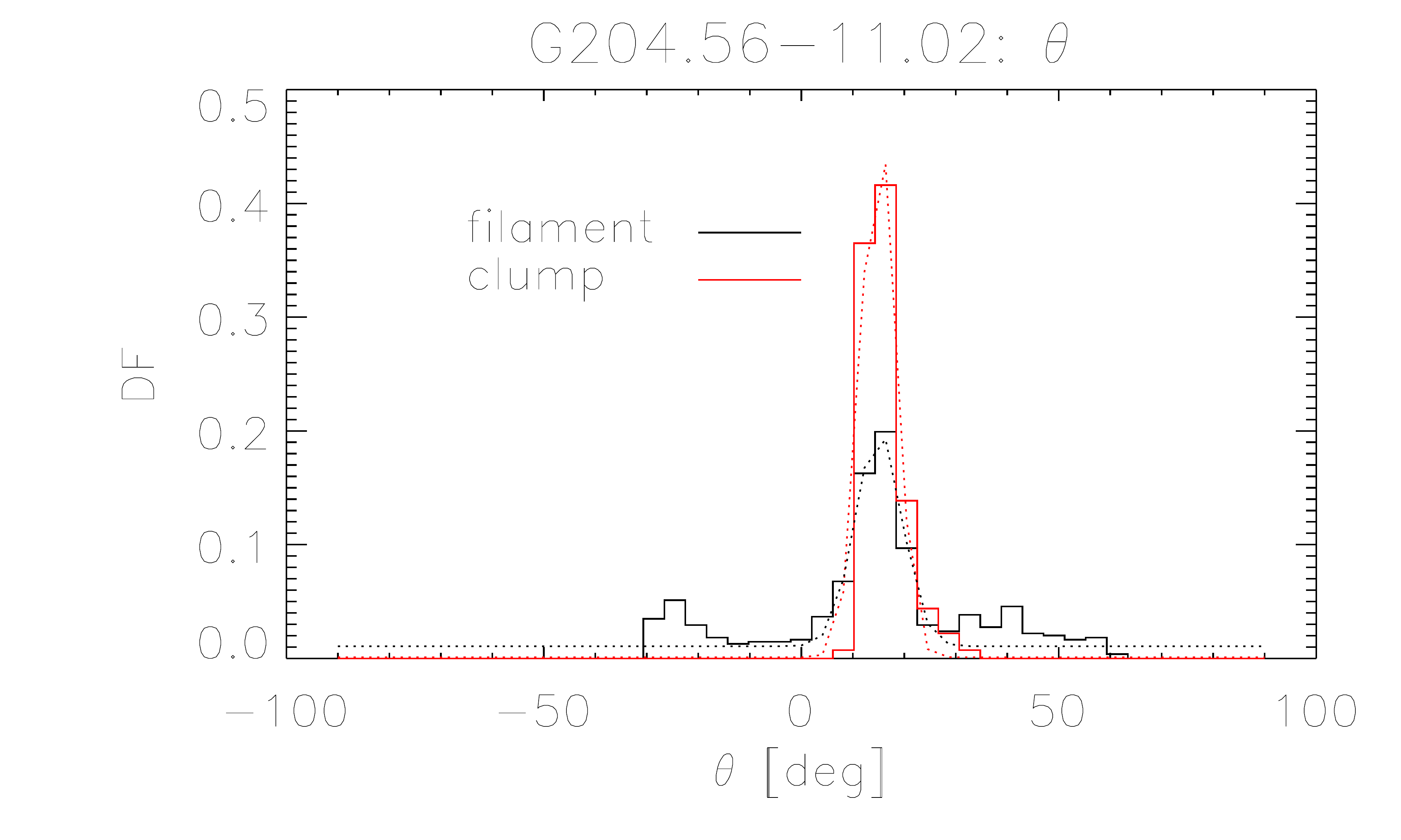} \\ 

\includegraphics[width=4.7cm]{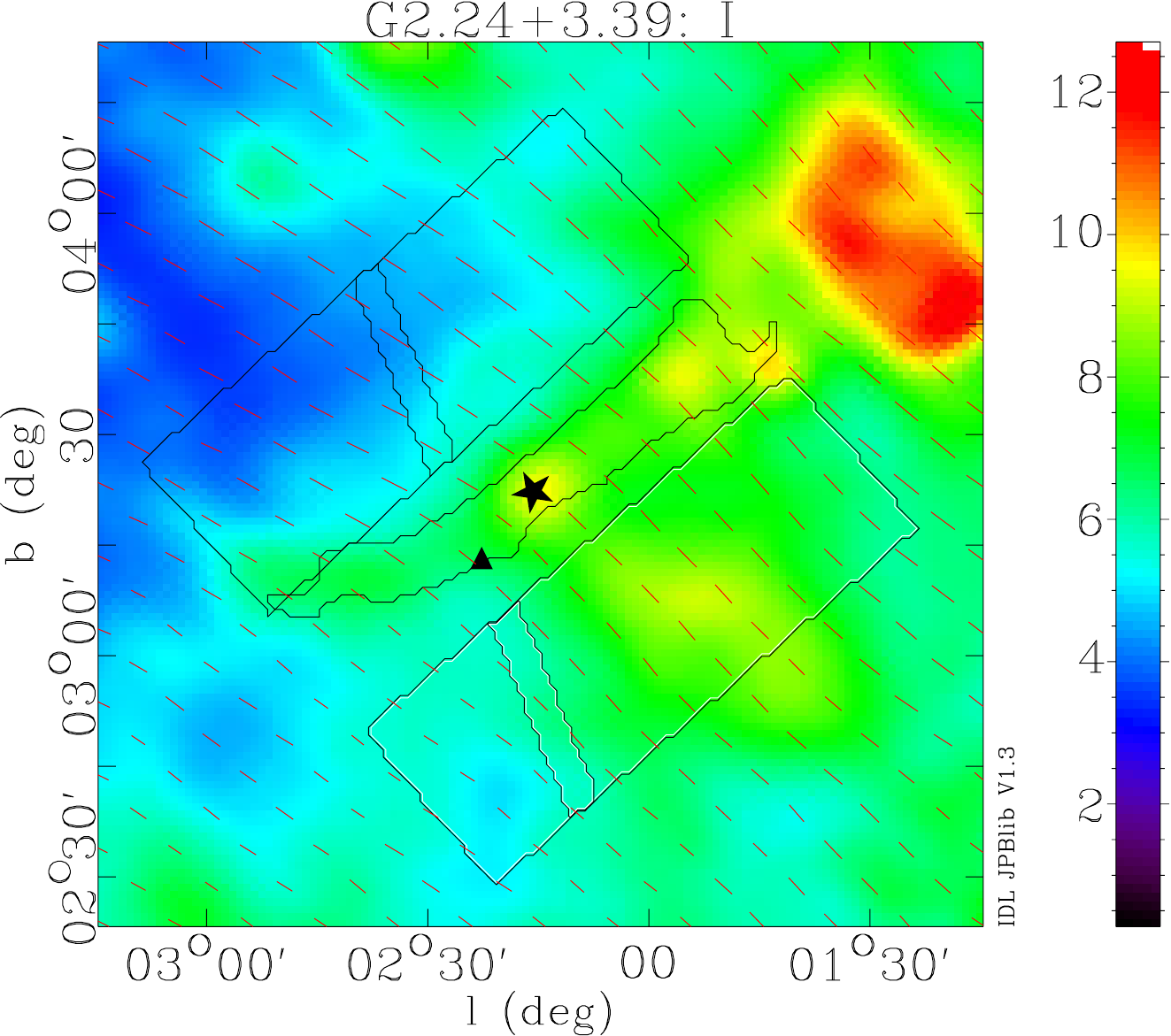} &
\includegraphics[width=5.4cm]{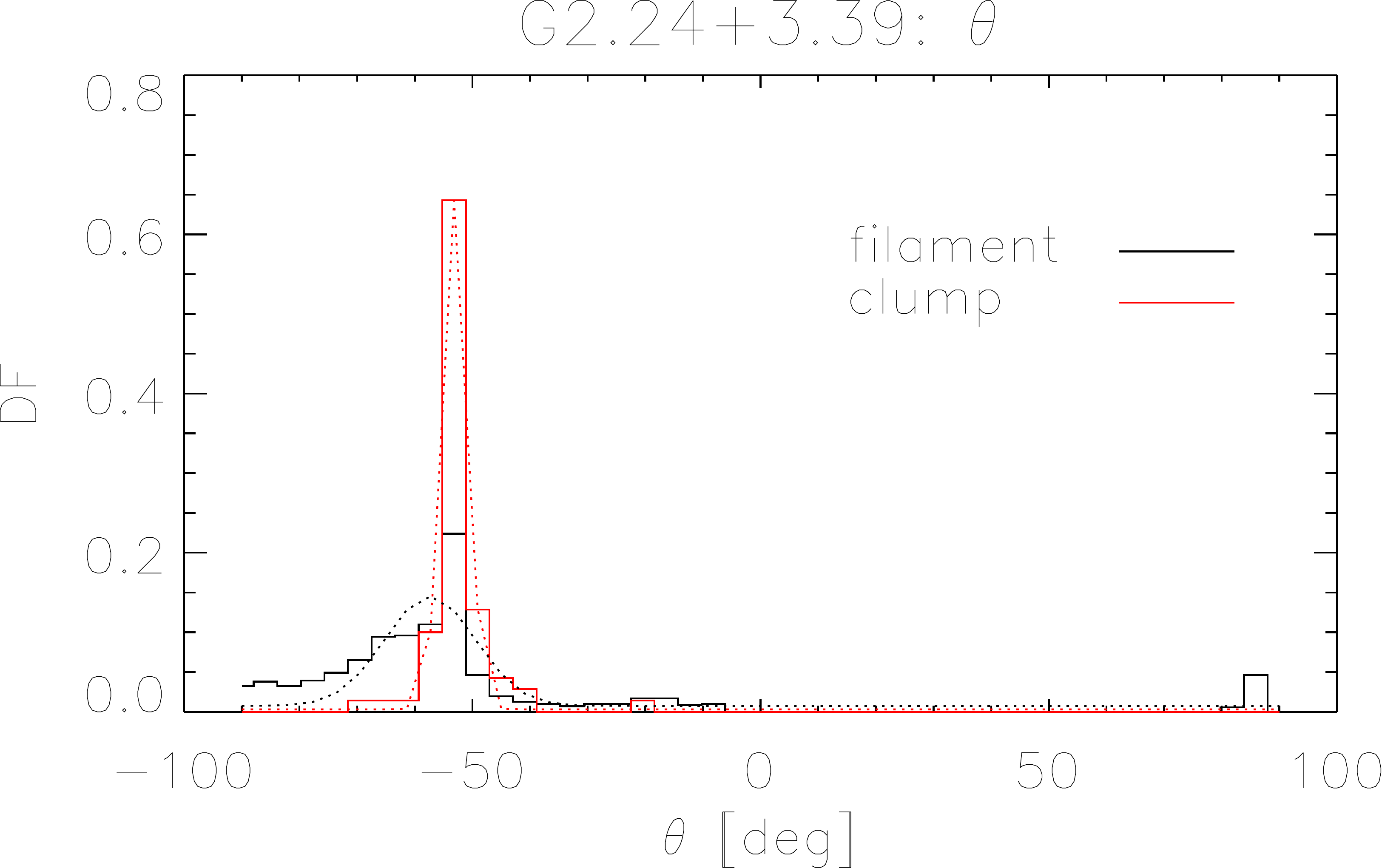} 

\end{tabular}
\caption{Examples illustrating the performed analysis. The left column shows the
minimaps centered on the PGCC positions, indicated by a star. The color scale
shows the 353 GHz intensity in MJy/sr, while the red line segments represent the
POS magnetic field orientation. In the first frame, the segment inside the red
rectangle represents the reference length corresponding to $10\%$ polarization.
The two black rectangles in each minimap delimit the regions used for determining
the uniformity of the background magnetic field orientation (cf.
Section~\ref{sec:uniformity}). The contours between the rectangles show the extent
of the detected $\irht$ filament. Small black parallelograms delimit regions that
correspond to the background area, for a pixel marked with the triangle, chosen
randomly for illustration. The right column shows the normalized histograms of the
filament orientation angles as well as the corresponding Gaussian fits, both for the entire filament (black line) and for the embedded clumps (red line).
}
\label{fig:regions_scheme}
\end{figure*}

As our study is focused on the relative orientations between magnetic fields and matter structures,
we seek filaments having uniform polarization angle in the background.
In each minimap, we define two rectangular regions on both sides of
the filament (see examples in Figure \ref{fig:regions_scheme}, left column). 
The side rectangles are parallel to each other, and
they are located at a distance equal to the width of the $\irht$ filament 
at the position of the clump. 
They have the same lengths, defined by the ends of the $\irht$ filament, 
and the same widths, set to $30\arcmin$.
Together, they constitute 
the background region of each minimap.

In a general case, the linear polarization angle is calculated using:
\begin{equation}
\polang_i = 0.5 \, {\rm atan}(-U_i, Q_i) \ ,
\label{eq:psi}
\end{equation}
and the corresponding average POS magnetic field orientation angle is obtained by rotation by $90$ degrees:
\begin{equation}
\angb_i = \polang_i + \frac{\pi}{2} \, . 
\label{eq:phi}
\end{equation}
In the side rectangles, we calculate the average Stokes $Q$ and $U$ parameters,
($\bar{Q}$, $\bar{U}$), from which we derive the average POS magnetic field orientation angle, $\bar{\angb}$, using Equations \ref{eq:psi}, \ref{eq:phi}.
As the angles in the $\pk$ data are counted positively in the direction from North
to West, we bring them to the IAU convention adopted for all the angles in this study by taking the negative value of the Stokes $U$ parameter.
Globally, the Stokes linear polarization parameters have similar values on both
sides of the detected filaments: at least $70\%$ of the filaments have Stokes
parameters that do not differ by more than $50\%$ between both sides.
However, in some cases, the difference between the two sides as well as the dispersion of the Stokes parameters on each side can be high.
The dispersion of the Stokes parameters individually does not represent the angles dispersion.
We calculate the standard deviation of the magnetic field 
orientation, $\sigangb$, in each region using circular statistics
\citep{berens2009}. 
This allows us to characterize the coherence of the magnetic field in the
background and to refine the minimaps selection for our analysis further in Section \ref{sec:uniform}.

%--------------------------------------------------------------------------%
\subsection{Separation of filament and background polarization parameters}
%--------------------------------------------------------------------------%
In order to investigate the contributions from the filament and its background in the polarization angle maps, we separate their respective contributions when possible.
For each pixel $i$ in a filament, we estimate the average background
Stokes polarization parameters, $\qbkg$ and $\ubkg$, by averaging $Q$ and
$U$ over appropriate cuts.
If the uncertainty for the considered pixel is
smaller than $10^{\circ}$, we use cuts passing through the side rectangles
perpendicular to the filament at the pixel position.
Otherwise, the cuts are perpendicular to
the filament at the central pixel, that is, 
at the position of the corresponding PGCC.
See Figure \ref{fig:regions_scheme} for examples of cuts.
As a first approximation, in view of our assumption of optically thin medium (see Section
\ref{sec:uniformity}), the Stokes polarization parameters of the filament
at pixel $i$ are obtained by subtracting the background contributions from the total signal: 
\begin{eqnarray}
\qfil = Q _i - \qbkg \nonumber \\
\ufil = U_i - \ubkg \, .
\label{eq:qu_fil}
\end{eqnarray}
The background magnetic field angle $\phi_{\rm bkg,i}$ and the magnetic field angle of the filament $\phi_{\rm fil,i}$ are then calculated using Equation~(\ref{eq:phi}) for each pixel in the filament. 
Figure \ref{fig:qu_before_after} shows the average Stokes linear polarization parameters as well as the standard deviations for each filament before and after the estimated background level has been subtracted. 
On average, background subtraction does not critically change the linear polarization parameters.
After background subtraction, the values of $Q$ and $U$ are changed by less than $20\%$ on average for $55\%$ of the filaments, and by less than $30\%$ for $70\%$ of them.

The above method is applied to the filament selection described further in Section~\ref{sec:uniform}.
The selected filaments are embedded in an environment
where the magnetic field orientation is uniform, so that our estimation of the linear background contribution is valid.

%--------------------------------------------------------------------------%
\subsection{Column density}
\label{sec:nh_characteristics}
%--------------------------------------------------------------------------%
We determine the column density for each pixel using 
\begin{equation}
N_{\mathrm{H}_2} = \frac{I_{\nu} }{ \mu m_{\mathrm{H}}\,  B_{\nu}(T) \,  \kappa_{\nu}} \, ,
\label{eq:nh}
\end{equation}
where we adopt the dust opacity $\kappa_{\nu}$=$0.1\,(\nu/1\,\rm{THz})^{\beta}\,
\rm{cm^{2}g^{-1}}$ \citep{beckwith1990}, which is appropriate for the case of
dense clouds at intermediate densities. $\mu$ is the mean molecular weight, which
is equal to $2.8$ amu in our analysis. 
The color temperature $T$ is estimated using the brightness emission in the {\pk} 857, 545, 353\,GHz channels, and the IRIS 3\,THz (100 $\mu$m) band. 
The maps were convolved to a $7\arcmin$ resolution and, for each pixel, the SED was fitted with $B_{\nu}(T) \nu^{\beta}$ using $\beta=2$. 
This value is usually adopted for starless core studies and it is close to the
mean value of $\beta$ derived for Planck clumps \citep{planck2011-7.7b,planck2016-XXVIII}. 
%%KF
$I_{\nu}$ is derived from the flux density at $857$ GHz, the closest {\pk} band to the 1\,THz reference of \cite{beckwith1990}, which minimizes the impact of assuming a fixed emissivity spectral index.
This frequency also corresponds to the {\pk} HFI band where the SNR is the strongest. 

%%%%% ========== %%%%%
\section{Filament selection}
\label{sec:fil_selection}
%%%%% ========== %%%%%

%--------------------------------------------------------------------------%
\subsection{The "uniform background" selection}
\label{sec:uniform}
\begin{figure}
\begin{center}
\includegraphics[width = 3.5 cm, angle = 90]{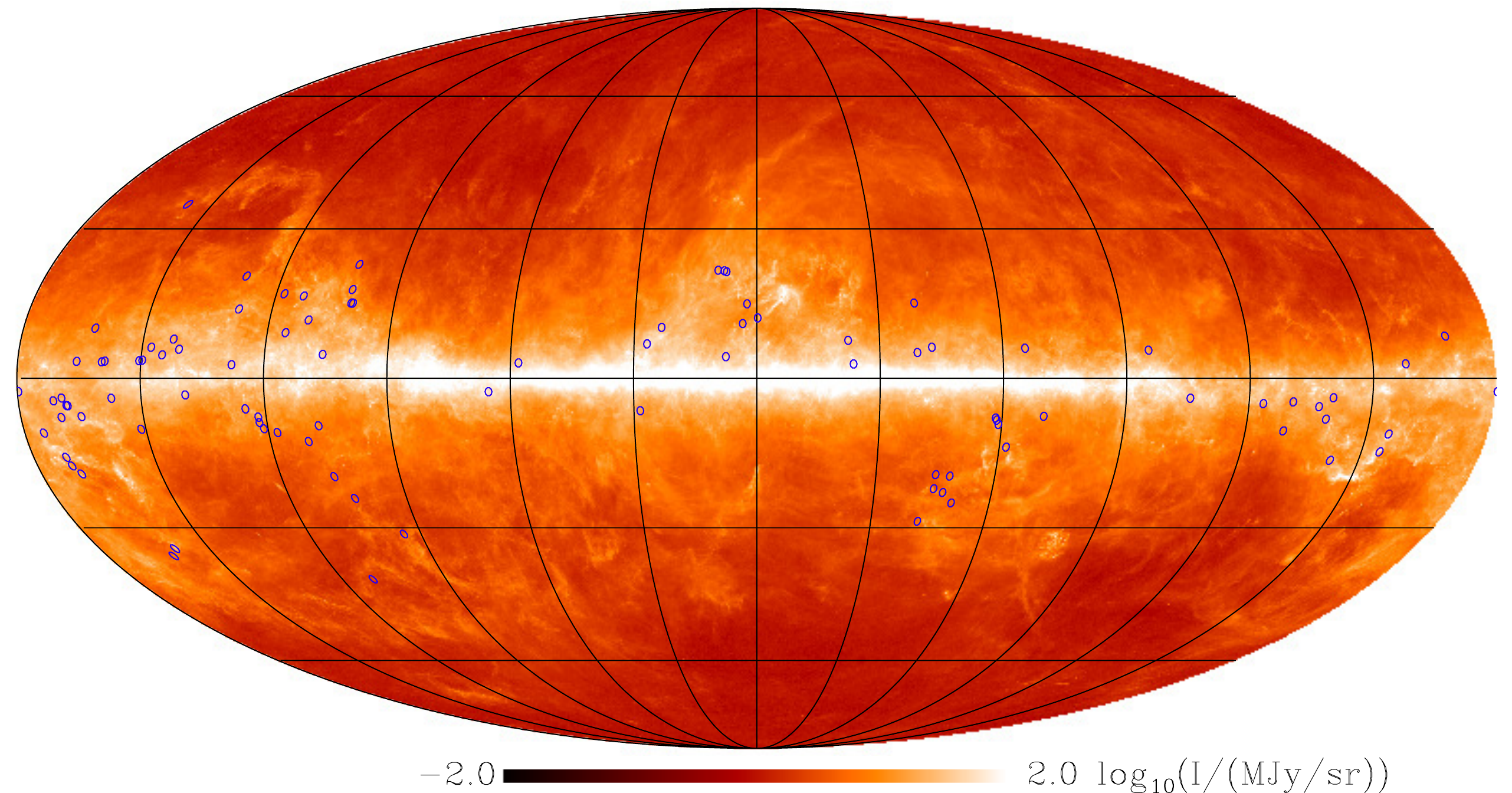}
\caption{Positions of the $\nuni$ PGCCs selected for this study. The background map shows the intensity map at the {\pk} at 857 GHz in logarithmic scale.}
\label{fig:allsky}
\end{center}
\end{figure}

%--------------------------------------------------------------------------%
About $\nhighstd$ of the pre-selected $\ndet$ filaments have the magnetic field
angle dispersion larger than $20\degr$ in both background rectangular regions introduced in Section \ref{sec:uniformity}.
Thus, most filaments are located in regions where the
magnetic field orientation seen by {\pk} is coherent.
To define the filaments that have a uniform magnetic field orientation
in the background, 
we require that $\bar{\angb}$ does not change by more than $20{\degr}$ between
both sides of the filament.  
This yields a selection of $\nuni$ objects distributed over the sky (cf. Figure \ref{fig:allsky}).
The rest of our analysis is focused on this selection of filaments. 
Note that only 19 filaments out of the selected $\nuni$ fall into
regions studied previously using {\pk} data \citep{planck2014-XXXV,planck2014-XXXIII}.
Thus, most of the selected minimaps have not been the subjects of
previous statistical studies on the relative orientation of magnetic fields and 
matter structures in the {\pk} data.

%--------------------------------------------------------------------------%
\subsection{Characteristics of the selected filaments}
\label{sec:characteristics}
%--------------------------------------------------------------------------%
Temperatures are reported in the PGCC catalogue for $80 \%$ of the selection and span between $9$ K and $19$ K with a mean value of $13.7$ K.
Distance estimates are provided for about $45 \%$ of the PGCCs and are mostly lower than 1 kpc.
The corresponding histogram, presented in Figure \ref{fig:dist_hist}, shows a main
peak at $\sim 0.2$ kpc , and a second peak at $\sim 0.4$ kpc.
Masses are determined for 38 clumps.
The masses of the 5 most distant clumps exceed $100 \, \rm{M}_{\odot}$ and approach the masses of molecular clouds. 
However, 21 clumps, which represent $23\%$ of the selection, have their
masses found to be smaller than $5 \,
\rm{M}_{\odot}$ with a mean value of $2.02 \, \rm{M}_{\odot}$.
The average gas column densities calculated according to Equation~(\ref{eq:nh}) range from $2.9 \times 10^{20}$\,cm$^{-2}$ to $2.5 \times 10^{22}$\,cm$^{-2}$ with a mean value of $3 \times 10^{21}$\,cm$^{-2}$.

We estimate the average sizes of the clumps as:
\begin{equation}
C = 2\, \sqrt[]{a b} \, ,
\label{eq:size}
\end{equation}
where $a$ and $b$ are the semi-major and semi-minor axes. 
The average sizes range from 10$\arcmin$ to 17$\arcmin$ with an average of $13 \arcmin$.
The corresponding histogram is presented in Figure \ref{fig:size_hist}.
The lengths of the selected filaments,
which are estimated as the sum of the two line segments connecting the center of
the clump to the ends of the filament, range from $34\arcmin$ to $104\arcmin$ with an average value of $66\arcmin$.
The widths of the filaments at the center of the
clumps range from $4\arcmin$ to $20\arcmin$ with an average value of $8\arcmin$.
The ratio of the longest to shortest axes of the filaments varies from 3 to 18 with an average of 9.
We emphasize that the lengths and widths given here are not the physical dimensions, 
but just detection dimensions.
These parameters depend on the detection method and the kernel size used in the SupRHT method. 

A conservative estimate of the dispersion of the magnetic field angle in the filaments, 
calculated using circular statistics, ranges from $6\degr$ to $73\degr$ 
with a mean value of $40\degr$.

%--------------------------------------------------------------------------%
\subsection{Column density subsamples}
\label{sec:low_high}
%--------------------------------------------------------------------------%
Following the observation that the preferential orientation of matter in filaments
with respect to the background magnetic field changes
with gas column density \citep{planck2015-XXXV}, we divide our selection of
filaments into two subsamples according to the column density of their
environment. We separate these two using a threshold column density
equal to the median background column density of $\nhbkglim$
$\rm{cm^{-2}}$. The corresponding subsamples are
called $\nhbkghigh$ and $\nhbkglow$.

We also distinguish between filaments that are more or less
contrasted with respect to their environment by calculating 
the column density contrast:
\begin{equation}
\Delta N_{\rm H} = \bar{N}_{\rm H} - \bar{N}_{\rm H, \, bkg} \, , 
\end{equation}
where $\bar{N}_{\rm H}$ is the average column density 
over the {\filament} area and $\bar{N}_{\rm H, \, bkg}$ is 
the average column density in the background area.
We denote $\dnhlow$ and $\dnhhigh$ the subsamples of
filaments that have differential column densities lower and higher than 
the median $\deltanh$ of $\dnhlim$ $\rm{cm^{-2}}$.

\begin{figure}
\begin{center}
\includegraphics[width = 8.3 cm, trim=0 120 0 0]{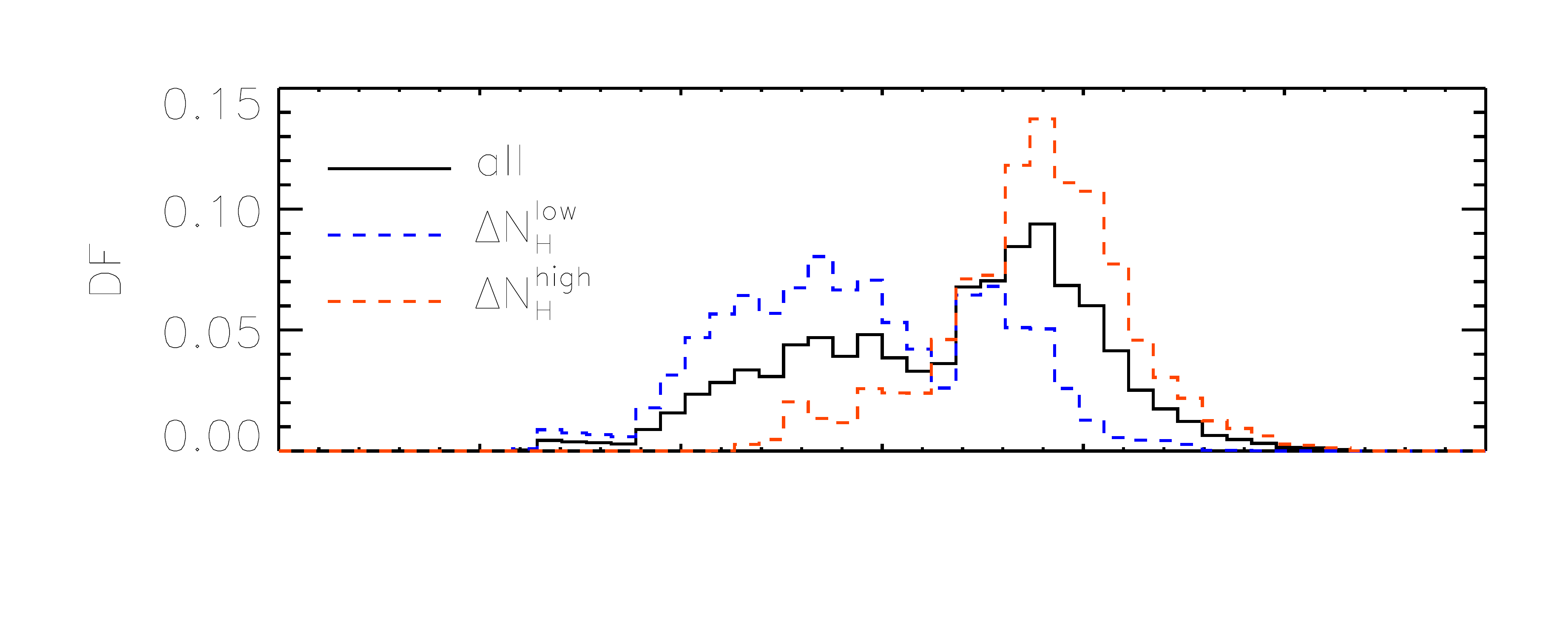} \\
\includegraphics[width = 8.3 cm]{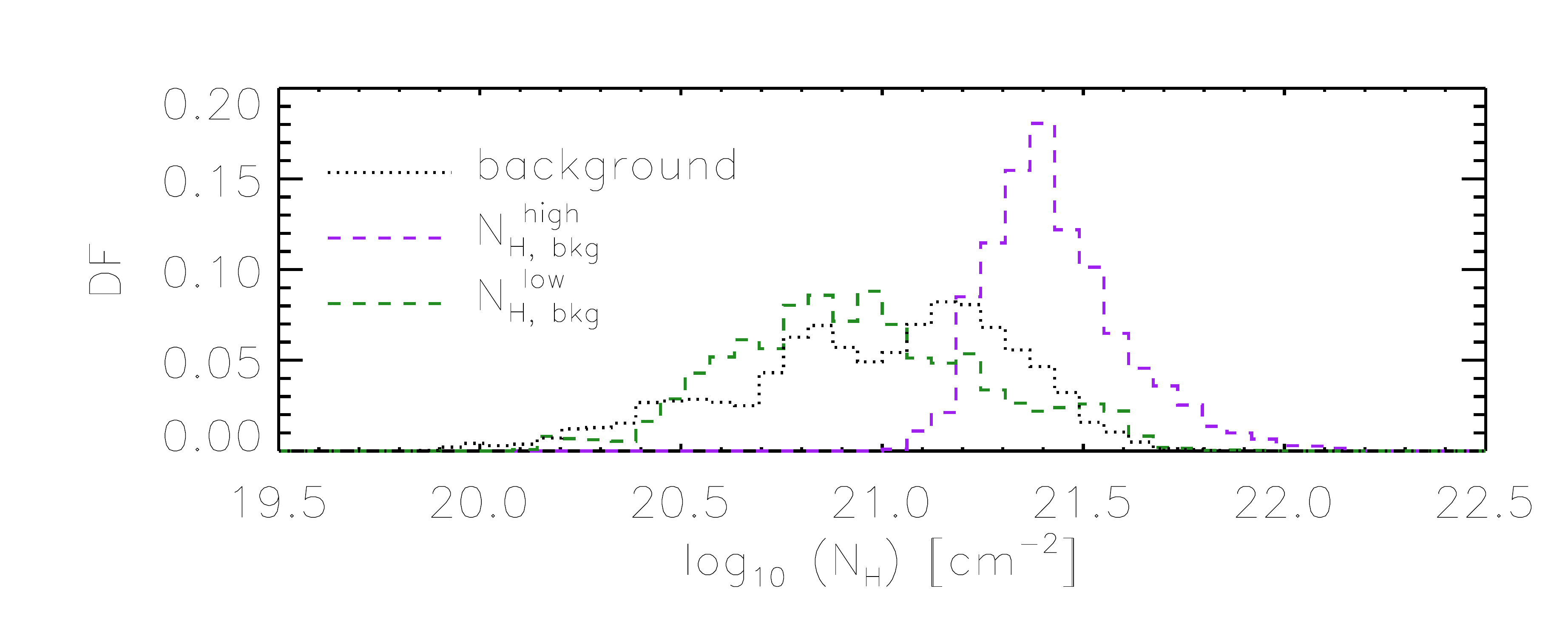}
\caption{Distribution functions of the column density in the pixels of the detected filaments. In the upper panel,  the black solid line represents all filaments, the blue and red dashed lines represent the low and high differential column density subsamples. In the lower panel, the gray dotted line represents the background regions, the green and purple lines show the low and high background column densities subsamples.} 
\label{fig:nh_hist}
\end{center}
\end{figure}

The histograms of the column density normalized to one for the various
subsamples are shown in Figure \ref{fig:nh_hist}.
The number of detected filaments in each subsample is reported in Table \ref{tab:nhsubsamples}.
The distribution of the number of pixels in the {\filament} and clump areas in the different subsamples are roughly 
the same and follow the general distribution shown in Figure \ref{fig:pix_hist}.
We emphasize that the PGCCs chosen for this study are outside of the Galactic
plane and the observed column densities are more probably due to local structures
rather than averaging along long LOS.
We observe in Figure \ref{fig:nh_hist} that with our selection we probe
environments with different column densities, shown by the gray dotted
line in the lower panel.
We also note that filaments embedded in high column density environments
tend to have higher column densities.
The same behavior was observed in \textit{Herschel} data at smaller scales
\citep{rivera-ingraham2017}, where dense filaments tend to be
embedded in dense environments.

\begin{table}
\caption{Number of filaments in background and differential column density
subsamples (cf. Section \ref{sec:low_high}). The 
threshold column densities between subsamples are $N_{\rm H, \, bkg} = \nhbkglim \, \rm{cm^{-2}}$ and $\Delta N_{\rm H} = \dnhlim \, \rm{cm^{-2}}$. In parentheses we report the number of filaments with known distances in each of the subsamples.}
\begin{center}
\begin{tabular}{|c|c|c|c|}
\hline 
 & $\nhbkglow$ &  $\nhbkghigh$ & total \\ \hline
 $\dnhlow$ & 30 & 15 & 45 (17) \\ \hline
$\dnhhigh$ & 15 & 30 & 45 (9) \\ \hline
total & 45 (9) & 45 (17) \\ \hline
\end{tabular}
\end{center}
\label{tab:nhsubsamples}
\end{table}%

%%%%% ========== %%%%%
\section{Results}
\label{sec:results}
%%%%% ========== %%%%%
Above, we obtained the orientation angles of matter in all the pixels of our filaments ($\angrht_i$), the average orientation angles of matter in the {\filament} and clump areas ($\bar{\angrht}_{\rm fil}$ and $\bar{\angrht}_{\rm clump})$, and the magnetic field angles in all the pixels of the filaments and their background ($\phi_{\rm fil,i}$ and $\phi_{\rm bkg,i}$).
We separated the filaments into subsamples depending on the column density of the background and on the column density contrast.

In what follows, we investigate the relative orientation between matter and magnetic fields depending on the filament characteristics such as the background column density and the column density contrast. We omit the subscript $i$ to simplify the expressions.

%--------------------------------------------------------------------------%
\subsection{Relative orientation between filaments and embedded clumps}
%--------------------------------------------------------------------------%
We calculate the absolute difference between the average orientation angles within the {\filament} and clump areas $| \bar{\angrht}_{\rm fil}-\bar{\angrht}_{\rm clump} |$ (cf. Section~\ref{sec:method}). 
For this purpose, only elongated clumps with ellipticities larger than $1.5$
 are used.
As an orientation angle corresponds to two possible directions, we consider only the smallest angle between the two line segments, which is in the range from $0$ to $90{\degr}$.
Most ($\sim98\%$) of the clumps have their major axes aligned with their hosting filament within 20$\degr$.
The same is observed when clumps with ellipticities smaller than $1.5$ are included in the analysis.

In order to check that this result is independent of the filament detection method, we compare the average clump position angles $\bar{\angrht}_{\rm clump}$ given by SupRHT to the position angles $\angcat$ provided in the PGCC catalogue, which is based on the cold residual maps. 
In the latter case, $90\%$ of the clumps are on average aligned with the filament within 20$\degr$.
Both methods agree on the position angle of the clumps within $20\degr$ for more than $70\%$ of the selection.
If we consider clumps with all ellipticities, this fraction goes down to $62\%$.
We note that these differences may also come from uncertainties in the detection method used to build the PGCC catalogue.

%--------------------------------------------------------------------------%
\subsection{Relative orientation between filaments and the background magnetic field}
%--------------------------------------------------------------------------%
In this section, we investigate the statistics of the relative orientation between filaments and the background magnetic field.
To this end, we calculate the absolute difference between the average background magnetic field angle and the position angle for each pixel ( $|\phi_{\rm bkg} - \angrht |$ ).
This quantity is also defined in the range from $0\degr$ to $90\degr$.
We do not calculate this quantity per filament because the size of our sample is not sufficient for significant statistics.

\begin{figure}
\begin{center}
\includegraphics[trim = 25 0 0 0, width = 7.8 cm]{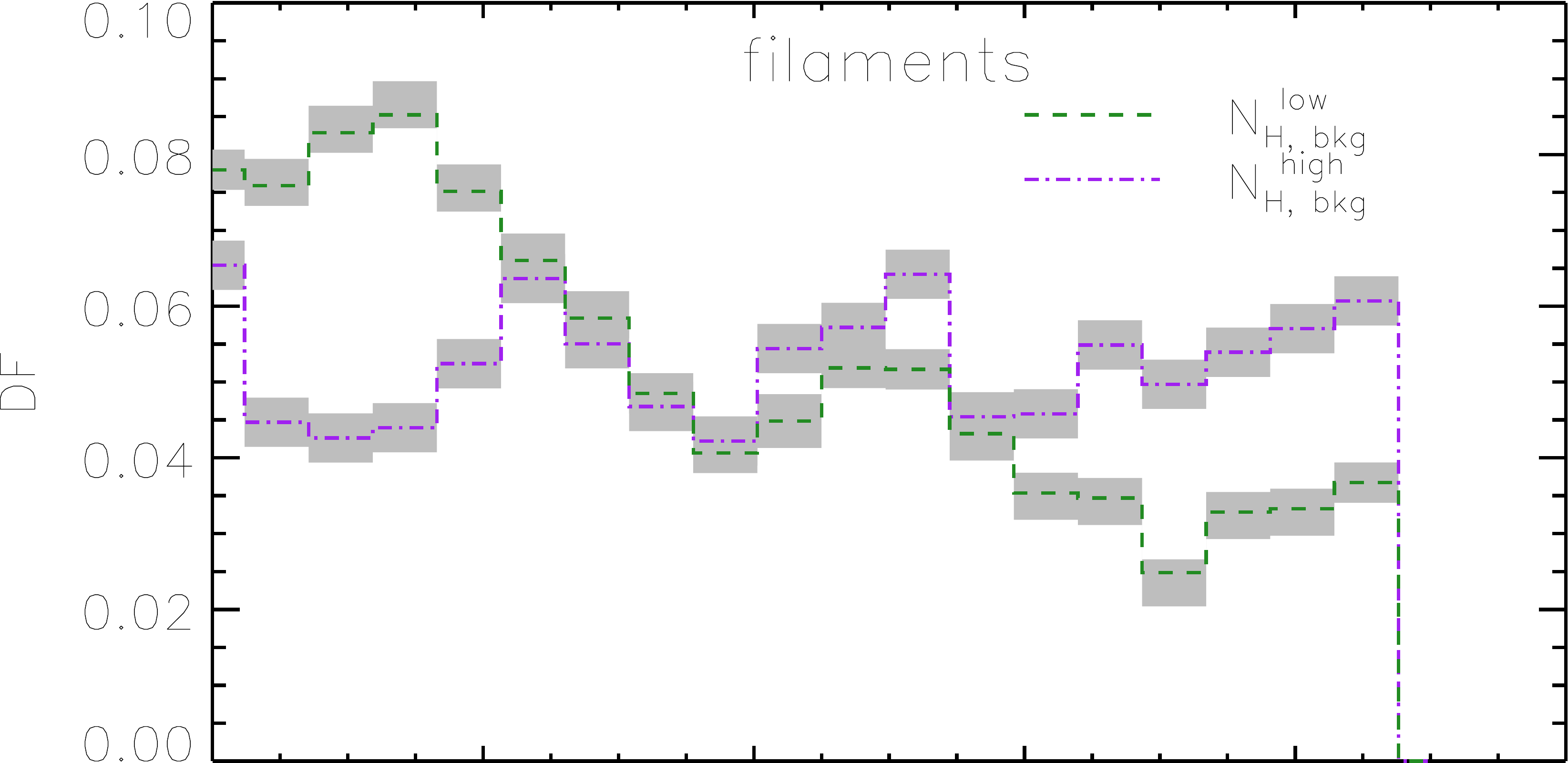}
\includegraphics[width = 8.25 cm]{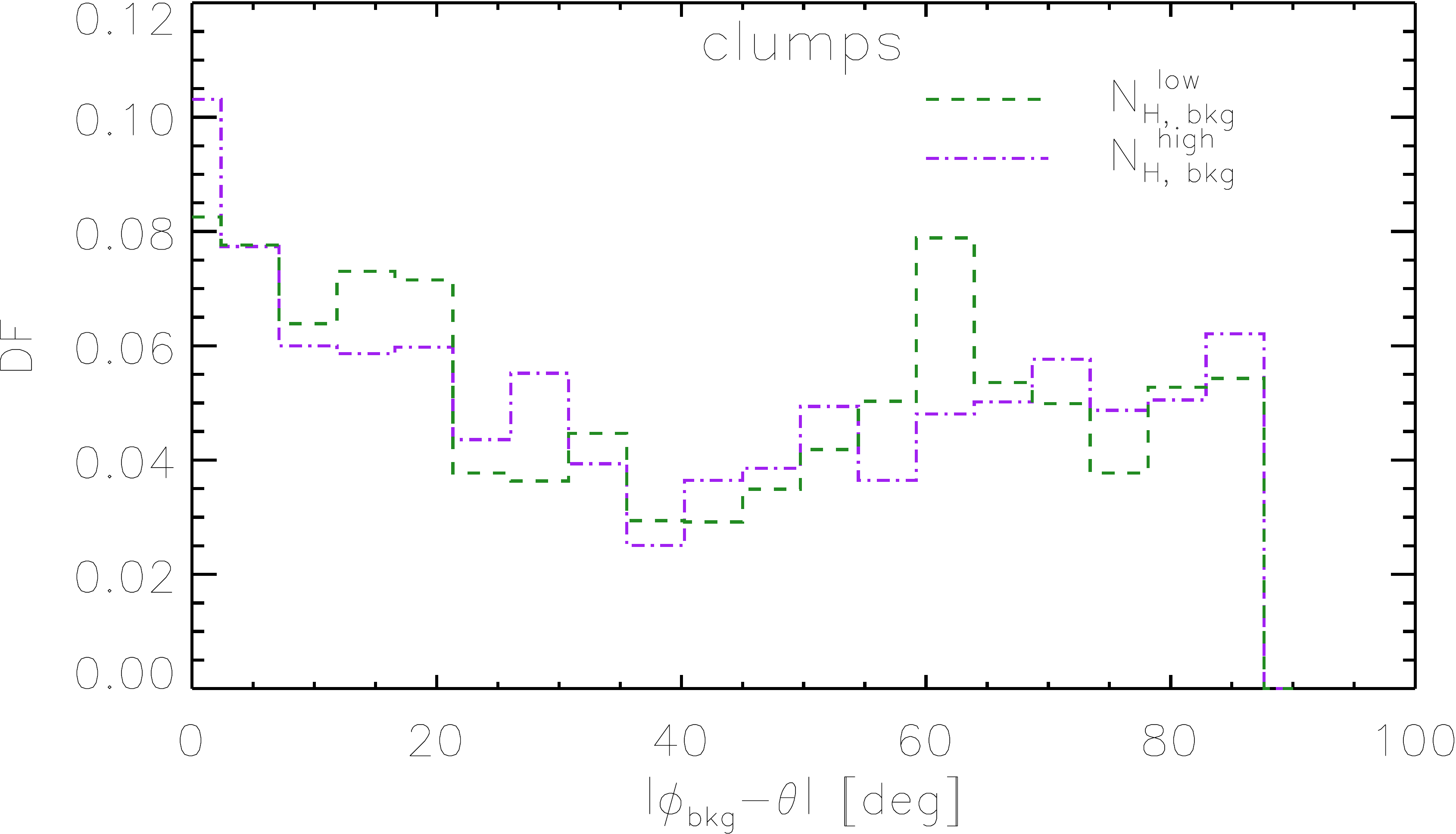}
\caption{DFs of the absolute difference between the orientations of the background magnetic field and the matter structures. The green dashed and purple dot-dashed lines are for low and high background column density subsamples, respectively. Upper panel: filament areas; lower panel: clump areas. For illustration purpose uncertainties are reported in grey shaded areas in the upper panel.}
\label{fig:beta_df}
\end{center}
\end{figure}

We show in the upper panel of Figure \ref{fig:beta_df} the DFs over pixels in the {\filament} area belonging to the low and high background column density subsamples.
In low-density environments the number of pixels decreases by a factor of two when increasing the value of $|\phi_{\rm bkg} - \angrht |$, that is from parallel to perpendicular relative orientation, whereas in dense environments filaments have no preferential relative orientation with respect to the background magnetic field.
In the lower panel of Figure \ref{fig:beta_df}, featuring the same analysis for the clump areas, we observe strong peaks at $0\degr$ but also a significant distribution at large $|\phi_{\rm bkg} - \angrht |$.
The two clumps DFs look similar but the $\nhbkghigh$ subsample has a slightly larger probability for either parallel and perpendicular relative orientations.

We carry out a non-parametric Mann-Whitney U-test (called U-test hereafter) in order to quantify the difference between the two pairs of distributions.
The U-test probability for the DFs of the {\filament} areas is $10^{-7}$, indicating that the distributions of $|\phi_{\rm bkg} - \angrht |$ in the low and high column density background subsamples are statistically different.
The U-test probability for the DFs of clump areas is $0.21$ which confirms that the distribution of $|\phi_{\rm bkg} - \angrht |$ in the low and high background column density are similar.
The U-test on the DFs of the {\filament} areas versus clump areas yields $0$ probability of similarity between both subsamples.
The summary of the trends observed in the above DFs is provided in Table \ref{tab:summary2}.

\begin{figure}
\begin{center}
\includegraphics[trim = 25 0 0 0, width = 7.85 cm]{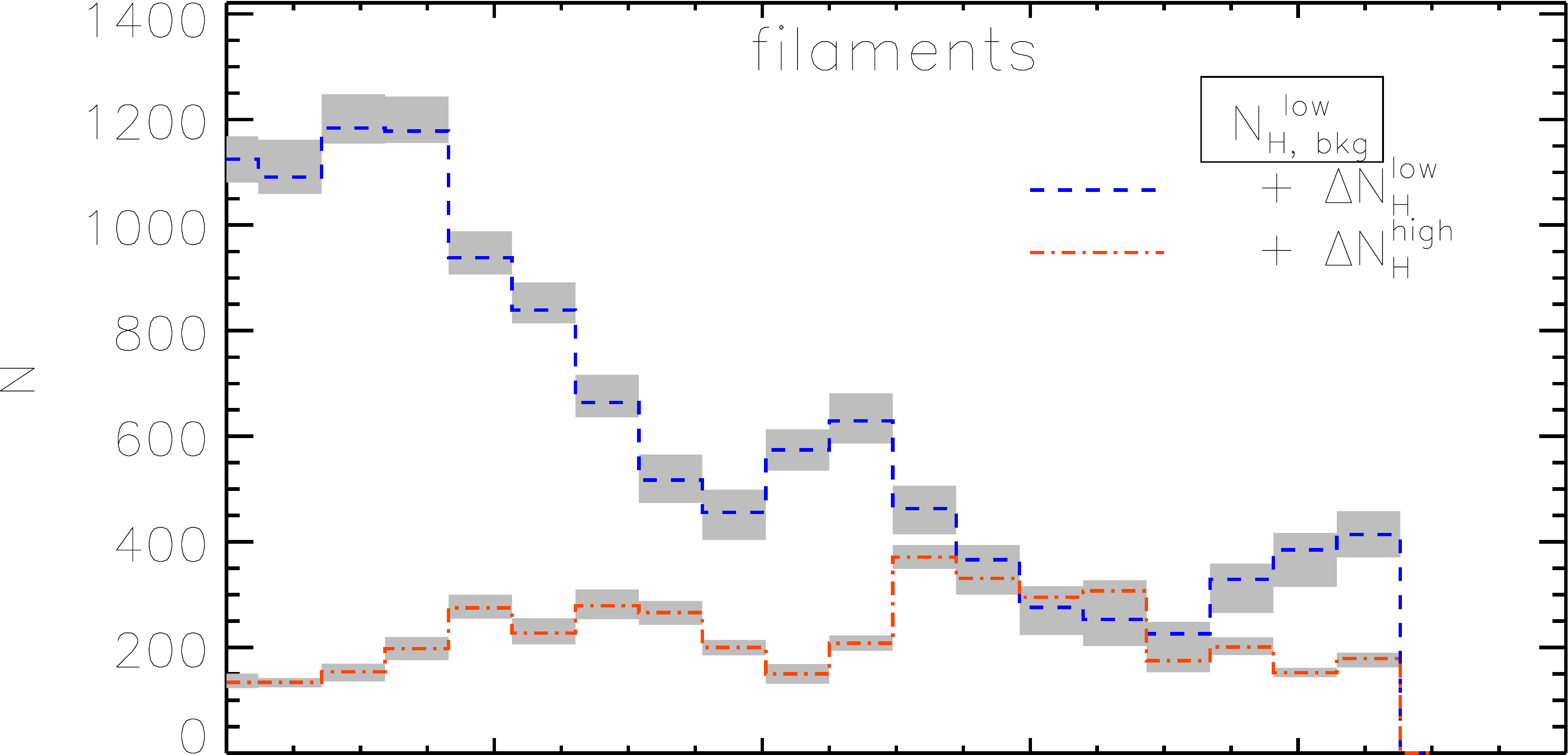} 
\includegraphics[width = 8.3 cm]{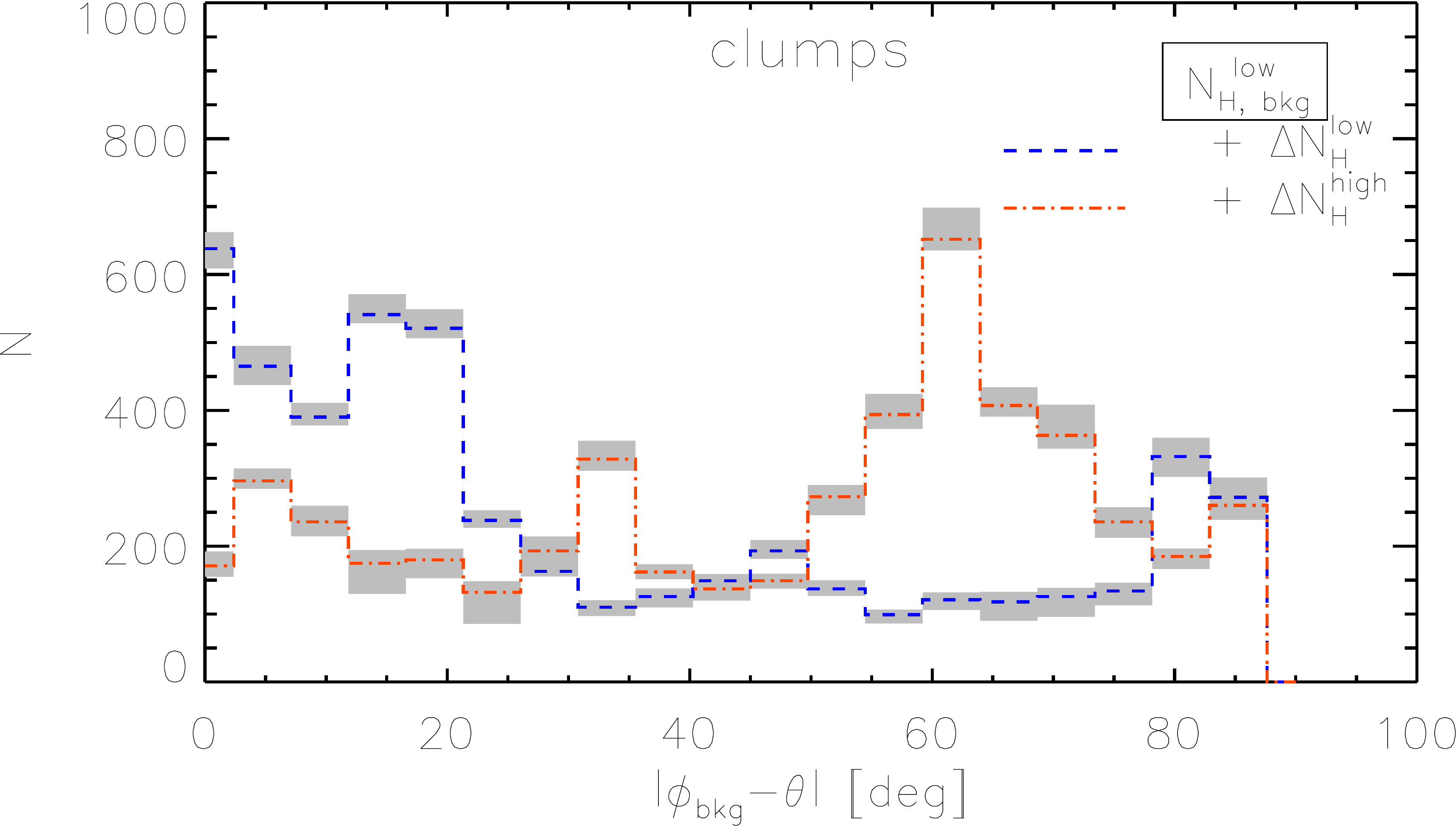}
\caption{Histograms of the absolute difference between the orientation of the filamentary structures and the background magnetic field orientation over the pixels in the filaments only (upper panel) and in the clumps only (lower panel) for the low background column density subsample. Blue and orange dashed lines correspond to the contributions of the low and high contrast filaments to each subsample. The uncertainties are shown in gray shaded areas.}
\label{fig:beta_df2}
\end{center}
\end{figure}

For each subsample, we further separate the contributions from the filaments depending on their column density contrast with respect to their environment.
Figure \ref{fig:beta_df2} shows the DFs over the {\filament} and clump areas in the low background column density subsample ($\nhbkglow$).
We observe that, in the {\filament} areas (upper panel of Figure \ref{fig:beta_df2}), there is much less parallel relative orientation in the high column density contrast subsample than in the low column density contrast subsample.
The corresponding DFs over the clump areas in the lower panel of Figure \ref{fig:beta_df2} shows that, again, there is a tendency for parallel relative orientation in the low contrast filaments. In the DF of high contrast filaments there is a peak around $60^{\circ}$. 
The peak is not conclusive as it can not be interpreted as perpendicular relative orientation. 
This could arise from projection effect of the perpendicular relative orientation or be a trace of the oblique orientations of the clumps in 3D due to the rotation of the clumps during the filament evolution.
In an attempt to investigate the origin of these two peaks we separate the clumps inside the low background column density subsample into two parts depending on their average column density.
The corresponding histograms are shown in Figure \ref{fig:ad_hoc_clumps}.
We recover the similar kind of figure as shown in the lower panel of Figure \ref{fig:beta_df2} for $N_H$ equal around $2 \times 10^{21}$ cm$^{-2}$. There is a clear dichotomy depending on the column density: high column density clumps show a preferential perpendicular relative orientation whereas the lower column density clumps show parallel relative orientation with respect to the background magnetic field. 

\begin{figure}
\begin{center}
\includegraphics[trim = 25 0 0 0, width = 7.8 cm]{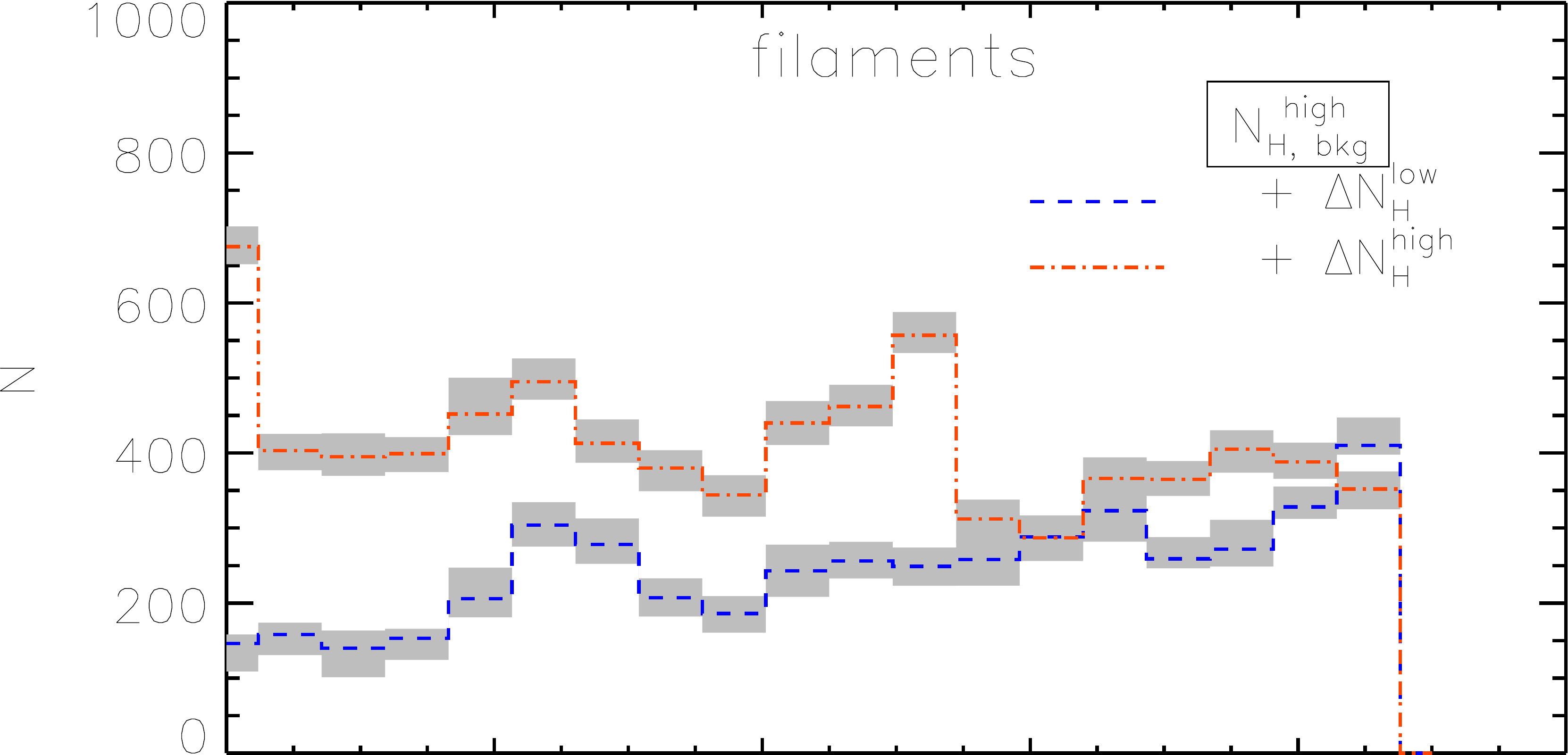}
\includegraphics[width = 8.25 cm]{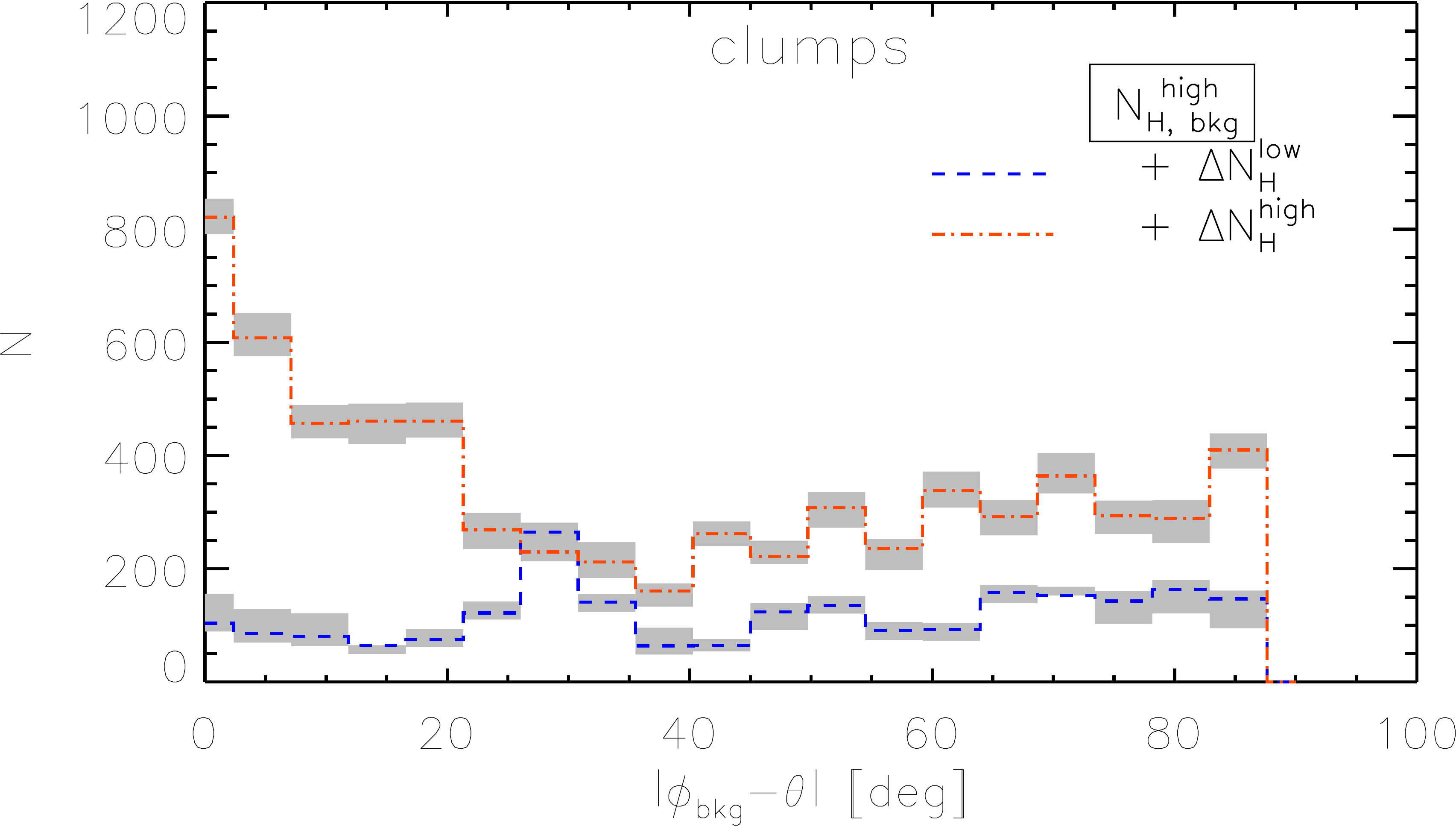}
\caption{Histograms of the absolute difference between the orientation of the filamentary structures and the background magnetic field orientation over the pixels in the filament areas (upper panel) and in the clump areas (lower panel) for the high background column density subsample. Blue dashed and orange dot-dashed lines correspond to the contributions of the low and high contrast filaments to each subsample. The uncertainties are shown in gray shaded areas.}
\label{fig:beta_df3}
\end{center}
\end{figure}

Figure \ref{fig:beta_df3} shows the DFs of $|\phi_{\rm bkg} - \angrht |$ over the {\filament} (upper panel) and clump (lower panel) areas in the low background column density subsamples.
For the {\filament} areas there is a tendency for perpendicular relative orientation in the $\dnhlow$ subsample whereas in the $\dnhhigh$ subsample there is a preferential parallel relative orientation with respect to the background magnetic field (upper panel of Figure \ref{fig:beta_df3}).
For the clump areas, the DF of the $\dnhlow$ subsample shows a uniform distribution of $|\phi_{\rm bkg} - \angrht |$ while the clumps inside the $\dnhhigh$ filaments tend to be aligned parallel to the background magnetic field.
The U-test probability for the four pairs of distributions are below $10^{-7}$, which means that the DFs of these column density subsamples are statistically different.

The uncertainties of the DFs are estimated using the bootstrap technique consisting of random sampling with replacement \citep{efron1993}.
The details on the uncertainties calculation using independent pixels are described in Appendix \ref{app:sigma}.
For illustration purpose, uncertainties are reported for some of the DFs.
The rest of the uncertainties are of similar magnitude.
\begin{table}
\begin{center}
\begin{tabular}{|l|l|c|c|r|}
& $N_{\rm H,bkg}$ & both $\Delta N_{\rm H}$ & $\dnhlow$ & $\dnhhigh$ \\ \hline \hline
\rotatebox[origin=c]{90}{{\filament} areas} & low & mostly $\parallel$ & mostly $\parallel$ & no tendency  \\ 
& high & no tendency & mostly $\perp$ & slightly $\parallel$  \\ \hline \hline
\rotatebox[origin=c]{90}{clump areas} & low & $\parallel$ and $\perp$  & mostly $\parallel$ & peak at $\simeq 60^{\circ}$ \\
& high & $\parallel$ and $\perp$  & no tendency & mostly $\parallel$ \\ \hline
\end{tabular}
\caption{Summary of the results shown in Figures \ref{fig:beta_df}, \ref{fig:beta_df2} and \ref{fig:beta_df3}. The observed tendencies of the relative orientation between the filaments and the background magnetic field. $N_{H,bkg}$ stands for the filaments in the environments with average column densities lower or higher than $\nhbkglim$ cm$^{-2}$. $\dnhlow$ and $\dnhhigh$ stands for the filaments that are contrasted with respect to their environment by more or less than $\dnhlim$ cm$^{-2}$. }
\label{tab:summary2}
\end{center}
\end{table}

%--------------------------------------------------------------------------%
\subsection{Relative orientation between filaments and their internal 
magnetic fields}
\begin{figure}
\begin{center}
\includegraphics[trim = 5 0 0 0, width = 7.9 cm]{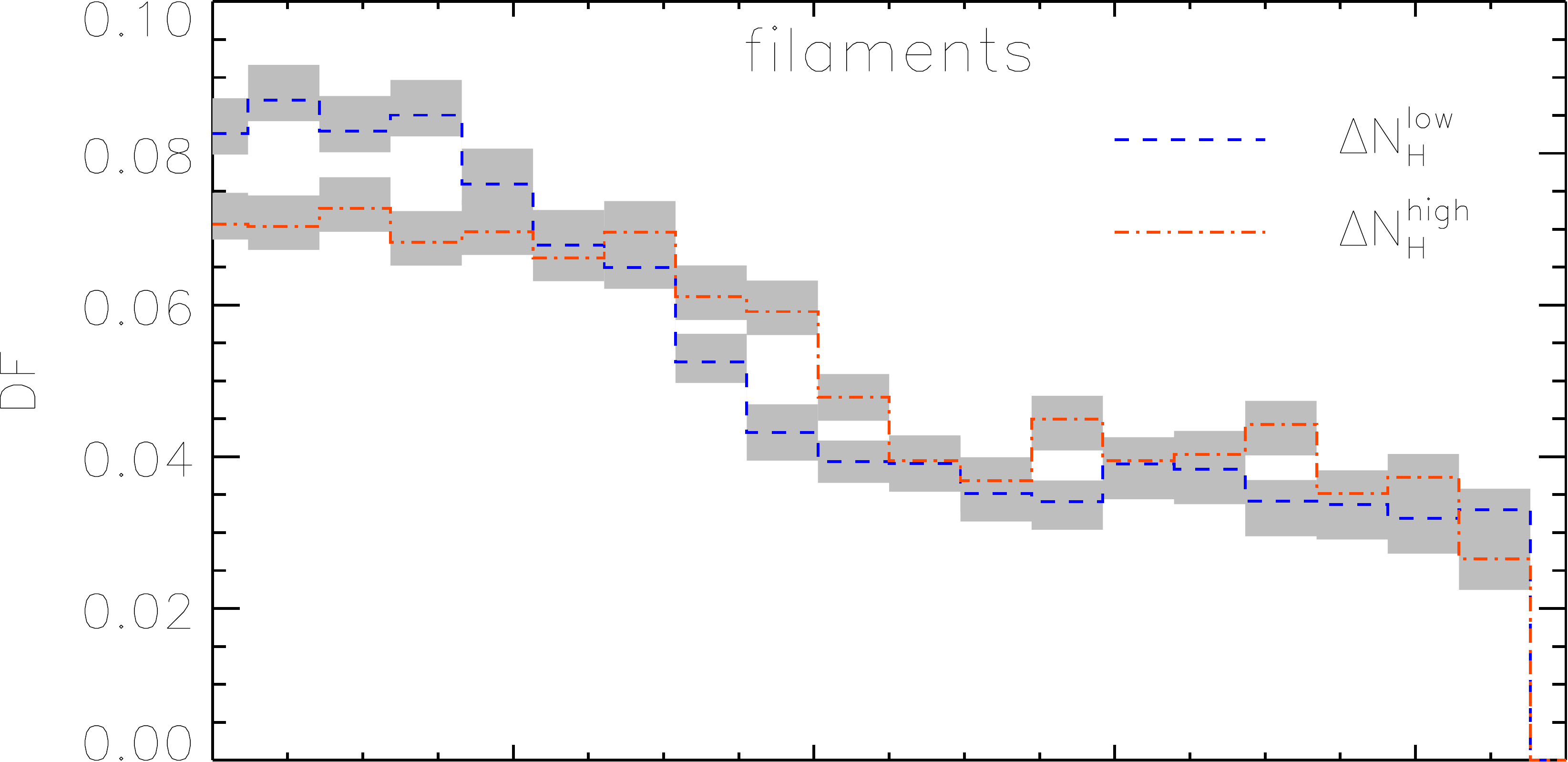}
\includegraphics[width = 7.95 cm]{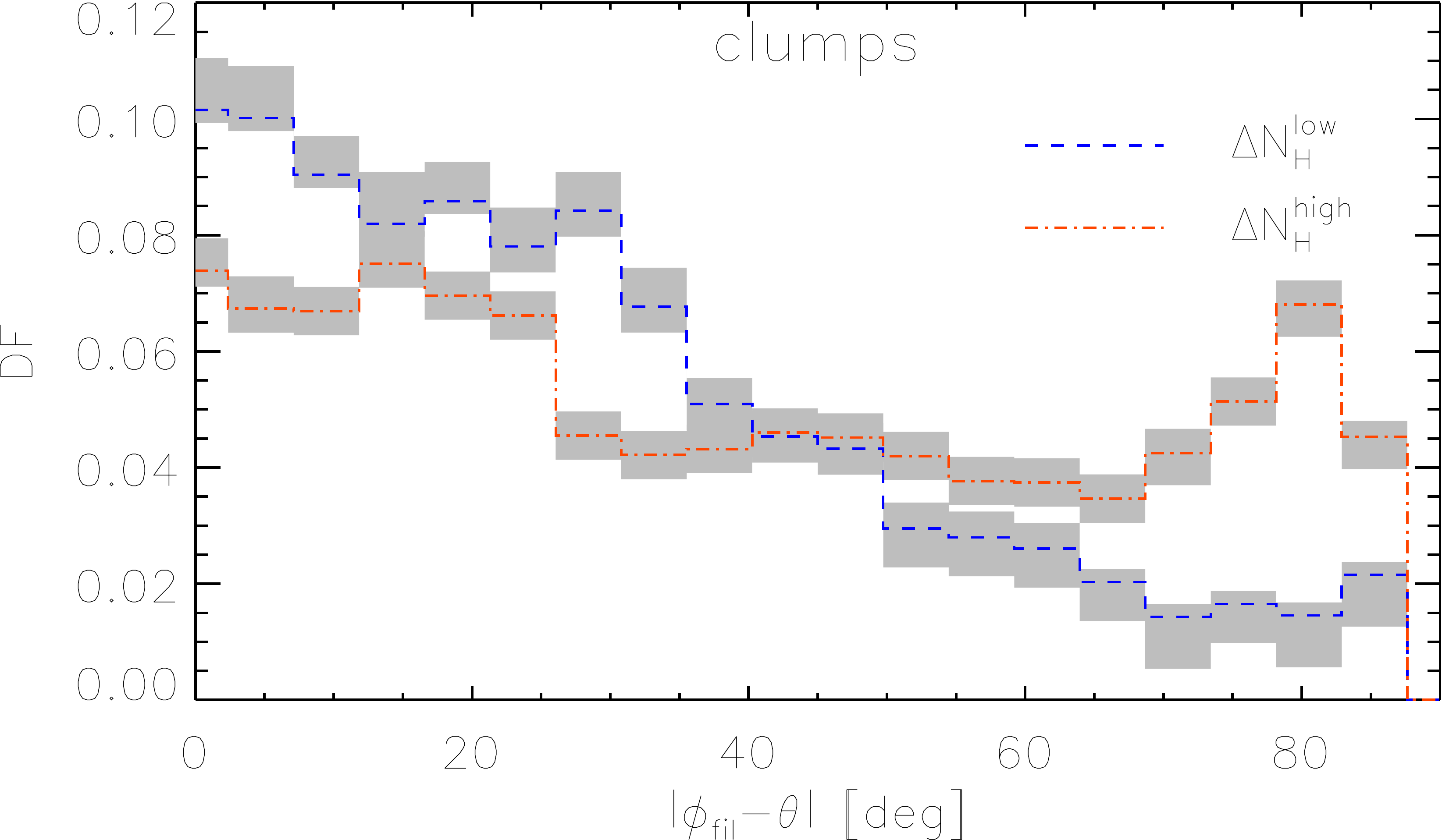}
\caption{DFs of the absolute difference between the orientation of the filaments and the magnetic field angle in the filaments. 
In each panel, the blue and orange dashed lines represent the DFs over the pixels in the low and high density contrast subsamples. Upper panel: filament areas; lower panel: clump areas. 
The uncertainties are shown as grey shaded areas.
}
\label{fig:nu_df}
\end{center}
\end{figure}

%--------------------------------------------------------------------------%
In order to analyse whether there is any coherence between the filaments orientation and the magnetic field orientation inside filaments, we calculate the absolute difference between the respective angles in each pixel ($\angrht$ and $\phi_{\rm fil}$). 
Figure \ref{fig:nu_df} shows the DFs over the pixels in both the {\filament} and clump areas.
The DFs show strong peaks at $0 \degr$ regardless of the column density contrast (Figure \ref{fig:nu_df}).
%The resulting distribution function for the differential column density subsamples are presented in Figure \ref{fig:nu_df}.
%All the DFs show strong peaks at $0 \degr$.
However, in the {\filament} areas (upper panel of Figure \ref{fig:nu_df}), the distribution is flatter for the high column density contrast subsample than for the low contrast subsample.
Clumps inside these filaments (lower panel of Figure \ref{fig:nu_df}) also tend to be less parallel to the filament magnetic field.
In addition, the corresponding DF shows a strong peak near perpendicular relative orientation.
Clumps inside the low contrast filaments globally show a preferential parallel relative orientation with respect to the filament magnetic field.
%Such a behavior is not observed in the $\dnhlow$ subsample which shows a preferential parallel alignment with respect to the magnetic field in both filaments and clumps areas.
The U-test probability for the two pairs of distributions are equal to $0$, which means that the DFs are statistically different.
The summary of the tendencies observed in the above DFs is provided in Table \ref{tab:summ3}.

\begin{figure}
\begin{center}
\includegraphics[width = 8.1 cm]{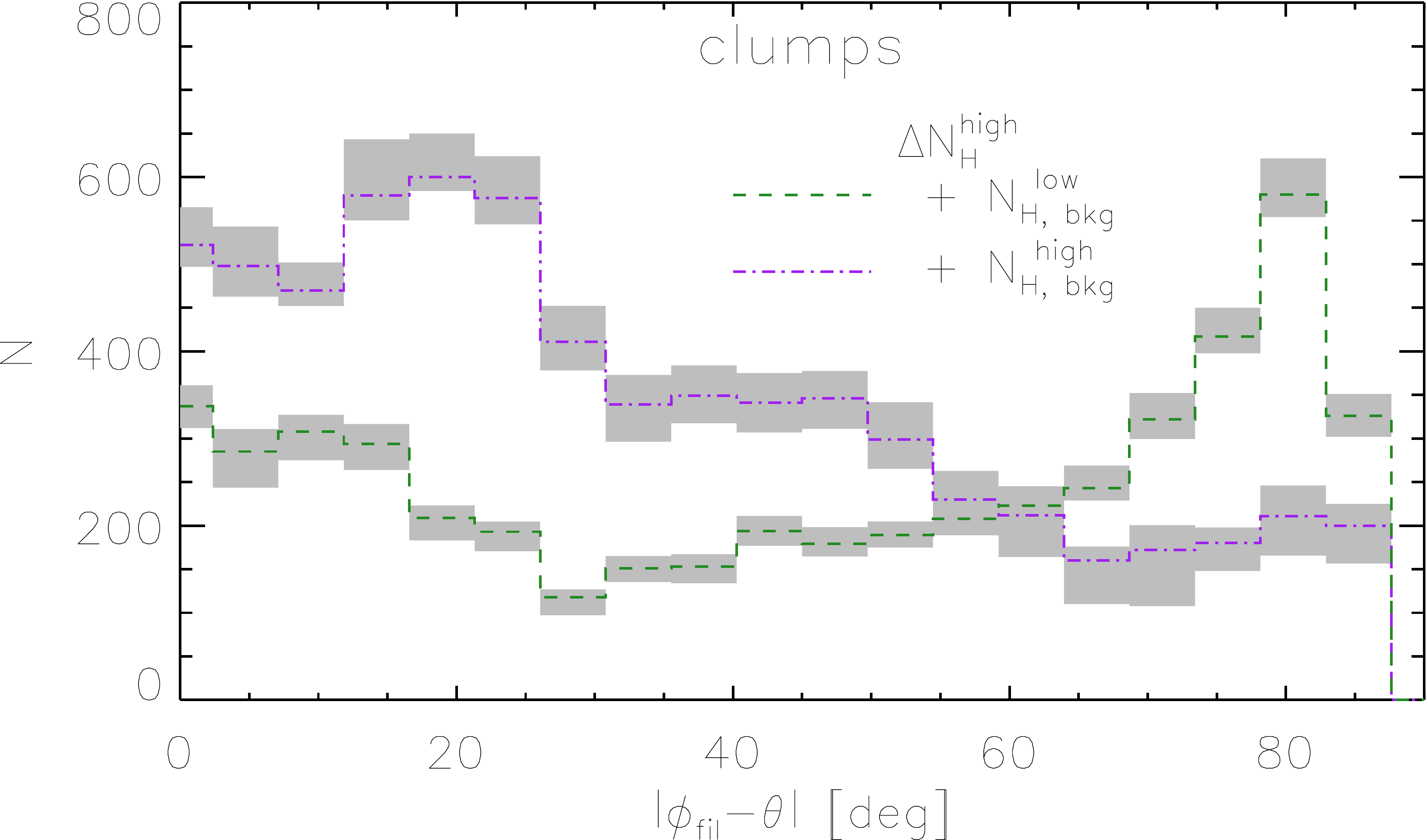}
\caption{Histograms of the absolute difference between the orientation of the filamentary structures and the magnetic field direction in the filaments of the high column density contrast subsample in the clump areas. Green and purple dashed lines correspond to the structures embedded into a low and high-density environment, respectively. The uncertainties are shown in grey shaded areas.}
\label{fig:nu_df2}
\end{center}
\end{figure}

To further investigate the bimodal distribution observed for clumps counterpart in the $\dnhhigh$ subsample, we separate the contributions from the low and high background column density subsamples, although the statistics are low when dividing the subsamples.
The corresponding histograms are shown in Figure \ref{fig:nu_df2}.
The peak near perpendicular relative orientation seen in the lower panel of Figure \ref{fig:nu_df} also appears in the low column density subsample, but not in the high column density subsample. This could arise from the component separation method. Indeed, subtracting a high background level strongly reduces the measured polarization degree and, therefore, makes the magnetic field angle estimation uncertain.

\begin{table}
\begin{center}
\begin{tabular}{|l|c|r|}
$\Delta N_{\rm H}$ & filaments & clumps \\ \hline \hline
low & mostly $\parallel$ & mostly $\parallel$  \\ \hline
high & mostly $\parallel$ & $\parallel$ and $\perp$  \\ \hline
\end{tabular}
\caption{Summary of the results shown in Figure \ref{fig:nu_df}. The observed tendencies of the relative orientation between the filaments and the magnetic field in the filaments. $\Delta N_H$ stands for the column density contrast which is lower or higher than $\dnhlim$ cm$^{-2}$.}
\label{tab:summ3}
\end{center}
\end{table}

%--------------------------------------------------------------------------%
\subsection{Relative orientation between magnetic fields inside the filaments and in the background}
\begin{figure}
\begin{center}
\includegraphics[width = 8 cm]{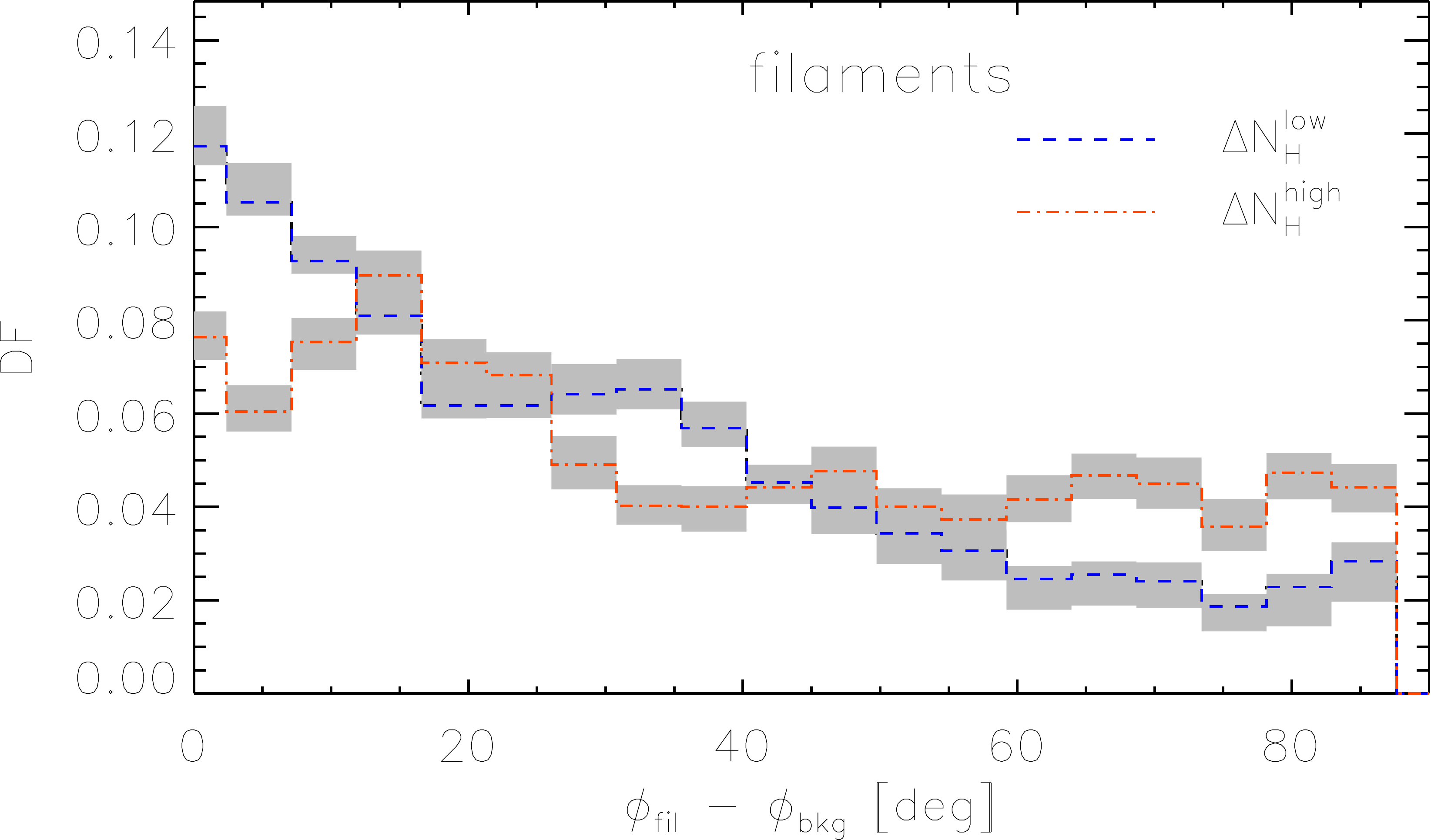}
\caption{DF of the average difference between the magnetic field angles in the filaments and in the background. The blue and red lines show the DFs for the low and high contrast filaments subsamples respectively. The uncertainties are shown as grey shaded areas.}
\label{fig:b_angles_diff}
\end{center}
\end{figure}

%--------------------------------------------------------------------------%
We calculate the difference between magnetic field angles inside the {\filament} areas and in the background using the Stokes parameters of both regions \citep{planck2014-XIX} and is equal to the difference between the respective polarization angles:
\begin{equation}
 \small { \psi_{\rm fil} - \psi_{\rm bkg} =}   \small {\frac{ \rm{atan}(Q_{\rm fil} U_{\rm bkg}  - U_{\rm fil} Q_{\rm bkg} ,Q_{\rm fil} Q_{\rm bkg}  + U_{\rm fil} U_{\rm bkg} ) } {2} }
\label{qu_diff}
\end{equation}
Figure \ref{fig:b_angles_diff} shows the corresponding DFs for the column density contrast subsamples.
For the low contrast filaments, the relative orientation is mostly parallel.
For the high contrast filaments, there is a slight tendency for parallel relative orientation with globally no preferential relative orientation.

Figure \ref{fig:app_b_angles_diff} shows the DFs of the difference between the magnetic fields in the filament and in the background for the background column density subsamples.
Even though the fields are mostly parallel in both, the DF of the $\nhbkghigh$ subsample is flatter than the DF of the $\nhbkglow$ subsample. 

\cite{planck2014-XXXIII} estimated the average difference between the magnetic field orientation in the filaments and in their background for three filaments: Musca, Taurus L1506 and B211. They used polynomial fits to the $Q$ and $U$ parameters after subtraction of the background contribution.
The first two filaments (Musca and L1506) are included in our final selection.
For these two clouds, we obtain average values of the difference between magnetic fields in the filament and in the background of $8.6^{\degr} \pm 2^{\degr}$and $55^{\degr} \pm 2^{\degr}$, which are comparable to respectively $10.6^{\degr}$ and $57.1^{\degr}$ obtained by \cite{planck2014-XXXIII}. 

%--------------------------------------------------------------------------%
\subsection{Nearby filaments}
%--------------------------------------------------------------------------%
We calculate rough estimates of the mean gas volume densities for the clumps and filaments with known distances $d$ in nearby molecular clouds.
In our selection, there are 26 PGCCs situated within $500$ pc from the Sun.
The clumps can be approximated by spheres of radius $R=d \, C$ and the mean gas volume density of a clump is given by:
\begin{equation}
\left\langle n_{H} \right\rangle_{clump} = \frac{3\left\langle N_{H} \right\rangle_{clump}}{4R} \, .
\end{equation}
Assuming that the filament depth is equal to the filament width, the mean gas volume density of a filament is given by:
\begin{equation}
\left\langle n_{H} \right\rangle_{fil} = \frac{\left\langle N_{H} \right\rangle_{fil}}{2dw} \, ,
\end{equation}
where $w$ is the filament angular width.
This yields only a rough estimate for the volume densities of the filaments since the width taken here is the detection width which underestimates the filament width as seen in the intensity maps. 
Also, the beam dilution prevents the detection of the densest regions.
The average POS diameters of the clumps (widths of the filaments) in parsecs varies from 0.4(0.9) to 2.1(4.9) with an average of 0.9(2.2).
The respective histograms are shown in Figure \ref{fig:size_pc_hist}. 
The majority (25 out of 26) of the clumps have an average radius smaller than 1 pc.
The histograms of the mean volume densities of the clumps and associated filaments are shown in Figure \ref{fig:nh_vol_hist}.
$n_H$ ranges between $400$ and $1.2\,10^4$ $\rm cm^{-3}$ with an average value of $2500$ $\rm cm^{-3}$ for the clumps, and between $150$ and $4000$ $\rm cm^{-3}$ with an average value of $1000$ $\rm cm^{-3}$ for the filaments.
Among the 26 filaments we identify those belonging to the $N_H$ subsamples, and the resulting numbers are reported in Table \ref{tab:nhsubsamples}.

We test the results shown previously on the PGCCs with known distances.
Globally we observe the same behavior as for the whole selection with few additional features detailed below.
In the clumps, the contrast between the two column density contrast subsamples is slightly enhanced at low angles with the clumps associated to the filaments in high column density environment being more parallel with the background magnetic field.
Inside the filament areas, the magnetic field and matter show mostly no preferential relative orientation in the $\dnhhigh$ subsample.
Inside the clumps, perpendicular alignment is now also observed in the $\dnhlow$ subsample. 

\begin{figure}
\begin{center}
\includegraphics[width = 8 cm]{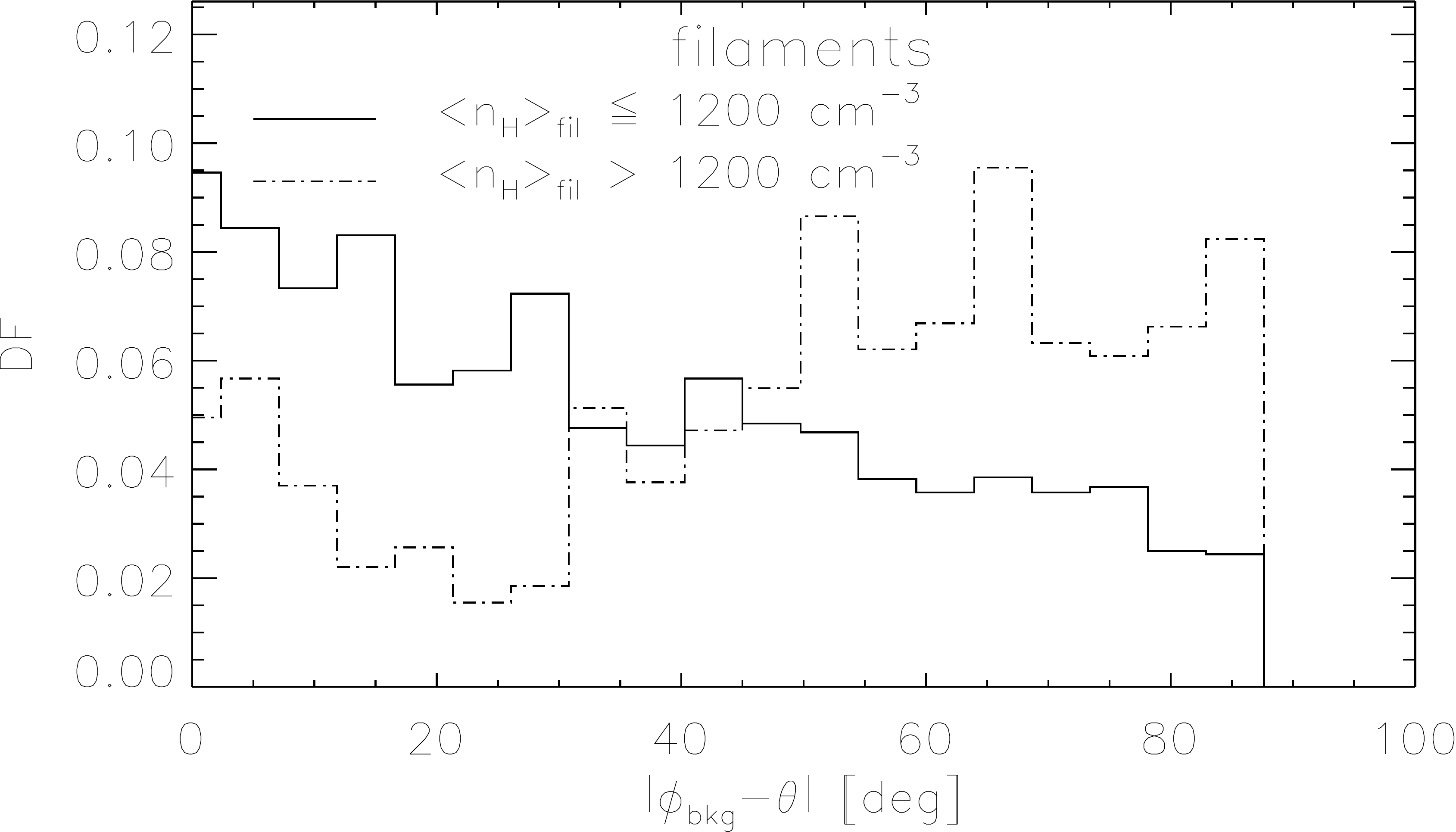}
\caption{The DFs of the relative orientation between the background magnetic field and filaments only for the PGCC with known distances.}
\label{fig:vol_dens_beta}
\end{center}
\end{figure}

Thresholding in $N_{H,bkg}$ and $\Delta N_H$ does not yield a threshold in $n_H$: there is a mix of low and high $n_H$ filaments and clumps in the 4 subsamples considered previously.
However, the average, minimal and maximal values are shifted towards lower or larger values according to the $N_H$ subsamples. 
For example, the average $n_H$ value is equal to $\sim 700 \rm cm^{-3}$ in the $\dnhlow$ and $\sim 1500 \rm cm^{-3}$ in the $\dnhhigh$ subsamples.
We find a preferential perpendicular alignment in the filament areas located in filaments (6 out of 26) with $\left\langle n_H \right\rangle_{fil} > \voldenslim$ while in those with lower gas volume densities there is few perpendicular alignment.
The corresponding DF is shown in Figure \ref{fig:vol_dens_beta}.
In the clumps we do not observe any particular difference compared to the results shown in Figure \ref{fig:beta_df}.

\begin{figure}
\begin{center}
\includegraphics[width = 8 cm]{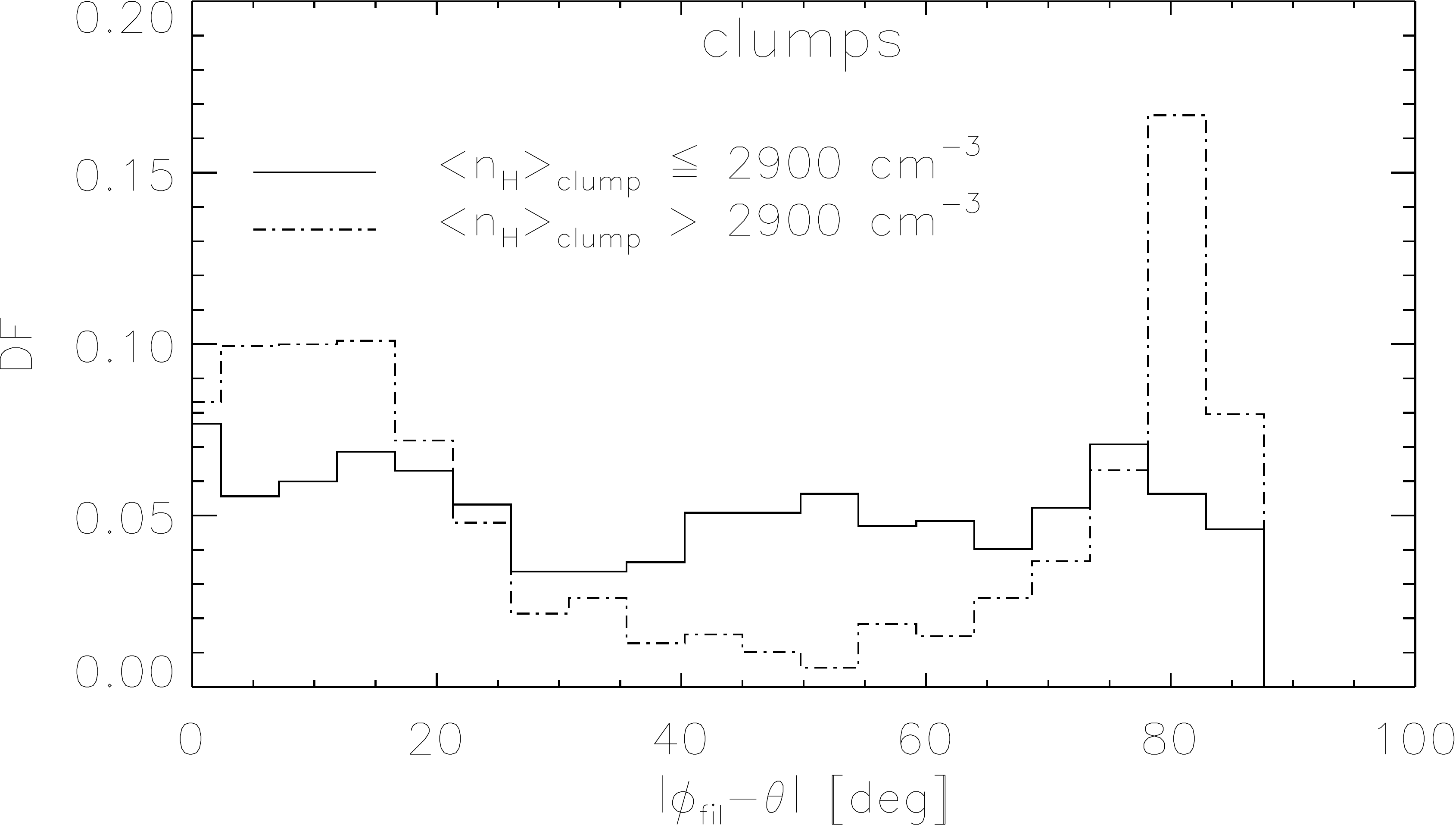}
\caption{The DFs of the relative orientation between the inner magnetic field and clumps only for the PGCC with known distances.}
\label{fig:vol_dens_nu}
\end{center}
\end{figure}

We perform the same analysis to study the relative orientation between the orientation of matter in the filament areas and magnetic field inside the filaments.
As it could be expected from Figure \ref{fig:nu_df}, there is a rather parallel relative orientation of matter in the filaments areas with respect to inner magnetic field. 
A difference in the DFs is observed when considering the clumps areas.
The behavior is found to change around $\left\langle n_H \right\rangle_{clump} = \voldenslimtwo$ and the corresponding DF is shown in Figure \ref{fig:vol_dens_nu}.
Above this density value, which applies for 7 clumps out of 26, the perpendicular counterpart of the relative orientation between the inner magnetic field and the clumps is important.
However, below this value the DF is flat and is not as conclusive as in the case of the column density contrast subsamples.

%%%%% ========== %%%%%
\section{Discussion}
\label{sec:discussion}
%%%%% ========== %%%%%
We have presented a statistical analysis of the relative orientations in
the POS between interstellar magnetic fields  
and filaments that were detected in the 353 GHz {\pk} data and
found to host Planck Cold Clumps.
We  selected 90 such filaments all
over the sky, excluding the Galactic plane, and
 retaining only filaments embedded in an environment with uniform magnetic field orientation.
The selected filaments span a range of background gas column density $N_{\rm H}$
from $1.2\times 10^{20}$\,cm$^{-2}$ to $3.8 \times 10^{21} \, \rm{cm^{-2}}$
(averaged values at an angular resolution of $7\arcmin$).

The two main novelties of our study are the following. First,
we separated the respective contributions of the filaments and their background,
assuming optically thin emission.
This separation enabled us to study the effects of the environment 
and of the evolutionary stages of the filaments through the background column density 
and column density contrast, respectively.
Second, in each filament, we drew a distinction between the area covered by clumps
(clump area) and the remaining area of the filament ({\filament}), 
in order to investigate possible different links between these structures relatively to the magnetic fields.

An advantage of our approach is in its applicability to large samples. Having
estimated the background contribution at some POS distance from the
filament, 
we could have missed the contribution of the immediate surroundings. 
However, the comparison with the results from 
detailed modeling of three individual filaments \citep{planck2014-XXXIII}
shows that this assumption is valid and introduces only a few degrees bias, 
which is negligible for the qualitative conclusions 
of our analysis. 
In addition, we did not account for projection effects.
As pointed out by \cite{planck2014-XXXII}, 
projection effects can produce a broadening of the distribution
functions at parallel relative orienation, whereas an observed perpendicular trend indicates 
a true perpendicular relative orientation in 3D.

The process of filament formation is generally governed by the interplay
between turbulence, self-gravity, and magnetic fields.
A statistical analysis of the relative orientations between filaments
and magnetic fields can bring new insights into the impact of magnetic fields.
Our large sample of filaments within different Galactic 
environments offers a distribution of viewing angle that may be assumed mainly random.

Below, we discuss the results presented in this work and compare them to previous studies.

    \begin{itemize}
    \item At the filament scale :  \\
    The {\pk} all-sky survey offers a unique opportunity for statistical studies of the magnetic field and density structures down to the scale of filaments and clumps. 
    The previous studies of the relative orientation between magnetic fields and filamentary structures have been performed at large scales, from intermediate and high-Galactic latitudes diffuse ISM \citep{planck2014-XXXII} to nearby molecular clouds of the Gould Belt \citep{planck2014-XXXV}. 
 
  The observed change of relative orientation with increasing $N_H$, from mostly parallel or having no preferred orientation to mostly perpendicular, has emphasized the significant role played by the magnetic field in shaping the structures in the diffuse ISM and molecular clouds. These observations were found to be in agreement with the predictions from numerical simulations of sub- or trans-Alfv\'enic MHD turbulence, where the magnetic field imposes an asymmetry in the formation of condensations from the diffuse ISM \citep{li2014}. 
  In this scenario, as discussed by \cite{soler2017}, low-density filaments could be produced by velocity shear induced by turbulence, stretching matter and magnetic field lines simultaneously, and resulting in parallel relative orientations between both.

More generally, in simulations with strong magnetic fields, the gas tends to be channelled along the field lines, and therefore converging flows driven by either large scale turbulence or self-gravity naturally produce dense filaments perpendicular to the magnetic field lines \citep{nakamura2008,price2008}. These converging flows could also explain the presence of less dense filaments (such as striations) along the magnetic field lines (\cite{stone1998,li2016}).
Magnetic fields lines may also be orientated along compressed gas regions at the fronts of the shocks between expanding bubbles which are supernova remnants or expanding HII regions \citep{inutsuka2015}.

 The contraction in the direction of the field lines that results in the formation of denser filaments and/or cores with their main axes perpendicular to the magnetic field direction was already suggested in the observational studies from \cite{palmeirim2013} in Taurus, \cite{matthews2013} in Lupus regions, or by \cite{cox2016} in Musca and \cite{malinen2016} in {L1642}.\\ 

Our results cannot be directly compared to those from \cite{planck2014-XXXV} where the derived polarization angles at the location of the filaments include the background contribution and the clumps counterparts, that can both have an impact in the histograms of relative orientation as shown in our study. Moreover, the alignment parameter used in \cite{planck2014-XXXV} is sampled within large angular range ($20\degr$) which fosters either parallel or perpendicular relative orientations and cannot be directly compared to our analysis based on the trends observed in the histograms.

However, we confirm the commonly observed parallel relative orientation for the filaments located in low-density environments, though with further interesting results.
By distinguishing subsamples according not only to the environment density but also to the filament density contrast, we observe different trends. In low-density environments, although there is a clear parallel alignment for low-contrast filaments (upper panel of Figure \ref{fig:beta_df2}), the ones with high-contrast do not show any preferential alignment with respect to the background magnetic field. These results suggest that even if the low-density regions are mostly dominated by turbulence and magnetic fields, self-gravity must play a significant role for higher density structures, resulting in an almost flat histogram of the relative orientations with the magnetic fields. 

In high-density regions, the orientation of the high contrast filaments do not show any dependence on the magnetic field orientation (upper panel of Figure \ref{fig:beta_df3}). This may indicate that gravity is also the dominant factor in this case. In fact, these structures with high density contrast can also be directly interpreted as the signature of regions where self-gravity dominates. On the contrary, the low contrast filaments exhibit a preferential perpendicular relative orientation which could be explained by a more important role of the magnetic field at this stage of evolution.\\ 

The limited angular resolution of the Planck data doesn't allow to probe the inner structure of the filament magnetic field, but still enables to estimate its mean orientation, as shown in \citep{planck2014-XXXIII}. 
We observe that the mean field is mainly parallel to the filament in both column density contrast subsamples, but shows slightly different trends. This may indicate that the low contrast filaments are formed mostly by turbulent motions or large scale dynamics that tend to compress simultaneously the matter and the magnetic field. The flatter distribution for the high contrast filaments suggests that the magnetic field is altered during the gravitational contraction.

    \item  On the transition in $N_H$: \\
    The selection of the two background column density subsamples corresponds to a division at the median value of our initial sample $N_{H,bkg} = \nhbkglim$ $\rm{cm^{-2}}$. It can however be compared to the threshold value in \cite{planck2014-XXXV} who derives $N_H \sim 5 \times 21$ $\rm{cm^{-2}}$, that gives $N_H \sim 2 \times 21$ $\rm{cm^{-2}}$ when using the same opacity $\kappa$ value (there is a factor of $\sim 2.5$ between the dust models opacity values adopted for the diffuse and the dense medium). It is also close to the threshold for the transition from aligned to perpendicular configuration determined by \cite{malinen2016} in the L1642 high-latitude star forming cloud, with $N_H \sim 1.6 \times 21$ $\rm{cm^{-2}}$.   

    To investigate the relative orientation as a function of the evolutionary stage of the filaments  more quantitatively, one would need to determine the linear mass of the filaments to assess whether  they are super-critical or not. 
    In our current study, supRHT is used to determine the direction of the filamentary structures but the method does not allow us to determine with accuracy the physical boundaries of the filaments and hence their linear masses. 

However, focussing our analysis on a subsample of filaments for which we have a distance estimate (in the nearby clouds, at distances lower than $500$ pc), we have shown that the change of relative orientation happens in the dense medium, with a transition in density around $ n_H  \sim 10^{3} \mathrm{cm^{-3}}$. Although this value is approximate, it reveals that the formation of dense filamentary structure is affected by the cloud-scale magnetic field. This result and the transition value estimate is consistent with the recent analysis by \citep{fissel2018} in Vela-C based on observations with BLAST-pol balloon-borne experiment and molecular lines tracing different gas densities. 

    \item At the clump scale. \\
    We observe, for the densest clumps of our sample, a bimodal distribution of the relative orientation between the matter and the magnetic field in the filaments. This is similar to the findings of \cite{zhang2014} in dense cores using SMA polarimetric measurements and JCMT continuum observations.
    \cite{ching2017} showed that within massive cores the magnetic field structure is complex, in contrast to the ordered parsec-scale magnetic fields in the filament. They also find that the major axes of the cores are either parallel or perpendicular to the magnetic field of the filament, suggesting that the parsec-scale magnetic fields play an important role in the formation of massive cores providing support against gravity.
    Nevertheless, \cite{poidevin2014} observed no preferential alignment between the distribution of cores observed with {\herschel} in the Lupus I region and the orientation of the magnetic fields on the 
    parsec-scale derived from the BLAST polarimetric measurements.\\
We emphasize that the different relative orientations observed (perpendicular, parallel or no preferential) may be related to the variety of the clumps in terms of nature and evolutionary stage.
The Planck clumps have been detected with a method using the color signature of cold sources, and based on the detection of the cold residuals emission after removal of the warmer background \citep{planck2011-7.7b}. Such a method results in the detection of extended cold components with a more complex morphology compared to methods identifying compact sources bright in their total intensity. The {\herschel} follow-up  of a representative sample of 350 PGCCs (in the frame of the Galactic Cold Cores Key Program) has revealed a rich and complex substructure within the Planck clumps, with sources at different evolutionary stages, from dense and cold cores (down to $7$ K) to protostellar objects \citep{planck2011-7.7a,juvela2012,montillaud2015}. The Planck clumps, being elongated and more extended than the Planck beam, could be considered as a "cold matrix" linking dense substructures to each other over a broad range of scales. More recently, the SCOPE JCMT legacy survey performed with the SCUBA-2 instrument at JCMT to map about 1000 PGCCs in 850 $\mathrm{\mu m}$ continuum emission, has  detected thousands of dense cores, identified as either starless cores or protostellar cores with young (Class 0/I) sources \citep{liu2016,tatematsu2017,juvela2017}.

\end{itemize}

%%%%% ========== %%%%%
\section{Conclusion and perspectives}
\label{sec:conclusion}
%%%%% ========== %%%%%
In this statistical analysis, our goal was to investigate whether the relative orientations in the POS between interstellar magnetic fields and filaments hosting Planck cold clumps may vary with the density of the environment and 
may also depend on the filaments evolutionary stage. 
For this purpose, 
we considered subsamples of filaments embedded in different background column densities and showing different density contrast with their environment. Furthermore, we drew a distinction between the area covered by clumps (clump area) and the remaining area of the filament ({\filament} area).

Our main results are summarized below:
\begin{enumerate}

\item Filaments versus background magnetic field
\begin{itemize}
\item The filaments embedded in low-density environment, i.e. below $N_{H,bkg} = \nhbkglim$ $\rm cm^{-2}$ (median value in our sample) show a clear parallel relative orientation with the background magnetic field (upper panel of Figure \ref{fig:beta_df}). 
However, while this preferential relative orientation is significant for the filaments with low density contrast, it disappears for high-contrast filaments (upper panel of Figure \ref{fig:beta_df2}). 
\item The filaments embedded in dense environment corresponding to column densities above $N_{H,bkg} = \nhbkglim$ $\rm cm^{-2}$, do not globally show any preferential relative orientation with respect to the background magnetic field (upper panel of Figure \ref{fig:beta_df}). 
\item Our column density limit between parallel and either perpendicular or no preferential relative orientation is close to the threshold values derived by \cite{malinen2016} and \cite{planck2014-XXXV} when using the same opacity as in our study (respectively $1.6 \, 10^{21}$ and $2 \, 10^{21}$ $\rm cm^{-2}$).
These results support the idea of a magnetic field strong enough to influence the formation of filaments. However, even if the low-density regions are mostly dominated by turbulence and magnetic fields, leading to preferential parallel alignments, self-gravity must play a significant role for the higher density structures. Although a perpendicular alignment is observed for low density contrast filaments in high-density environment, no preferential relative orientation is prevailing for the high-contrast subsamples, whatever the density of their environment. \\
\item In the clump areas, both perpendicular and parallel relative orientations are observed regardless of the environment column density (lower panel of Figure \ref{fig:beta_df}). However, in the low column density environment, the relative orientation turns from parallel to mostly perpendicular as one moves from the lowest to highest density contrast (lower panel of Figure \ref{fig:beta_df2}). The finding indicates that the large scale (background) magnetic fields play an important role not only during the formation of filaments but also of their embedded clumps.
\item We also find that the relative orientation between the background magnetic field and the filaments strongly depends not only on the density, but also on the nature of the structures such as the embedded clumps or the hosting filaments.
\end{itemize}

\item Filaments versus their inner magnetic field
\begin{itemize}
\item For the filaments, we find a preferential parallel relative orientation in both low and high density contrast subsamples which is consistent with the study from \cite{planck2014-XXXIII}. However, we note that for the high density contrast filaments, the distribution is flatter (upper panel of Figure \ref{fig:nu_df}).
\item For the clumps, in both subsamples we also find preferential parallel relative orientation. However, in the high column density contrast subsample, there is a significant peak of the DF at angles larger than $70^{\circ}$ which indicates perpendicular relative orientation (lower panel of Figure \ref{fig:nu_df}).
This could be a signature of the presence of dense structures that were formed by contraction along the magnetic field lines with their major axes perpendicular to it.
\end{itemize}

\item Filament magnetic field versus background magnetic field
\begin{itemize}
\item The magnetic fields in the background and in the filaments are mostly parallel to each other for the low column density contrast filaments. For the high column density contrast filaments the parallel relative orientation between both fields is rare while globally we observe no preferential relative orientation (Figure \ref{fig:b_angles_diff}). This relative alignment depends also on the environment column density, being weaker for the filaments embedded in denser environments (Figure \ref{fig:app_b_angles_diff}). This could point to a coupling between the matter and the magnetic field during the filament formation process, as emphasized in the study by \cite{planck2014-XXXIII}.
\end{itemize}

\item Relative orientation in the nearby filaments \\
We have looked for a possible correspondence between the relative orientation of filaments and magnetic fields (both background and inner) depending on the gas volume density using filaments of our sample with a distance estimate and associated to nearby molecular clouds.
\begin{itemize}
\item Filaments with densities larger (respectively lower) than $\sim \voldenslim $ are mostly perpendicular (respectively parallel) with respect to the background magnetic field (Figure \ref{fig:vol_dens_beta}).
\item Densest clumps (densities larger than $\voldenslimtwo$) contribute to either parallel or perpendicular relative orientation with respect to the inner magnetic field while no preferential relative orientation is established for clumps with lower densities (Figure \ref{fig:vol_dens_nu}). This could disentangle the similar behavior observed in the clumps in the low and high column density environments, shown in the lower panel of Figure \ref{fig:beta_df}. Denser clumps show tendency to be perpendicular to the filament magnetic field.
\end{itemize}
We emphasize that these values of volume densities are subject to limitations due to uncertainties in the distance estimates and to the assumptions on the filaments and clumps shape and geometry, and thus are only tentative values.
However, these results show that the change of relative orientation happens in the dense medium, and that our transition value estimate is consistent with the recent analysis by \cite{fissel2018} in Vela-C based on observations with BLAST-pol balloon-borne experiment and molecular lines tracing different gas densities.
\end{enumerate}

This study brings information on the role of the magnetic fields in the evolution of the dense ISM matter at the intermediate scale between molecular clouds and dense cores.
It reveals several regimes in the evolution of the filaments where the magnetic field is one of the dominant factors.
Our results also indicate that the evolution of the filament with respect to its environment as well as the environment itself is the key in studying the role played by the magnetic field in structuring the matter, and can already be investigated with the {\pk} data.
A dedicated analysis on known molecular clouds could bring new insights on the origin of the parallel or perpendicular alignment depending on the evolution of the clouds and/or efficiency of star formation in the clouds.
We would like to emphasize that the contribution of the Cold Clumps should be considered in future studies of filamentary structures.

A more precise study of the interplay between matter and magnetic fields in the Cold Clumps at higher resolution could disentangle the dependence of the relative orientation on the geometry of the clump and of its evolutionary stage.
Recent attempts in this field were performed with, e.g., the {\pk} follow-ups of the PGCCs with the SCUBA-2/POL-2 polarimeter at JCMT (\cite{liu2018}, \cite{juvela2018}). 
We have also started a statistical analysis of the polarization properties of a large sample of clumps from the PGCCs in order to investigate the dependency of the polarized emission with the column density, considering the respective impacts of the dust alignment efficiency and the magnetic field geometry (Ristorcelli et al. in prep.).

%%%%% ========== %%%%%
\section*{Acknowledgements}
%%%%% ========== %%%%%

Funding from MES RK state-targeted program BR05236454 is acknowledged.
Funding from NU ORAU grant SST 2015021 is acknowledged. 
D.A. and E.A. acknowledge the NU Faculty Development Competitive Program (grant No. 110119FD4503).
M.J. and E.R.M. acknowledge the support of the Academy of Finland Grant No. 285769. 
I.R, K.F., J.-P.B. and L.M. thank the support from the Programme National Physique et Chimie du Milieu Interstellaire (PCMI) of CNRS/INSU with INC/INP co-funded by CEA and CNES.

D.A. thanks Professor Anvar Shukurov for his suggestions on the statistical aspects of the study.

The authors thank the anonymous referee for the valuable comments that helped to improve the quality of the paper.

\bibliographystyle{mnras}
\bibliography{thebibliography.bib}

%%%%%%%%%%%%%%%%% APPENDICES %%%%%%%%%%%%%%%%%%%%%

\appendix
\onecolumn

\begin{footnotesize}
\begin{landscape}
\begin{longtable}{cccccccccccccccc}
\caption{Characteristics of the filaments used in the study. Columns: number in
the PGCC catalogue, Galactic longitude, Galactic latitude, length, width, average
position angle given by the SupRHT method using Gaussian fit, its uncertainty,
average column density over the filament and its uncertainty, average column
density over the filament excluding clumps areas and its uncertainty, average
column density over the background region and its uncertainty, average magnetic
field angle over the filament using Gaussian fit, average magnetic field angle
over the filament using circular statistics and its uncertainty.} \\
\hline
Num & \textit{l} & \textit{b} & \textit{s} & \textit{w} & $\langle\theta \rangle_{\rm fil,2}$ & $\sigma(\langle \theta \rangle_{\rm fil,2})$ & $ \langle N_{\rm H,tot} \rangle $ & $\sigma(N_{\rm H,tot})$ & $ \langle N_{\rm H,fil} \rangle $ & $\sigma(N_{\rm H,fil})$ & $ \langle N_{\rm H,bkg} \rangle $ & $\sigma(\rm N_{H,bkg})$ & $\langle\psi\rangle_2$ & $ \langle \psi \rangle_1$ & $\sigma(\psi)_1$ \\
&&&&&&&&&&&&&&& \\
$$ & $[\degr]$ & $[\degr]$ & $[']$ & $[']$  & $[\degr]$ & $[\degr]$ & $[\times 10^{20}$ & $[\times 10^{20}$ & $[\times 10^{20}$ & $[\times 10^{20}$ & $[\times 10^{20}$ & $[\times 10^{20}$ & $[\degr]$ & $[\degr]$  & $[\degr]$  \\
$$ & $$ & $$ & $$ & $$ & $$ & $$ & $\rm{cm^{-2}}]$ & $\rm{cm^{-2}}]$ & $\rm{cm^{-2}}]$ & $\rm{cm^{-2}}]$ & $\rm{cm^{-2}}]$ & $\rm{cm^{-2}}]$ & $$ & $$  & $$  \\
\hline
\endfirsthead
\hline
Num & \textit{l} & \textit{b} & \textit{s} & \textit{w} & $\langle\theta \rangle_{\rm fil,2}$ & $\sigma(\langle \theta \rangle_{\rm fil,2})$ & $ \langle N_{\rm H,tot} \rangle $ & $\sigma(N_{\rm H,tot})$ & $ \langle N_{\rm H,fil} \rangle $ & $\sigma(N_{\rm H,fil})$ & $ \langle N_{\rm H,bkg} \rangle $ & $\sigma(\rm N_{H,bkg})$ & $\langle\psi\rangle_2$ & $ \langle \psi \rangle_1$ & $\sigma(\psi)_1$ \\
&&&&&&&&&&&&&&& \\
$$ & $[\degr]$ & $[\degr]$ & $[']$ & $[']$  & $[\degr]$ & $[\degr]$ & $[\times 10^{20}$ & $[\times 10^{20}$ & $[\times 10^{20}$ & $[\times 10^{20}$ & $[\times 10^{20}$ & $[\times 10^{20}$ & $[\degr]$ & $[\degr]$  & $[\degr]$  \\
$$ & $$ & $$ & $$ & $$ & $$ & $$ & $\rm{cm^{-2}}]$ & $\rm{cm^{-2}}]$ & $\rm{cm^{-2}}]$ & $\rm{cm^{-2}}]$ & $\rm{cm^{-2}}]$ & $\rm{cm^{-2}}]$ & $$ & $$  & $$  \\
\hline
\endhead
%\hline
27  &  1.395  &  20.933  &  92.0  &  11.0  &  84.0  &  1.0  &  20.9  &  13.2  &  17.9  &  6.71  &  20.9  &  13.2  &  N/A  &  N/A  &  N/A \\ 
54  &  300.797  &  -9.101  &  81.0  &  7.0  &  9.0  &  1.0  &  35.6  &  5.79  &  34.7  &  4.95  &  35.6  &  5.79  &  -71.0  &  -72.3  &  10.1 \\ 
67  &  355.324  &  14.725  &  103.0  &  14.0  &  -57.0  &  12.0  &  66.0  &  13.0  &  63.9  &  11.5  &  66.0  &  13.0  &  N/A  &  N/A  &  N/A \\ 
75  &  161.671  &  -35.917  &  59.0  &  7.0  &  52.0  &  7.0  &  11.0  &  6.07  &  9.49  &  4.48  &  11.0  &  6.07  &  67.0  &  56.0  &  36.8 \\ 
80  &  110.627  &  -12.496  &  60.0  &  7.0  &  60.0  &  4.0  &  11.2  &  2.18  &  10.7  &  1.65  &  11.2  &  2.18  &  56.0  &  48.3  &  30.3 \\ 
86  &  110.418  &  11.532  &  55.0  &  6.0  &  24.0  &  2.0  &  25.5  &  5.51  &  23.2  &  3.37  &  25.5  &  5.51  &  29.0  &  25.7  &  21.0 \\ 
88  &  301.267  &  -8.25  &  84.0  &  9.0  &  32.0  &  3.0  &  32.5  &  6.71  &  31.1  &  6.06  &  32.5  &  6.71  &  -68.0  &  -57.9  &  47.2 \\ 
119  &  301.598  &  -7.834  &  55.0  &  8.0  &  34.0  &  5.0  &  31.3  &  7.37  &  33.2  &  7.57  &  31.3  &  7.37  &  -72.0  &  -59.9  &  50.7 \\ 
124  &  28.46  &  -6.401  &  47.0  &  5.0  &  53.0  &  12.0  &  15.8  &  7.66  &  12.5  &  1.99  &  15.8  &  7.66  &  47.0  &  40.6  &  37.4 \\ 
138  &  142.645  &  7.738  &  84.0  &  7.0  &  15.0  &  20.0  &  25.1  &  4.63  &  23.6  &  2.72  &  25.1  &  4.63  &  81.0  &  33.5  &  56.2 \\ 
158  &  314.958  &  -21.971  &  90.0  &  8.0  &  20.0  &  5.0  &  10.3  &  3.33  &  9.36  &  2.65  &  10.3  &  3.33  &  38.0  &  36.6  &  38.2 \\ 
185  &  140.983  &  5.733  &  43.0  &  6.0  &  -33.0  &  2.0  &  46.9  &  11.7  &  41.5  &  6.24  &  46.9  &  11.7  &  -64.0  &  -44.6  &  42.8 \\ 
210  &  312.501  &  -22.741  &  77.0  &  9.0  &  42.0  &  12.0  &  10.0  &  4.37  &  9.39  &  3.82  &  10.0  &  4.37  &  27.0  &  26.8  &  6.0 \\ 
219  &  7.657  &  21.185  &  50.0  &  7.0  &  52.0  &  22.0  &  27.0  &  7.44  &  23.6  &  3.94  &  27.0  &  7.44  &  -56.0  &  -35.6 &  47.5 \\ 
233  &  205.458  &  -14.561  &  63.0  &  7.0  &  -36.0  &  6.0  &  82.8  &  38.4  &  75.5  &  35.4  &  82.8  &  38.4  &  -72.0  &  -30.5  &  64.7 \\ 
235  &  325.567  &  6.034  &  48.0  &  7.0  &  7.0  &  5.0  &  25.9  &  10.7  &  23.5  &  10.6  &  25.9  &  10.7  &  N/A  &  N/A  &  N/A \\ 
238  &  159.412  &  -34.371  &  41.0  &  7.0  &  58.0  &  3.0  &  48.8  &  20.8  &  50.7  &  22.9  &  48.8  &  20.8  &  -50.0  &  -45.6  &  18.4 \\ 
266  &  121.929  &  -7.668  &  52.0  &  7.0  &  36.0  &  13.0  &  20.0  &  3.16  &  19.1  &  2.5  &  20.0  &  3.16  &  -55.0  &  -50.9  &  24.7 \\ 
366  &  317.275  &  6.132  &  49.0  &  4.0  &  13.0  &  11.0  &  26.3  &  8.67  &  21.7  &  7.16  &  26.3  &  8.67  &  19.0  &  22.8  &  25.6 \\ 
403  &  298.318  &  -13.616  &  66.0  &  9.0  &  -201.0  &  50.0  &  15.3  &  6.62  &  13.8  &  5.49  &  15.3  &  6.62  &  -66.0  &  -8.5  &  53.5 \\ 
448  &  160.528  &  -19.727  &  47.0  &  8.0  &  65.0  &  7.0  &  26.6  &  4.61  &  23.9  &  2.31  &  26.6  &  4.61  &  N/A  &  N/A  &  N/A \\ 
498  &  337.688  &  7.485  &  55.0  &  7.0  &  -26.0  &  12.0  &  27.2  &  7.68  &  23.5  &  5.04  &  27.2  &  7.68  &  78.0  &  55.7  &  46.6 \\ 
504  &  173.118  &  2.355  &  53.0  &  6.0  &  75.0  &  4.0  &  93.5  &  23.7  &  83.9  &  17.4  &  93.5  &  23.7  &  N/A  &  N/A  &  N/A \\ 
516  &  201.717  &  -11.22  &  71.0  &  8.0  &  50.0  &  14.0  &  29.0  &  4.97  &  28.4  &  4.3  &  29.0  &  4.97  &  N/A  &  N/A  &  N/A \\ 
520  &  91.299  &  -38.159  &  70.0  &  10.0  &  -23.0  &  8.0  &  5.02  &  1.63  &  4.44  &  1.12  &  5.02  &  1.63  &  N/A  &  N/A  &  N/A \\ 
533  &  169.964  &  -18.991  &  88.0  &  7.0  &  17.0  &  12.0  &  29.2  &  10.6  &  27.5  &  9.59  &  29.2  &  10.6  &  -86.0  &  -41.4  &  59.4 \\ 
562  &  28.714  &  3.882  &  66.0  &  6.0  &  73.0  &  2.0  &  176.0  &  73.9  &  138.0  &  33.9  &  176.0  &  73.9  &  N/A  &  N/A  &  N/A \\ 
633  &  127.882  &  2.675  &  74.0  &  8.0  &  -35.0  &  27.0  &  33.0  &  4.36  &  32.0  &  2.21  &  33.0  &  4.36  &  -64.0  &  -36.8  &  50.6 \\ 
673  &  171.411  &  -17.376  &  86.0  &  7.0  &  52.0  &  6.0  &  40.6  &  15.7  &  41.7  &  16.9  &  40.6  &  15.7  &  53.0  &  47.9  &  24.3 \\ 
680  &  158.225  &  -20.287  &  78.0  &  6.0  &  -16.0  &  6.0  &  78.6  &  57.9  &  78.8  &  60.1  &  78.6  &  57.9  &  N/A  &  N/A  &  N/A \\ 
684  &  202.364  &  2.504  &  77.0  &  8.0  &  -74.0  &  3.0  &  63.6  &  20.2  &  60.7  &  18.4  &  63.6  &  20.2  &  N/A  &  N/A  &  N/A \\ 
715  &  107.483  &  -9.376  &  37.0  &  12.0  &  46.0  &  4.0  &  13.1  &  2.93  &  12.7  &  2.65  &  13.1  &  2.93  &  57.0  &  50.6  &  38.3 \\ 
747  &  117.871  &  -10.719  &  69.0  &  6.0  &  -190.0  &  43.0  &  4.94  &  0.646  &  4.85  &  0.608  &  4.94  &  0.646  &  28.0  &  22.0  &  42.7 \\ 
748  &  220.738  &  -8.055  &  37.0  &  9.0  &  -32.0  &  2.0  &  31.1  &  6.37  &  35.5  &  6.11  &  31.1  &  6.37  &  -84.0  &  -61.8  &  58.0 \\ 
751  &  93.614  &  -4.456  &  72.0  &  8.0  &  -45.0  &  11.0  &  49.3  &  18.9  &  40.1  &  19.2  &  49.3  &  18.9  &  N/A  &  N/A  &  N/A \\ 
790  &  82.369  &  9.768  &  44.0  &  4.0  &  -1.0  &  5.0  &  7.1  &  1.83  &  5.55  &  0.711  &  7.1  &  1.83  &  N/A  &  N/A  &  N/A \\ 
794  &  170.0  &  -16.132  &  74.0  &  7.0  &  73.0  &  8.0  &  68.5  &  18.6  &  69.0  &  19.2  &  68.5  &  18.6  &  N/A  &  N/A  &  N/A \\ 
830  &  159.517  &  3.252  &  75.0  &  6.0  &  -148.0  &  32.0  &  31.3  &  4.43  &  30.7  &  4.25  &  31.3  &  4.43  &  67.0  &  55.2  &  40.1 \\ 
834  &  157.592  &  -8.898  &  74.0  &  6.0  &  -80.0  &  5.0  &  44.9  &  14.2  &  45.5  &  15.2  &  44.9  &  14.2  &  N/A  &  N/A  &  N/A \\ 
846  &  165.157  &  -7.566  &  82.0  &  6.0  &  67.0  &  10.0  &  33.6  &  8.86  &  32.5  &  8.66  &  33.6  &  8.86  &  88.0  &  39.2  &  70.6 \\ 
852  &  100.328  &  14.847  &  73.0  &  9.0  &  67.0  &  7.0  &  16.1  &  3.42  &  15.9  &  3.54  &  16.1  &  3.42  &  -43.0  &  -42.6  &  11.8 \\ 
895  &  154.957  &  -15.184  &  39.0  &  9.0  &  66.0  &  2.0  &  25.5  &  6.94  &  25.7  &  7.71  &  25.5  &  6.94  &  N/A  &  N/A  &  N/A \\ 
937  &  121.036  &  -9.959  &  72.0  &  7.0  &  80.0  &  5.0  &  9.29  &  3.23  &  9.25  &  3.35  &  9.29  &  3.23  &  -41.0  &  -21.9  &  48.1 \\ 
943  &  157.242  &  -3.933  &  65.0  &  9.0  &  -56.0  &  7.0  &  28.1  &  5.63  &  27.8  &  7.02  &  28.1  &  5.63  &  -30.0  &  -23.3  &  36.2 \\ 
966  &  27.709  &  -21.028  &  84.0  &  9.0  &  53.0  &  12.0  &  7.61  &  2.86  &  7.3  &  2.8  &  7.61  &  2.86  &  N/A  &  N/A  &  N/A \\ 
996  &  216.0  &  -15.978  &  92.0  &  9.0  &  45.0  &  8.0  &  43.5  &  21.5  &  42.8  &  22.7  &  43.5  &  21.5  &  N/A  &  N/A  &  N/A \\ 
1019  &  165.622  &  3.342  &  56.0  &  5.0  &  50.0  &  8.0  &  18.8  &  4.0  &  17.4  &  2.63  &  18.8  &  4.0  &  88.0  &  3.2  &  66.4 \\ 
1025  &  9.833  &  21.5  &  86.0  &  7.0  &  53.0  &  2.0  &  16.9  &  4.88  &  16.6  &  4.87  &  16.9  &  4.88  &  17.0  &  12.2  &  34.7 \\ 
1128  &  139.186  &  -3.289  &  66.0  &  6.0  &  -67.0  &  5.0  &  32.2  &  5.1  &  31.4  &  4.59  &  32.2  &  5.1  &  -19.0  &  -21.9  &  17.0 \\ 
1136  &  317.678  &  -28.655  &  46.0  &  6.0  &  -33.0  &  7.0  &  5.68  &  1.13  &  5.24  &  0.557  &  5.68  &  1.13  &  -5.0  &  -12.5  &  34.2 \\ 
1230  &  171.92  &  -15.66  &  70.0  &  3.0  &  84.0  &  1.0  &  43.6  &  12.9  &  43.2  &  13.0  &  43.6  &  12.9  &  -46.0  &  -11.8  &  49.7 \\ 
1269  &  170.824  &  -15.913  &  75.0  &  8.0  &  70.0  &  6.0  &  48.4  &  13.1  &  48.2  &  13.6  &  48.4  &  13.1  &  N/A  &  N/A  &  N/A \\ 
1270  &  302.488  &  -4.706  &  63.0  &  12.0  &  13.0  &  3.0  &  18.8  &  3.79  &  16.9  &  1.6  &  18.8  &  3.79  &  N/A  &  N/A  &  N/A \\ 
1342  &  175.333  &  -10.828  &  35.0  &  5.0  &  -77.0  &  9.0  &  19.2  &  1.99  &  18.1  &  0.413  &  19.2  &  1.99  &  -10.0  &  -13.7  &  18.1 \\ 
1466  &  7.548  &  4.261  &  79.0  &  8.0  &  14.0  &  7.0  &  25.1  &  5.26  &  24.1  &  4.46  &  25.1  &  5.26  &  36.0  &  36.2  &  16.3 \\ 
1480  &  266.964  &  -4.641  &  35.0  &  8.0  &  -30.0  &  14.0  &  35.5  &  5.4  &  35.0  &  5.42  &  35.5  &  5.4  &  N/A  &  N/A  &  N/A \\ 
1496  &  58.039  &  3.037  &  62.0  &  7.0  &  2.0  &  5.0  &  40.0  &  5.98  &  37.9  &  5.06  &  40.0  &  5.98  &  27.0  &  30.8  &  19.1 \\ 
1519  &  128.213  &  13.701  &  72.0  &  12.0  &  72.0  &  6.0  &  28.2  &  4.36  &  27.8  &  4.18  &  28.2  &  4.36  &  87.0  &  -20.6  &  60.7 \\ 
1528  &  121.862  &  -8.766  &  79.0  &  7.0  &  23.0  &  7.0  &  10.8  &  2.21  &  10.2  &  2.05  &  10.8  &  2.21  &  -3.0  &  14.7  &  33.4 \\ 
1535  &  185.412  &  -8.962  &  55.0  &  6.0  &  -63.0  &  9.0  &  15.5  &  3.2  &  15.5  &  3.37  &  15.5  &  3.2  &  N/A  &  N/A  &  N/A \\ 
1555  &  171.747  &  -15.627  &  70.0  &  5.0  &  84.0  &  1.0  &  44.1  &  11.2  &  39.1  &  8.4  &  44.1  &  11.2  &  N/A  &  N/A  &  N/A \\ 
1594  &  268.228  &  -3.238  &  68.0  &  10.0  &  44.0  &  8.0  &  48.9  &  9.21  &  46.5  &  6.4  &  48.9  &  9.21  &  N/A  &  N/A  &  N/A \\ 
1602  &  149.679  &  3.562  &  96.0  &  9.0  &  58.0  &  8.0  &  58.5  &  12.1  &  58.1  &  12.5  &  58.5  &  12.1  &  19.0  &  17.8  &  21.0 \\ 
1669  &  26.839  &  6.787  &  51.0  &  5.0  &  48.0  &  4.0  &  43.0  &  8.05  &  39.8  &  5.12  &  43.0  &  8.05  &  -88.0  &  -22.3  &  64.9 \\ 
1695  &  236.513  &  -4.999  &  50.0  &  8.0  &  -62.0  &  2.0  &  19.8  &  2.64  &  19.4  &  2.81  &  19.8  &  2.64  &  -82.0  &  -1.5  &  50.7 \\ 
1759  &  311.356  &  -19.419  &  69.0  &  6.0  &  -88.0  &  5.0  &  4.85  &  0.968  &  4.29  &  0.344  &  4.85  &  0.968  &  -68.0  &  -46.0  &  46.2 \\ 
1778  &  294.553  &  5.928  &  70.0  &  8.0  &  18.0  &  4.0  &  11.1  &  1.87  &  9.89  &  1.18  &  11.1  &  1.87  &  75.0  &  16.3  &  55.1 \\ 
1789  &  204.562  &  -11.02  &  72.0  &  6.0  &  15.0  &  4.0  &  30.5  &  12.5  &  30.2  &  13.3  &  30.5  &  12.5  &  13.0  &  12.4  &  16.9 \\ 
1835  &  124.843  &  -6.04  &  34.0  &  12.0  &  61.0  &  9.0  &  14.9  &  2.38  &  13.2  &  1.92  &  14.9  &  2.38  &  -53.0  &  -30.9  &  48.9 \\ 
1846  &  156.729  &  35.137  &  54.0  &  6.0  &  68.0  &  4.0  &  3.42  &  1.07  &  2.53  &  0.502  &  3.42  &  1.07  &  67.0  &  41.4  &  45.2 \\ 
1874  &  320.845  &  5.094  &  65.0  &  19.0  &  -5.0  &  27.0  &  18.7  &  1.61  &  18.5  &  1.37  &  18.7  &  1.61  &  -79.0  &  -31.5  &  59.2 \\ 
1955  &  210.97  &  -19.329  &  75.0  &  19.0  &  75.0  &  6.0  &  133.0  &  48.1  &  134.0  &  51.3  &  133.0  &  48.1  &  N/A  &  N/A  &  N/A \\ 
1959  &  101.592  &  22.68  &  80.0  &  7.0  &  26.0  &  6.0  &  4.87  &  0.707  &  4.52  &  0.508  &  4.87  &  0.707  &  22.0  &  21.4  &  11.6 \\ 
2048  &  181.842  &  -18.381  &  87.0  &  6.0  &  30.0  &  9.0  &  28.7  &  11.8  &  28.9  &  12.2  &  28.7  &  11.8  &  N/A  &  N/A  &  N/A \\ 
2050  &  111.153  &  -40.985  &  74.0  &  6.0  &  -80.0  &  5.0  &  2.23  &  0.707  &  2.15  &  0.651  &  2.23  &  0.707  &  -73.0  &  3.4  &  54.5 \\ 
2176  &  10.381  &  3.167  &  62.0  &  8.0  &  -30.0  &  11.0  &  46.3  &  9.1  &  43.0  &  8.32  &  46.3  &  9.1  &  N/A  &  N/A  &  N/A \\ 
2212  &  264.449  &  5.545  &  77.0  &  6.0  &  -20.0  &  6.0  &  16.8  &  1.97  &  16.8  &  2.08  &  16.8  &  1.97  &  60.0  &  36.0  &  44.5 \\ 
2246  &  301.148  &  -16.179  &  62.0  &  6.0  &  -78.0  &  4.0  &  19.1  &  7.96  &  20.1  &  9.06  &  19.1  &  7.96  &  N/A  &  N/A  &  N/A \\ 
2308  &  217.071  &  -16.255  &  45.0  &  10.0  &  65.0  &  10.0  &  17.8  &  3.62  &  15.4  &  2.85  &  17.8  &  3.62  &  -20.0  &  -22.1  &  13.3 \\ 
2315  &  171.435  &  -4.45  &  69.0  &  5.0  &  -36.0  &  4.0  &  17.5  &  3.72  &  16.8  &  4.07  &  17.5  &  3.72  &  -28.0  &  -7.6  &  47.8 \\ 
2346  &  222.834  &  -5.606  &  83.0  &  7.0  &  30.0  &  4.0  &  25.4  &  1.97  &  25.3  &  1.99  &  25.4  &  1.97  &  82.0  &  25.3  &  56.7 \\ 
2465  &  219.551  &  -3.85  &  39.0  &  5.0  &  4.0  &  16.0  &  17.9  &  2.73  &  15.8  &  1.19  &  17.9  &  2.73  &  52.0  &  17.1  &  44.0 \\ 
2511  &  147.822  &  6.124  &  66.0  &  8.0  &  -58.0  &  9.0  &  22.1  &  2.05  &  21.4  &  1.88  &  22.1  &  2.05  &  -87.0  &  -24.4  &  73.4 \\ 
2531  &  336.453  &  2.827  &  71.0  &  6.0  &  -86.0  &  6.0  &  53.0  &  5.61  &  52.1  &  5.12  &  53.0  &  5.61  &  -17.0  &  -16.4  &  23.2 \\ 
2539  &  94.629  &  -31.306  &  92.0  &  9.0  &  -49.0  &  29.0  &  4.63  &  1.11  &  4.66  &  1.13  &  4.63  &  1.11  &  -53.0  &  -42.7  &  33.9 \\ 
2583  &  314.948  &  -19.129  &  51.0  &  7.0  &  66.0  &  4.0  &  3.18  &  0.415  &  2.93  &  0.143  &  3.18  &  0.415  &  85.0  &  17.3  &  65.2 \\ 
2587  &  112.809  &  2.76  &  70.0  &  5.0  &  -77.0  &  3.0  &  53.1  &  14.0  &  52.6  &  14.3  &  53.1  &  14.0  &  N/A  &  N/A  &  N/A \\ 
2602  &  230.663  &  -10.392  &  85.0  &  7.0  &  -32.0  &  15.0  &  7.77  &  1.18  &  7.45  &  0.975  &  7.77  &  1.18  &  73.0  &  20.6  &  49.4 \\ 
2614  &  7.666  &  -7.856  &  86.0  &  7.0  &  -63.0  &  8.0  &  12.8  &  2.03  &  12.8  &  2.08  &  12.8  &  2.03  &  N/A  &  N/A  &  N/A \\ 
2627  &  179.744  &  -2.625  &  73.0  &  11.0  &  47.0  &  13.0  &  34.0  &  5.6  &  33.5  &  5.54  &  34.0  &  5.6  &  46.0  &  30.7  &  45.2 \\ 
2703  &  229.255  &  -4.656  &  54.0  &  4.0  &  -84.0  &  2.0  &  12.8  &  3.74  &  12.0  &  3.96  &  12.8  &  3.74  &  63.0  &  26.6  &  50.9 \\ 
2710  &  202.033  &  2.85  &  56.0  &  17.0  &  -74.0  &  3.0  &  60.7  &  22.5  &  65.6  &  21.5  &  60.7  &  22.5  &  -56.0  &  -56.1  &  13.1 \\ 
2760  &  144.957  &  4.62  &  42.0  &  7.0  &  18.0  &  9.0  &  27.0  &  1.7  &  26.8  &  1.35  &  27.0  &  1.7  &  32.0  &  35.8  &  17.1 \\ 
2778  &  359.781  &  11.933  &  80.0  &  6.0  &  -71.0  &  8.0  &  18.3  &  3.7  &  18.5  &  3.82  &  18.3  &  3.7  &  -57.0  &  -4.7  &  53.5 \\ 
2927  &  89.494  &  2.041  &  68.0  &  10.0  &  63.0  &  12.0  &  89.6  &  14.9  &  88.9  &  14.9  &  89.6  &  14.9  &  N/A  &  N/A  &  N/A \\ 
2953  &  309.837  &  -24.868  &  69.0  &  10.0  &  -65.0  &  8.0  &  3.69  &  0.48  &  3.56  &  0.364  &  3.69  &  0.48  &  -66.0  &  -63.7  &  22.7 \\ 
2964  &  8.264  &  21.383  &  60.0  &  6.0  &  22.0  &  3.0  &  27.6  &  8.03  &  28.4  &  8.36  &  27.6  &  8.03  &  0.0  &  10.7  &  31.9 \\ 
3115  &  103.313  &  -23.938  &  48.0  &  5.0  &  76.0  &  6.0  &  3.1  &  0.374  &  3.0  &  0.313  &  3.1  &  0.374  &  -78.0  &  -42.7  &  55.3 \\ 
3133  &  65.317  &  -2.652  &  48.0  &  6.0  &  78.0  &  7.0  &  52.9  &  9.96  &  49.6  &  13.8  &  52.9  &  9.96  &  -51.0  &  -44.8  &  34.4 \\ 
3134  &  23.384  &  10.052  &  54.0  &  7.0  &  -1.0  &  2.0  &  27.4  &  3.27  &  25.4  &  1.82  &  27.4  &  3.27  &  31.0  &  29.4  &  14.8 \\ 
3148  &  298.036  &  5.975  &  47.0  &  7.0  &  -24.0  &  8.0  &  12.8  &  1.86  &  11.8  &  1.12  &  12.8  &  1.86  &  N/A  &  N/A  &  N/A \\ 
3182  &  289.848  &  -7.503  &  98.0  &  10.0  &  45.0  &  13.0  &  7.85  &  1.17  &  7.62  &  1.05  &  7.85  &  1.17  &  -82.0  &  6.9  &  55.1 \\ 
3439  &  53.706  &  6.929  &  49.0  &  6.0  &  20.0  &  7.0  &  17.2  &  2.84  &  15.9  &  2.17  &  17.2  &  2.84  &  N/A  &  N/A  &  N/A \\ 
3488  &  129.124  &  20.312  &  71.0  &  9.0  &  66.0  &  9.0  &  6.52  &  1.24  &  6.48  &  1.33  &  6.52  &  1.24  &  74.0  &  60.6  &  38.2 \\ 
3561  &  29.121  &  2.224  &  59.0  &  5.0  &  61.0  &  6.0  &  54.2  &  5.99  &  53.2  &  5.34  &  54.2  &  5.99  &  N/A  &  N/A  &  N/A \\ 
3570  &  115.528  &  9.004  &  71.0  &  9.0  &  43.0  &  10.0  &  20.1  &  2.78  &  19.7  &  2.71  &  20.1  &  2.78  &  37.0  &  39.9  &  14.7 \\ 
3593  &  101.279  &  17.602  &  88.0  &  9.0  &  -37.0  &  4.0  &  12.2  &  2.69  &  12.4  &  3.07  &  12.2  &  2.69  &  -7.0  &  -5.6  &  27.6 \\ 
3700  &  254.402  &  -3.935  &  71.0  &  12.0  &  51.0  &  14.0  &  29.2  &  4.18  &  29.4  &  4.24  &  29.2  &  4.18  &  -70.0  &  -51.0  &  45.9 \\ 
3711  &  106.604  &  -19.538  &  51.0  &  6.0  &  -75.0  &  8.0  &  5.59  &  1.09  &  6.01  &  0.625  &  5.59  &  1.09  &  -66.0  &  -15.1  &  54.3 \\ 
3749  &  187.089  &  -33.345  &  76.0  &  6.0  &  -14.0  &  6.0  &  9.51  &  0.81  &  9.4  &  0.76  &  9.51  &  0.81  &  N/A  &  N/A  &  N/A \\ 
3831  &  150.424  &  3.475  &  53.0  &  5.0  &  41.0  &  2.0  &  38.0  &  6.84  &  35.1  &  5.7  &  38.0  &  6.84  &  73.0  &  56.1  &  51.8 \\ 
3888  &  158.791  &  3.374  &  51.0  &  6.0  &  -84.0  &  7.0  &  25.8  &  7.6  &  32.5  &  5.67  &  25.8  &  7.6  &  60.0  &  54.3  &  26.4 \\ 
3907  &  118.001  &  16.754  &  49.0  &  8.0  &  57.0  &  6.0  &  5.77  &  0.741  &  5.36  &  0.418  &  5.77  &  0.741  &  -13.0  &  -18.4  &  41.5 \\ 
3935  &  162.39  &  9.907  &  53.0  &  8.0  &  10.0  &  3.0  &  8.77  &  0.758  &  9.14  &  0.426  &  8.77  &  0.758  &  -69.0  &  -43.6  &  47.1 \\ 
3987  &  170.1  &  -7.731  &  69.0  &  8.0  &  -71.0  &  5.0  &  25.3  &  4.39  &  25.2  &  5.05  &  25.3  &  4.39  &  -29.0  &  -15.8  &  36.0 \\ 
4187  &  100.772  &  14.924  &  52.0  &  6.0  &  -61.0  &  4.0  &  9.89  &  1.45  &  9.73  &  1.65  &  9.89  &  1.45  &  -35.0  &  -29.6  &  32.6 \\ 
4192  &  191.502  &  8.333  &  81.0  &  9.0  &  -69.0  &  8.0  &  7.17  &  2.01  &  6.32  &  1.4  &  7.17  &  2.01  &  78.0  &  7.3  &  52.3 \\ 
4240  &  300.611  &  -2.901  &  38.0  &  3.0  &  -1.0  &  24.0  &  26.1  &  3.73  &  25.7  &  3.7  &  26.1  &  3.73  &  N/A  &  N/A  &  N/A \\ 
4258  &  168.081  &  -5.408  &  70.0  &  7.0  &  79.0  &  3.0  &  25.5  &  2.52  &  25.1  &  1.88  &  25.5  &  2.52  &  -84.0  &  -50.9  &  56.3 \\ 
4272  &  113.039  &  16.309  &  39.0  &  4.0  &  4.0  &  4.0  &  15.1  &  3.21  &  13.2  &  0.919  &  15.1  &  3.21  &  14.0  &  13.6  &  7.5 \\ 
4330  &  166.176  &  -16.496  &  63.0  &  9.0  &  -67.0  &  12.0  &  20.8  &  4.81  &  19.4  &  3.67  &  20.8  &  4.81  &  N/A  &  N/A  &  N/A \\ 
4397  &  2.244  &  3.387  &  75.0  &  7.0  &  -53.0  &  2.0  &  47.2  &  7.87  &  46.1  &  7.82  &  47.2  &  7.87  &  N/A  &  N/A  &  N/A \\ 
4406  &  2.432  &  14.73  &  79.0  &  6.0  &  -65.0  &  16.0  &  8.94  &  1.89  &  8.81  &  1.85  &  8.94  &  1.89  &  -69.0  &  -62.2  &  37.3 \\ 
4503  &  155.301  &  -23.076  &  59.0  &  5.0  &  -191.0  &  42.0  &  6.97  &  1.62  &  7.2  &  1.74  &  6.97  &  1.62  &  N/A  &  N/A  &  N/A \\ 
4610  &  3.472  &  10.813  &  88.0  &  7.0  &  -60.0  &  6.0  &  9.58  &  1.57  &  8.98  &  1.28  &  9.58  &  1.57  &  -62.0  &  -28.2  &  53.3 \\ 
4789  &  215.573  &  -15.685  &  38.0  &  5.0  &  -3.0  &  4.0  &  21.2  &  3.51  &  21.2  &  3.74  &  21.2  &  3.51  &  N/A  &  N/A  &  N/A \\ 
4806  &  168.421  &  -5.388  &  81.0  &  16.0  &  80.0  &  4.0  &  25.5  &  2.16  &  25.6  &  2.23  &  25.5  &  2.16  &  -64.0  &  -30.1  &  49.3 \\ 
4838  &  160.261  &  -19.793  &  60.0  &  7.0  &  66.0  &  8.0  &  27.0  &  4.72  &  28.2  &  5.11  &  27.0  &  4.72  &  N/A  &  N/A  &  N/A \\ 
4957  &  159.174  &  -9.002  &  60.0  &  4.0  &  -80.0  &  4.0  &  24.0  &  1.62  &  23.7  &  1.3  &  24.0  &  1.62  &  N/A  &  N/A  &  N/A \\ 
5459  &  320.924  &  14.906  &  45.0  &  7.0  &  -1.0  &  4.0  &  4.28  &  0.62  &  4.0  &  0.512  &  4.28  &  0.62  &  0.0  &  -3.7  &  44.5 \\ 
5482  &  212.171  &  4.301  &  69.0  &  7.0  &  2.0  &  3.0  &  12.9  &  1.36  &  12.7  &  1.41  &  12.9  &  1.36  &  N/A  &  N/A  &  N/A \\ 
5499  &  169.382  &  -3.89  &  61.0  &  12.0  &  -54.0  &  3.0  &  22.1  &  2.44  &  20.1  &  1.44  &  22.1  &  2.44  &  -56.0  &  -33.8  &  44.3 \\ 
5601  &  110.211  &  10.254  &  69.0  &  9.0  &  -47.0  &  18.0  &  13.7  &  1.91  &  13.7  &  1.97  &  13.7  &  1.91  &  N/A  &  N/A  &  N/A \\ 
5688  &  105.836  &  4.713  &  67.0  &  6.0  &  0.0  &  12.0  &  32.4  &  7.16  &  31.3  &  7.45  &  32.4  &  7.16  &  -23.0  &  -23.0  &  23.6 \\ 
5777  &  151.172  &  -10.066  &  75.0  &  7.0  &  84.0  &  1.0  &  6.79  &  0.804  &  6.66  &  0.725  &  6.79  &  0.804  &  -79.0  &  -30.2  &  57.3 \\ 
5811  &  115.762  &  22.701  &  75.0  &  7.0  &  15.0  &  6.0  &  4.57  &  0.527  &  4.3  &  0.402  &  4.57  &  0.527  &  N/A  &  N/A  &  N/A \\ 
5849  &  291.473  &  -4.201  &  62.0  &  10.0  &  78.0  &  4.0  &  21.4  &  2.45  &  20.8  &  1.97  &  21.4  &  2.45  &  N/A  &  N/A  &  N/A \\ 
13203  &  281.768  &  -30.794  &  68.0  &  2.0  &  73.0  &  4.0  &  5.39  &  1.08  &  5.25  &  1.26  &  5.39  &  1.08  &  -19.0  &  -10.4  &  44.2 \\ 
\hline
\label{tab:table}
\end{longtable}
\end{landscape}
\end{footnotesize}

\twocolumn

\appendix
\section{Kernel dimensions tests}
\label{sec:app_kernel}
We test kernels of different sizes: $k_{\rm s}= 11 \arcmin, \, 21 \arcmin, \, 31 \arcmin$ noted as K11, K21, and K31, respectively, and observe their effect on the detection of the filaments in the intensity minimaps. 
These selected sizes are motivated by the sizes of the elongated clumps and by the resolution of the data.
K11 yields positive detection for Bok globules.
K11 and K21 yield positive detections for cometary shape clumps.
K31 is the most restrictive in detecting linear filaments that span in two directions from the center of the minimaps.
The kernel width $k_{\rm w} = 6 \arcmin$ is found to eliminate the round-shape isolated clumps.
Larger widths do not allow us to detect the filaments at the position of the clumps because of the average size of the clumps, which is about $6'$.
Resulting maps of $\irht$ better reflect the variations of the structures compared to larger values of $k_{\rm w}$ whereas smaller values are not appropriate as they do not allow us to take the clump width into account.   

\section{Estimation of uncertainties}
\label{app:sigma}
In order to properly estimate the uncertainties, we proceed to bootstrap on independent pixels that corresponds to pixels situated at the minimum distances of 7' between each other, which is equal to the resolution of the data.
This diminishes the number of the pixels used. 
Thus, there are only $3\%$ of the pixels considered in each bootstrap round.
The histograms of uncertainties are calculated as the ratio of standard deviations of the $\nboot$ realizations of bootstrap over the values in each bin of the corresponding histograms ($\sigma_N/N$).
We have checked that the average bootstrapped DFs are compatible with the DFs obtained directly from the data.

\section{Extra figures}
\label{sec:app}
\begin{figure}

\includegraphics[width=8cm]{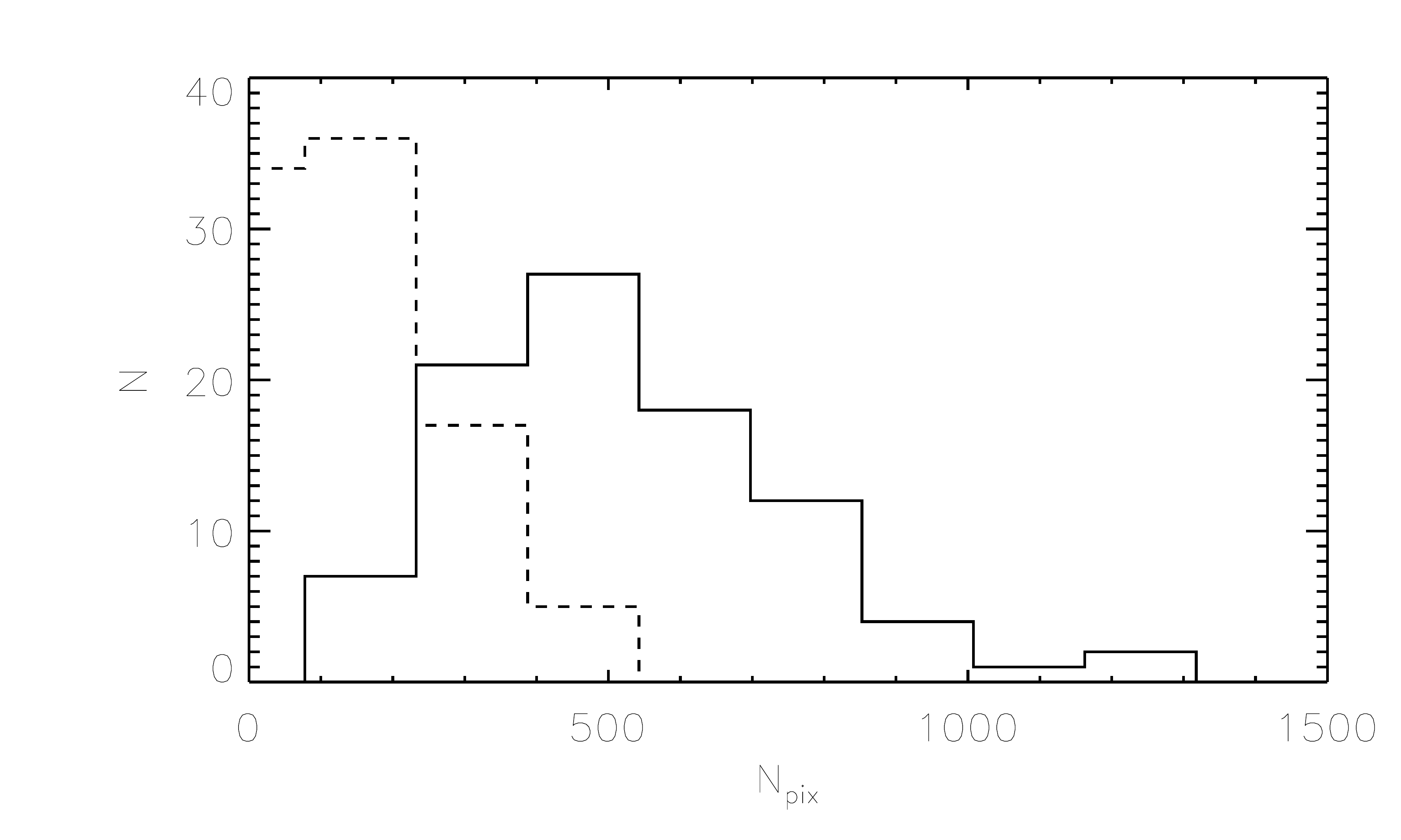}
\caption{Histogram of the number of pixels ($N_\mathrm{pix}$) in the filaments only (solid line) and in the clumps only (dashed line).}
\label{fig:pix_hist}

\end{figure}

\begin{figure}
\includegraphics[width = 8 cm]{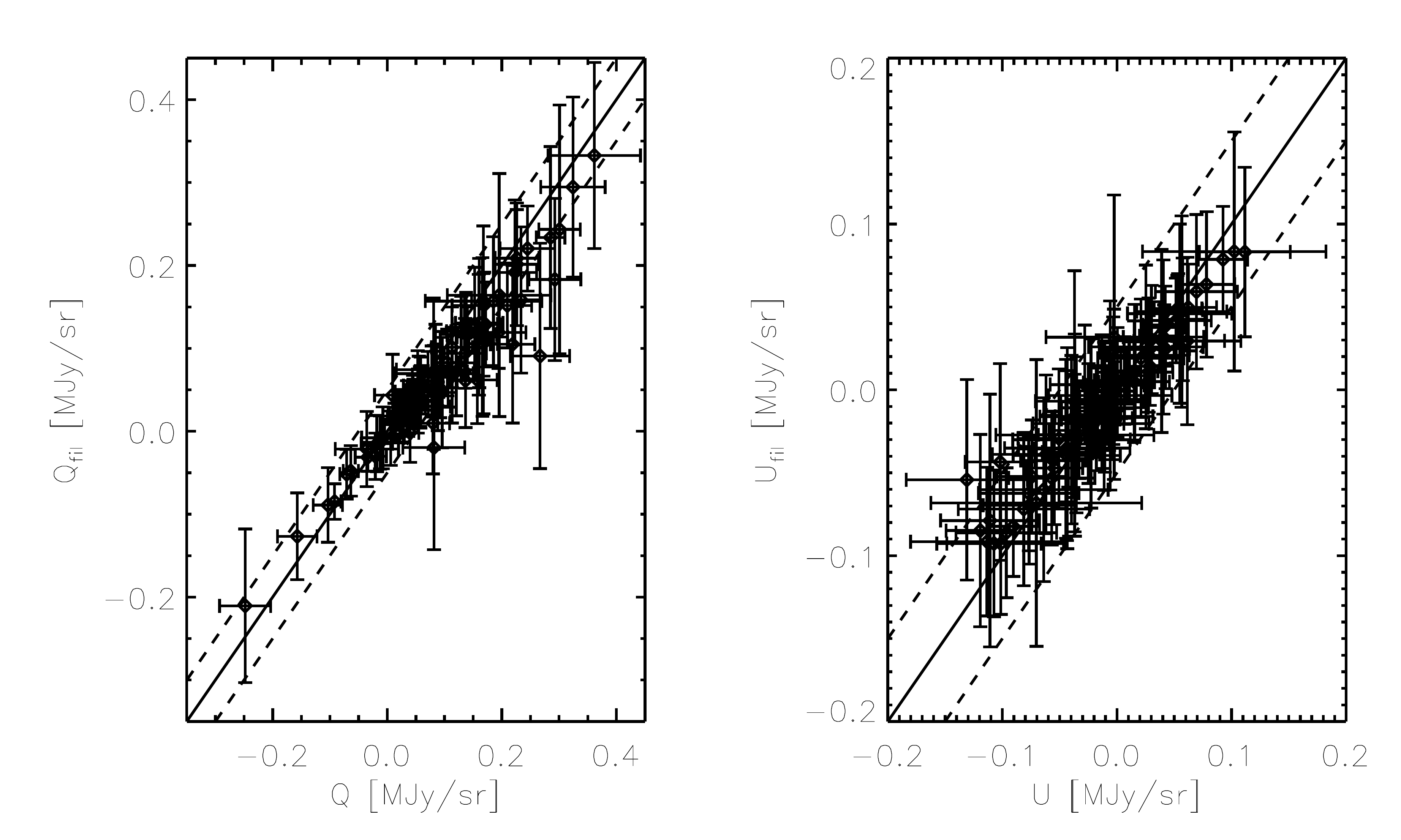}
\caption{Scatter plot between the average Stokes parameters for each filament before and after the background subtraction in the filament pixels ($Q$ is shown in the left panel and $U$ is shown in the right panel). The error bars show the standard deviation in each filament. The lines show the one-to-one correlation and the dashed lines delimit $50\%$ variation.}
\label{fig:qu_before_after}
\end{figure}

\begin{figure}
\includegraphics[width = 8 cm]{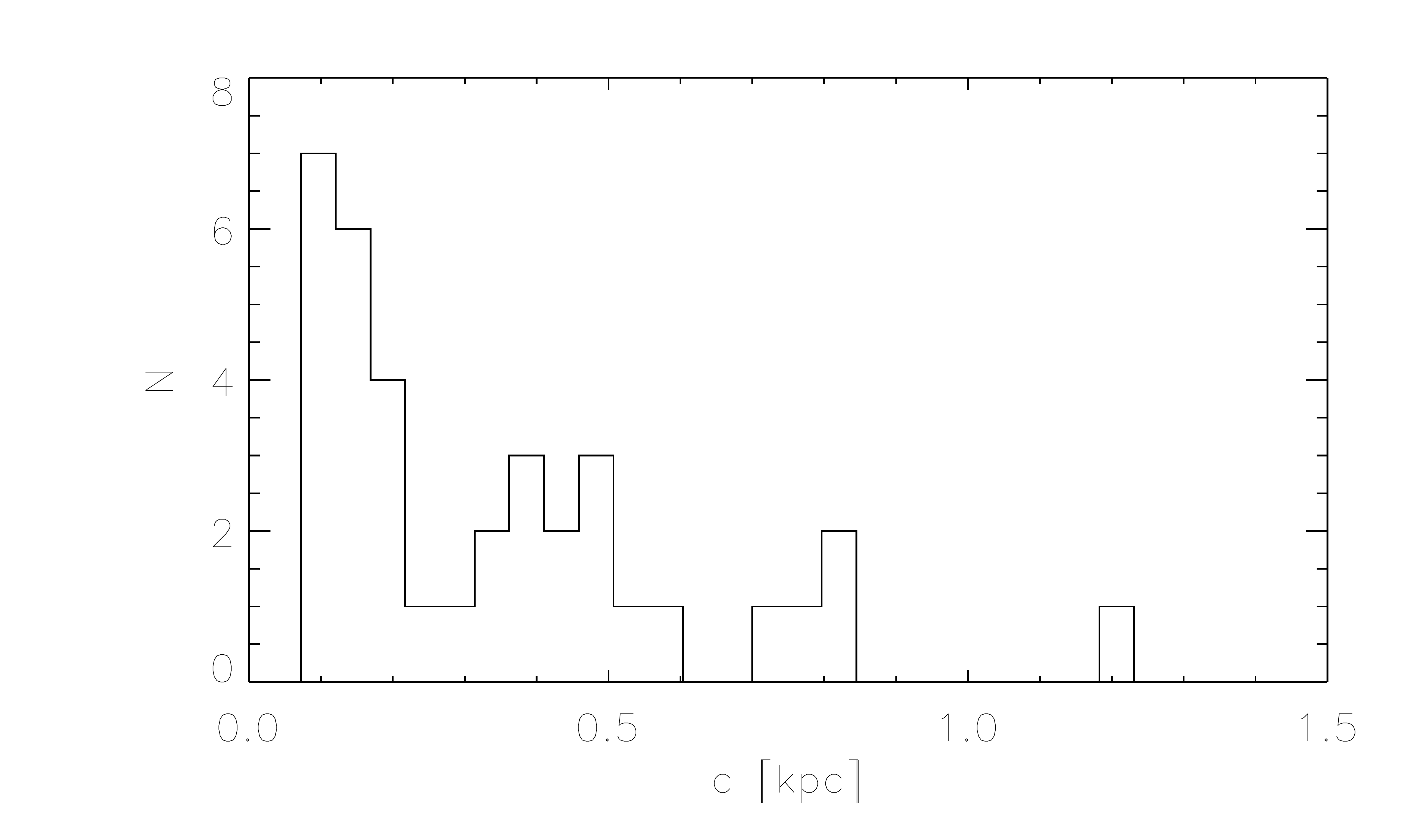}
\caption{Histogram of the distances provided in the PGCC catalogue for the filaments used in our study.}
\label{fig:dist_hist}
\end{figure}

\begin{figure}
\includegraphics[width = 8 cm]{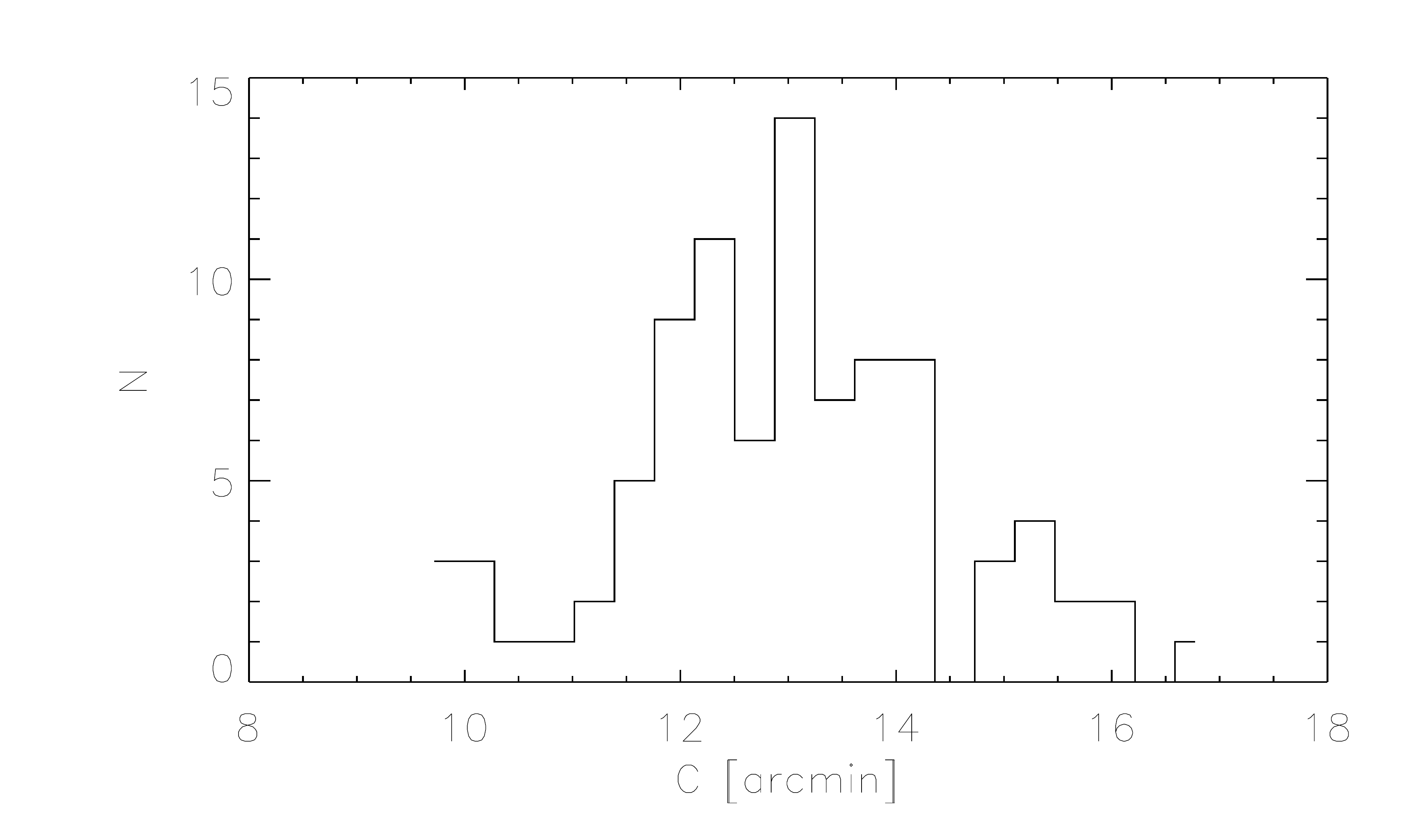}
\caption{Histogram of the average clump sizes.}
\label{fig:size_hist}
\end{figure}

\begin{figure}
\includegraphics[width = 7.5 cm]{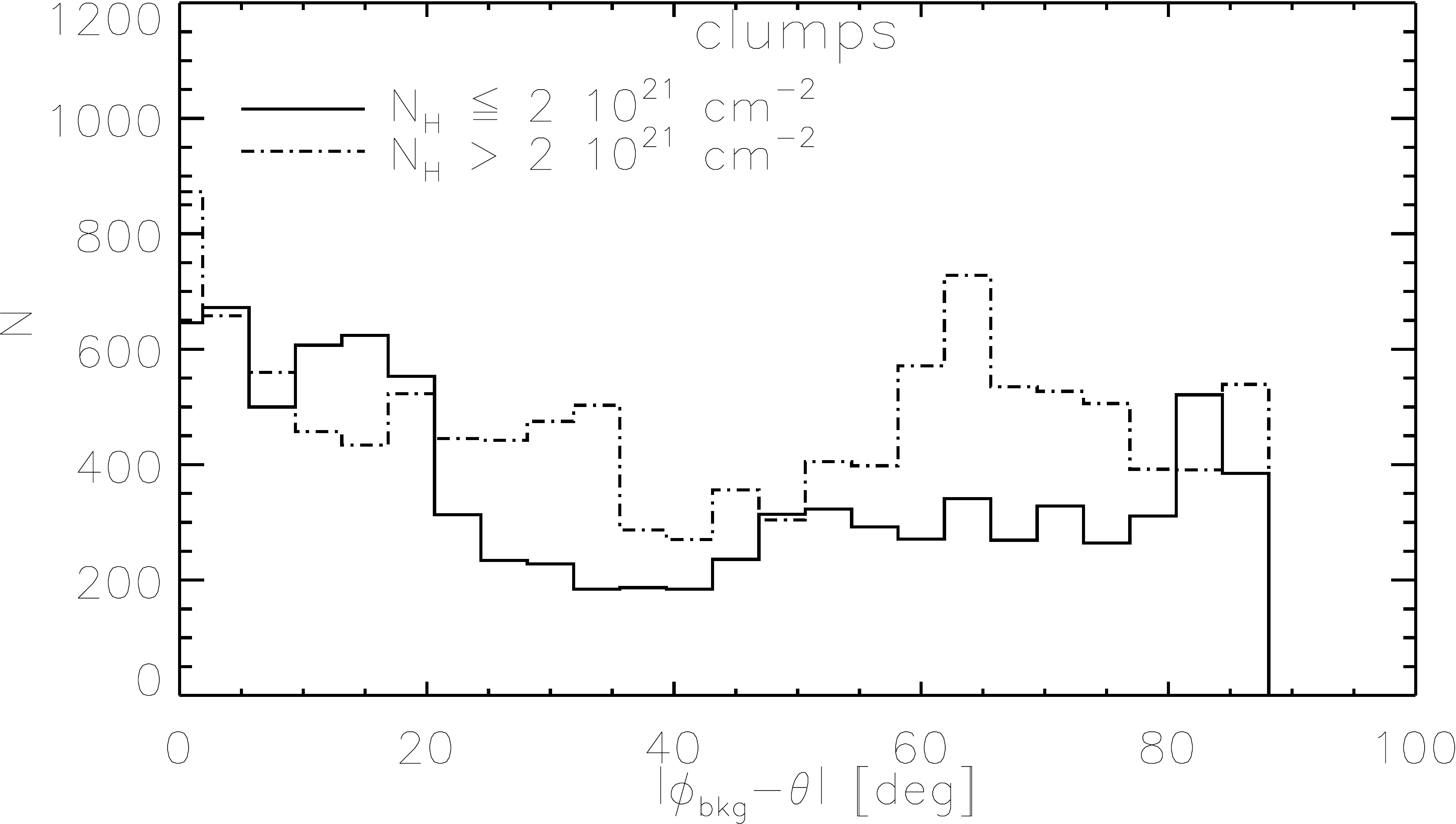}
\caption{Histograms of absolute relative orientation between background magnetic field and orientation of the clumps for the clumps with densities lower and higher than $2 \, 10^{21}$ $\rm cm^{-2}$ (solid and dash-dotted lines respectively).}
\label{fig:ad_hoc_clumps}
\end{figure}

\begin{figure}
\includegraphics[width=7.5  cm]{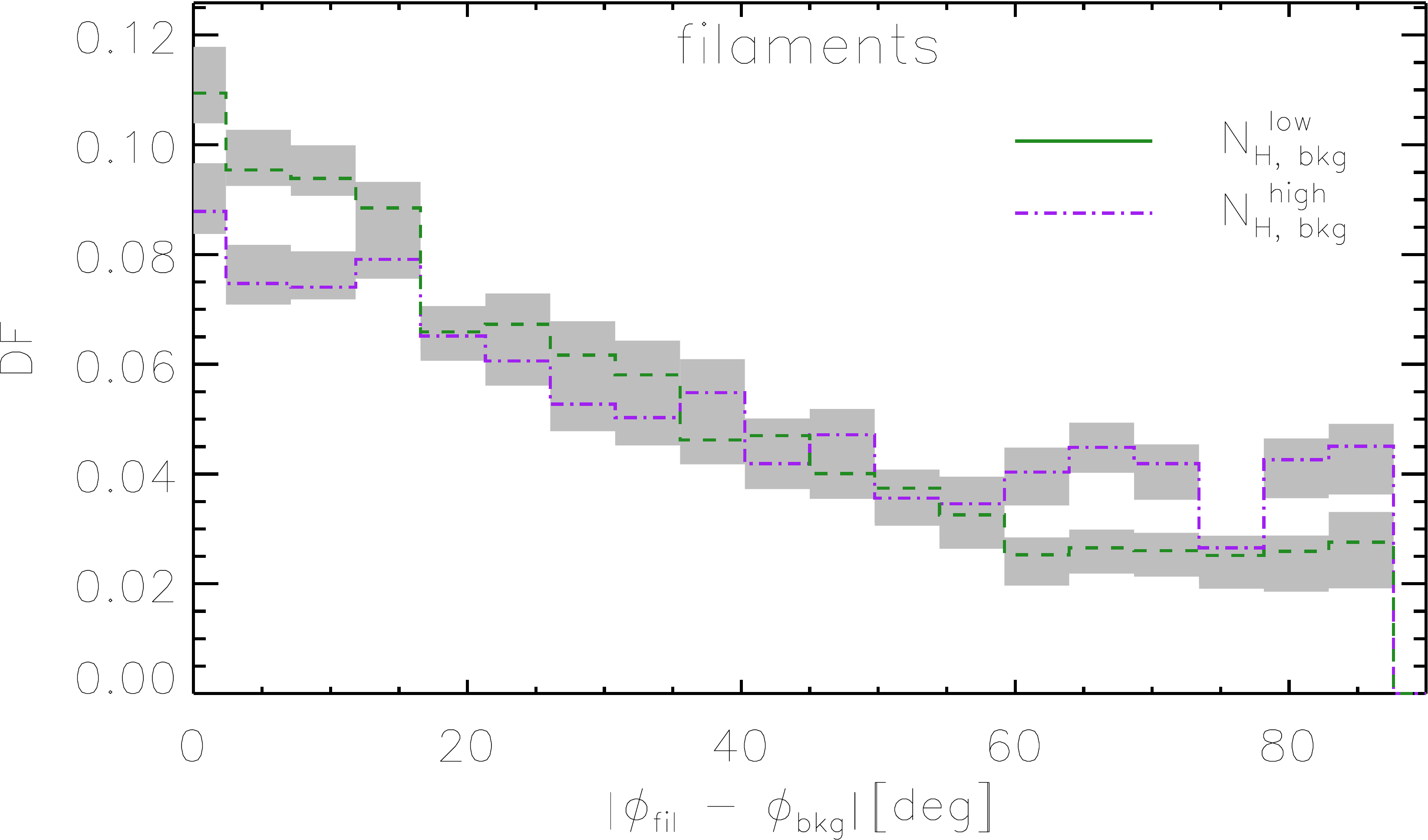}
    \caption{DF of the average difference between the magnetic field angles in the filaments and in the background. The green and purple lines show the DFs for the  $\nhbkglow$ and $\nhbkghigh$ subsamples respectively. The uncertainties are shown as gray shaded areas.}
 \label{fig:app_b_angles_diff}
\end{figure}

\begin{figure}
\includegraphics[width=8cm]{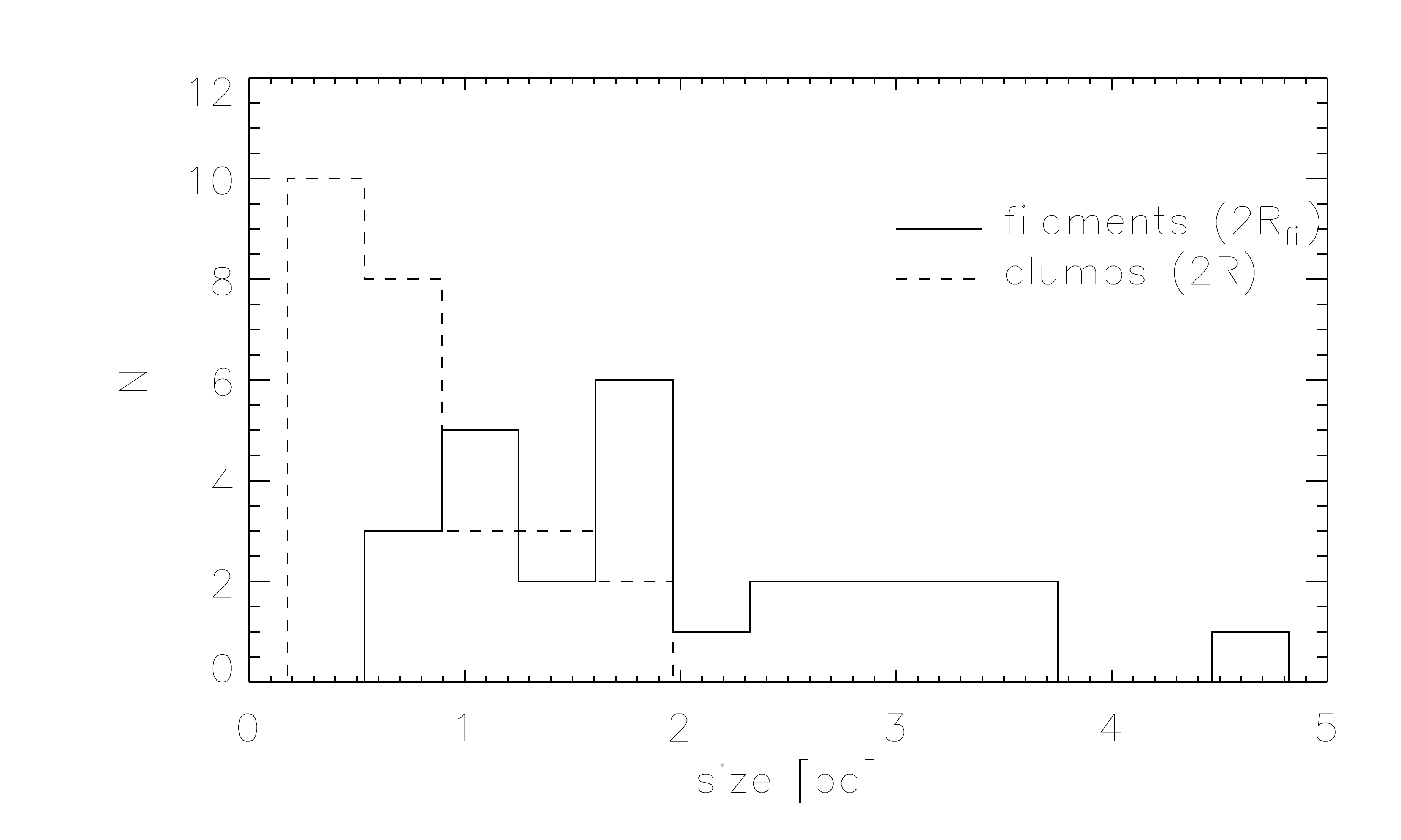}
\caption{Histograms of the average clump size and  filament width taken at the center of the corresponding clump for the PGCCs with the distances lower than $500$ pc.}
\label{fig:size_pc_hist}

\end{figure}

\begin{figure}
\includegraphics[width=8cm]{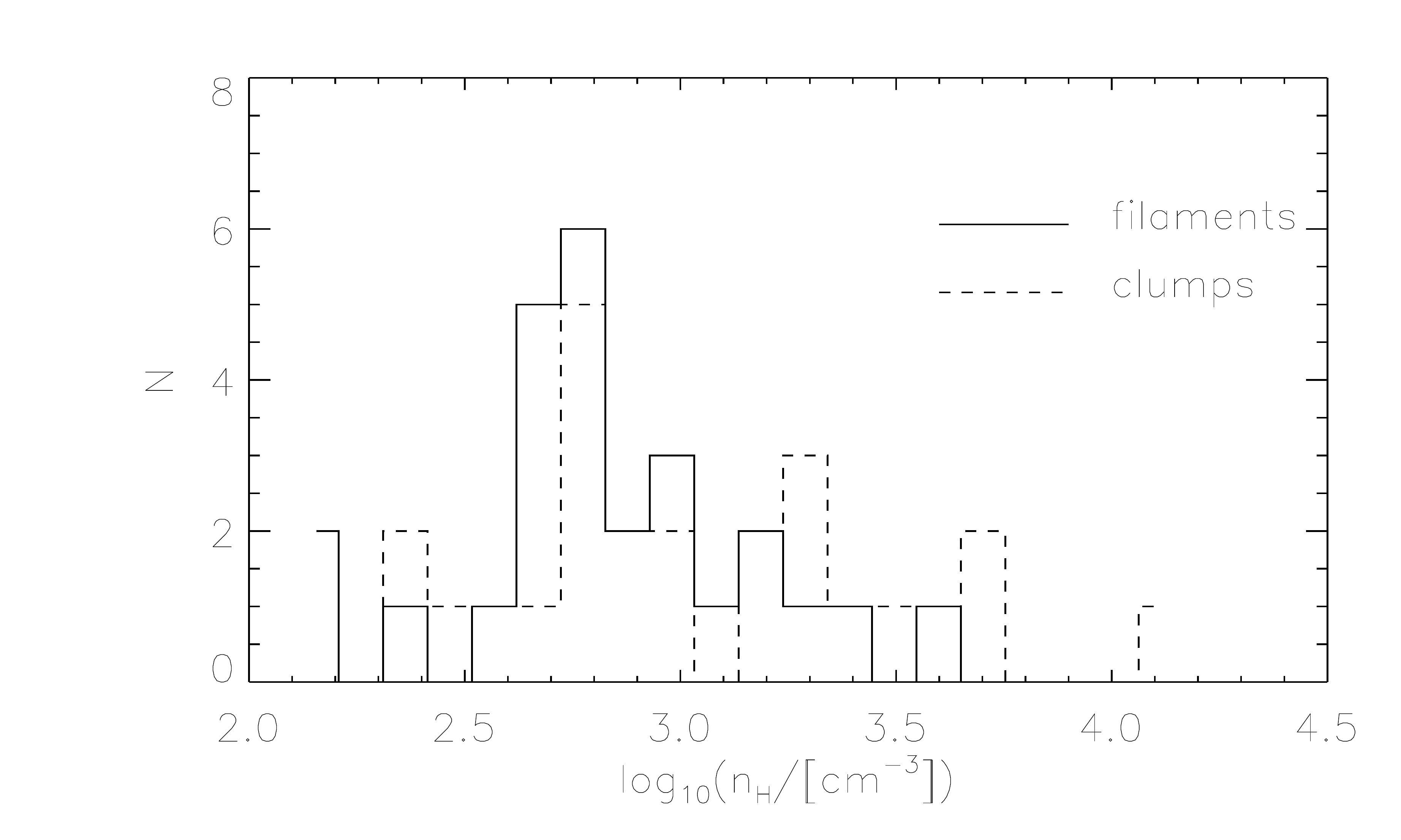}
\caption{Histograms of the estimated mean gas volume densities for clumps(dashed line) and filaments(solid line) for PGCCs with distances lower than $500$ pc.}
\label{fig:nh_vol_hist}

\end{figure}

%%%%%%%%%%%%%%%%%%%%%%%%%%%%%%%%%%%%%%%%%%%%%%%%%%

% Don't change these lines
\bsp	% typesetting comment
\label{lastpage}

\end{document}